\documentclass[11pt,a4paper]{article}
\pdfoutput=1

\usepackage{jcappub}

\usepackage{caption}
\usepackage{subcaption}
\usepackage{rotating}
\usepackage{color}
\usepackage{bm}
\usepackage{hyperref} 
\usepackage{url}
\usepackage{float}
\usepackage{lscape}
\usepackage{geometry}
\usepackage{amsmath}
\usepackage{color,soul}
\usepackage{longtable}
\usepackage[table]{xcolor}
\usepackage{verbatim}
\usepackage{longtable}
\usepackage{fontawesome}
\usepackage[frozencache,cachedir=minted-cache]{minted}

\newcommand{\be}{\begin{equation}}
\newcommand{\ee}{\end{equation}}

\numberwithin{equation}{section}

\definecolor{rossos}{rgb}{0.7,0,0.3}
\definecolor{violachiaro}{rgb}{1,0.6,1}
\definecolor{rossochiaro}{rgb}{1,0.6,0.6}
\definecolor{verdechiaro}{rgb}{0.6,1,0.6}
\definecolor{giallochiaro}{rgb}{1,1,0.6}
\definecolor{bluscuro}{rgb}{0.15, 0.2, 0.9}
\definecolor{verdes}{rgb}{0.1, 0.5, 0.1}
\definecolor{gold}{rgb}{1,0.84,0}
\definecolor{forestgreen}{rgb}{0.13,0.55,0.13}
\definecolor{oucrimsonred}{rgb}{0.6, 0.0, 0.0}
\definecolor{persianblue}{rgb}{0.11, 0.22, 0.73}
\definecolor{forestgreen}{rgb}{0.13,0.35,0.13}
 \hypersetup{colorlinks, citecolor=oucrimsonred, linkcolor=persianblue, urlcolor=persianblue}

\begin{document}

%\preprint{YITP-SB-2023-16}

\title{\href{https://github.com/vallima/PRyMordial}{\texttt{PRyMordial}}: The First Three Minutes, Within and Beyond the Standard Model}

\author[a]{Anne-Katherine Burns,}
\author[a]{Tim M.P. Tait,} 
\author[b,c]{Mauro Valli}

\affiliation[a]{\textit{Department of Physics and Astronomy, University of California, Irvine, CA 92697 USA}}
\affiliation[b]{\textit{C.N. Yang Institute for Theoretical Physics, Stony Brook University, Stony Brook, NY 11794 USA}}
\affiliation[c]{\textit{INFN Sezione di Roma, Piazzale Aldo Moro 2, I-00185 Rome, Italy}}

\emailAdd{annekatb@uci.edu, \ ttait@uci.edu, \ mauro.valli@roma1.infn.it}

\abstract{In this work we present \texttt{PRyMordial}: A package dedicated to efficient computations of  observables in the Early Universe with the focus on the cosmological era of Big Bang Nucleosynthesis (BBN). The code offers fast and precise evaluation of BBN light-element abundances together with the effective number of relativistic degrees of freedom, including non-instantaneous decoupling effects. \texttt{PRyMordial} is suitable for state-of-the-art analyses in the Standard Model as well as for general investigations into New Physics active during BBN. 
After reviewing the physics implemented in \texttt{PRyMordial}, we provide a short guide on how to use the code for applications in the Standard Model and beyond. The package is written in Python, but more advanced users can optionally take advantage of the open-source community for Julia. \texttt{PRyMordial} is publicly available on \faGithub \href{https://github.com/vallima/PRyMordial}{\,GitHub}.
}
	 

\maketitle

%------------------------------------------------------------------------------

\section{Introduction}\label{sec:Intro}

The snapshot of the Universe approximately three minutes after the Big Bang~\cite{Weinberg:1977ji} can be regarded as one of the most remarkable predictions of the Standard Model (SM) of Particle Physics in conjunction with the (so-called) concordance model of Cosmology, $\Lambda$CDM. 

While a theory for the origin of chemical elements based on an epoch of high-energy densities and pressures was already formulated by Alpher, Bethe, and Gamow more than seventy years ago~\cite{PhysRev.73.803}, the discovery of the quasi-black body spectrum of the Cosmic Microwave Background (CMB)~\cite{1965ApJ...142..419P,1965ApJ...142..414D} paved the road for the modern formulation of the theory of Big Bang Nucleosynthesis (BBN)~\cite{doi:10.1146/annurev.ns.27.120177.000345}. Indeed, thanks to the CMB, we know today that the SM particle species were in a thermal state during an epoch dominated by radiation. Extrapolating this cosmological picture back in time when the Universe was not yet transparent to light, within the standard lore of Cosmology and of Particle Physics we can accurately predict~\cite{Sarkar:1995dd,Olive:1999ij,Steigman:2007xt,Pospelov:2010hj,Cyburt:2015mya,Grohs:2023voo}: 
\begin{itemize}
\item[\textit{1)}] The evolution of the number of relativistic degrees of freedom until recombination, $N_{\rm eff}$; 
\item[\textit{2)}] The cosmological abundance of light nuclides synthesized from protons and neutrons, as a function of the number density of baryons relative to photons, $\eta_{B} \equiv n_{B}/n_{\gamma}$.
\end{itemize}
Regarding \textit{1)}, given the current knowledge of neutrino oscillations~\cite{Esteban:2020cvm}, $N_{\rm eff}$ is predicted in the SM via solving a set of integro-differential equations for the neutrino density matrix at finite temperature~\cite{Akita:2022hlx}, yielding $N^{\rm SM}_{\, \rm eff} = 3.044$ with an error estimated to be below the level of per mil~\cite{Bennett:2020zkv,Froustey:2020mcq,Akita20}. 

Concerning \textit{2)}, a detailed analysis of CMB anisotropies in temperature and polarization currently constrains $\eta_{B}$ with 1\% accuracy or better~\cite{Ade16}, anchoring the primordial asymmetry between baryons and anti-baryons to be $\mathcal{O}(10^{-10})$~\cite{Canetti:2012zc}. Assuming no large asymmetry in the lepton sector as well, see e.g.~\cite{Serpico:2005bc}, standard BBN turns into an extremely predictive theory, often dubbed ``parameter free''. 

On the observational side, multi-wavelength astronomical campaigns have been able to provide rich spectroscopic information about emission and absorption lines of gas clouds in metal-poor extra-galactic environments, see e.g.~\cite{RiemerSorensen:2017vx,Cooke:2017cwo,2020ApJ89677H,Aver:2020fon}, bringing us today to a percent-level determination of the abundance of  primordial deuterium and helium-4. 
Given the predictions of the standard theory and the precision of those measurements, together with the strong constraints on the thermal history provided by the CMB~\cite{Planck20,Yeh:2022heq}, the study of the Early Universe around the BBN epoch offers unique insight on New Physics (NP)~\cite{Boehm13,Hardy:2014mqa,Alvey:2019ctk,Sibiryakov:2020eir,Mahbubani:2020knq,Sabti:2019mhn,Depta:2020zbh,Giovanetti21,Chu:2022xuh,Burns:2022hkq}. 
Looking at the exciting prospects of next-gen CMB experiments~\cite{2019arXiv190704473A,2019BAAS51g147L,2019BAAS51g6S}, and at the expected future sensitivity in the field of observational astronomy~\cite{Grohs:2019cae,Lague:2019yvs}, it is therefore very timely to have tools at our disposal that allow for numerically efficient, yet precise computations that test the SM in the Early Universe, and that are flexible enough to broadly explore NP scenarios.

A few packages have already been developed to accurately investigate the BBN era. 
A publicly available version of the historical code of Ref.~\cite{Wagoner:1966pv} (whose most up-to-date version is currently adopted by the PDG~\cite{PDG:2022pth}) is described in~\cite{1992STIN9225163K}. At the same time, publicly released codes dedicated to state-of-the-art BBN analyses are also available; in particular:

\begin{itemize}
    \item \href{https://parthenope.na.infn.it}{\texttt{PArthENoPE}}~\cite{Pisanti:2007hk,Consiglio:2017pot,Gariazzo:2021iiu} is a code originally written in FORTRAN~77 that in its latest re-incarnation also enjoys a graphical user interface; it offers a very efficient evaluation of BBN light-element abundances based on fitting formulae worked out for both weak rates and nuclear cross sections.
    \item \href{http://www2.iap.fr/users/pitrou/primat.htm}{\texttt{PRIMAT}}~\cite{Pitrou:2018cgg,Pitrou:2020etk} is an user-friendly Mathematica package containing all the inputs and ingredients for an ab-initio computation of neutron freeze-out and of weak rates; moreover, it has tabulated the largest nuclear network at hand in order to track the abundance of heavy nuclides as well.
\end{itemize}

Both codes include a few built-in options to account for the study of some specific NP scenarios.  \href{https://alterbbn.hepforge.org}{\texttt{AlterBBN}}~\cite{Arbey:2011nf,Arbey:2018zfh} is a C++ open-source software developed for broad investigation of Physics Beyond the SM (BSM) in the BBN era. However, while allowing for fast numerical evaluations, {\texttt{AlterBBN}} does not implement the same level of detail and accuracy in its computation of light primordial abundances compared to \texttt{PArthENoPE} or \texttt{PRIMAT}. In fact, these two packages may currently represent the best tools to perform precision cosmological analyses~\cite{Planck20,ACT:2023kun}. 

While powerful and flexible, these public codes nevertheless suffer from a few
limitations and/or missing features. A precision tool for Cosmology, able to handle BSM Particle Physics should:
\begin{itemize}
\item[-] Allow for the evaluation of the physics of the thermal bath in a fast but precise way, following, e.g., the approach highlighted in~\cite{Escudero19,Escudero20,Chu:2022xuh}, and
implemented in the standalone code ~\href{https://github.com/MiguelEA/nudec_BSM}{\texttt{NUDEC\_BSM}}; 
\item[-] Interconnect a first-principle computation of the thermal background with an ab-initio precise calculation of the neutron-to-proton ($n \leftrightarrow p$) conversion, as the one implemented in {\texttt{PRIMAT}}~\cite{Pitrou:2018cgg};
\item[-] Render easily accessible exploration of the impact of the input parameters characterizing the BBN era and the uncertainties in the set of thermonuclear rates on the basis of more model-dependent/more data-driven approaches available in literature, see~\cite{Pitrou:2021vqr,Sabti:2021reh,Burns:2022hkq};
\item[-] Adopt a user-friendly, modern programming language compatible with numerical efficiency of the computations, while smoothly interfacing with standard libraries for statistically advanced analyses like Monte Carlo (MC) ones, see  e.g.~\cite{2013PASP..125..306F,Schulz:2021BAT}.
\end{itemize}

In this work, we introduce \href{https://github.com/vallima/PRyMordial}{\texttt{PRyMordial}}: A new public tool for the community of Particle Physics and Cosmology that precisely aims at filling in the above gaps for precision studies on the physics of the Early Universe both within and beyond the SM. 
The package is written and runs entirely with Python~3. Moreover, for the most advanced users, the resolution of the set of stiff differential equations for the BBN nuclear-reaction network can  be further optimized with the optional switch to some routines of the \href{https://sciml.ai}{SciML} kit~\cite{rackauckas2017differentialequations}, the open-source software for scientific machine learning in Julia. 

Our article is organized as follows: In \autoref{sec:PhysPRyM} we present all the key ingredients of the physics implemented in \texttt{PRyMordial}; In \autoref{sec:HowPRyM} we discuss in detail how \texttt{PRyMordial} is structured and we provide several examples on the usage of the code; In \autoref{sec:Concl} we comment on future directions for further development of \texttt{PRyMordial} along with possible interesting applications. We finally collect in \autoref{app:InstallPRyM} a set of instructions for the installation of the package and its dependencies.

\section{Physics in \href{https://github.com/vallima/PRyMordial}{\texttt{PRyMordial}}}\label{sec:PhysPRyM}

In this section we present the key equations present in {\tt PRyMordial}, which stand out as a reference for the physics implemented within the code as well as representing a guideline regarding its use (see \autoref{sec:HowPRyM}). We organize the presentation in three distinct topics: the thermodynamics of the plasma; the weak rates for $n \leftrightarrow p$ conversion; and the set of thermonuclear rates for the key reactions responsible of the non-zero primordial abundance of deuterium, helium-3 and -4, and lithium-7.

\subsection{Thermodynamics with Non-instantaneous Decoupling}\label{sec:thermodyn}

The description of the thermal background during the BBN era in $\Lambda$CDM follows from an isotropic, homogeneous Universe  modelled by the Einstein field equation:

\begin{equation}
\label{eq:Fried}
H^2 \equiv \left(\frac{d \log{a}}{dt} \right)^2 = \frac{8 \pi}{3 M^2_{\rm Pl}} \,  \rho_{\rm tot} \ , 
\end{equation}
where $H$ is the Hubble rate of space-time expansion, $a$ the scale factor of the Friedmann–Lema\^itre– Robertson–Walker metric, $\rho_{\rm tot}$ the total energy density present in the Universe, and $M_{\rm Pl} \equiv 1/\sqrt{G_{\rm N}}$, with $G_{\rm N}$ the Newton gravitational constant.

Within an axiomatic characterization of the Early Universe provided by \textit{local thermodynamic equilibrium}~\cite{Kolb:1990vq,Rubakov:2017xzr}, SM species are described according to the spin-statistics theorem and the temperature $T_{\gamma}$ of the thermal bath (provided chemical potentials $\mu$ can be neglected, i.e., $\mu/T_{\gamma} \ll 1$). Standard BBN takes place during radiation domination, and thus features contributions to $\rho_{\rm tot}$ largely from relativistic species, i.e. $\rho_{\rm tot} \simeq \rho_{\rm rad} \propto T_{\gamma}^4$. This observation dramatically simplifies the investigation of BBN, allowing one to decouple the study of the thermal background from the nucleon dynamics. Indeed, after the QCD crossover takes place~\cite{Borsanyi:2016ksw} protons and neutrons are already non-relativistic, i.e. they are highly Boltzmann-suppressed well before the MeV scale temperatures characteristic of the BBN era.

Hence, for temperatures $T_{\gamma} \lesssim \mathcal{O}(10)$~MeV, one can accurately describe $\rho_{\rm tot}$ in the SM as a sum of just three contributions:
\begin{equation}
\label{eq:rhotot}
\rho_{\gamma} = \frac{\pi^{2}}{15}  \, T_{\gamma}^4 \ \ , \ \ \rho_{e^\pm} =  \frac{2}{\pi^{2}} T^{4}_{\gamma}  \, \int_{x_{e}}^{\infty} d \tilde{x} \,  \frac{\tilde{x}^{2} \sqrt{\tilde{x}^2-x_e^2}}{\exp(\tilde{x}+1)}  \ \ , \ \ \rho_{\nu, \rm tot} = 3 \times \frac{7 \pi^{2}}{120}  \, T_{\nu}^4 \ ,
\end{equation}
where $x_e \equiv m_e/T_{\gamma}$ and we distinguish the temperature of the electron-positron-photon system, $T_\gamma$, from that of neutrinos, $T_\nu$.\footnote{While $T_{e} = T_{\gamma}$ follows from $e^{\pm}$ being tightly coupled to photons via fast QED processes, the approximation underlying  $T_{\nu}$, namely $ T_{\nu_{e}} \simeq T_{\nu_{\mu}} \simeq T_{\nu_{\tau}}$, can be motivated by the active flavor mixing of $\nu$ oscillations at $T_{\gamma}$ of few MeV~\cite{Dolgov:2002ab}.} Indeed, while the initial condition $T_\nu = T_\gamma$ must hold at early times for the two systems to be in thermal (more precisely, in chemical and kinetic) equilibrium, around the MeV scale neutrinos are expected to freeze out from the thermal bath as weakly-interacting relativistic species~\cite{Dolgov:2002wy}. Neglecting tiny departures from a Fermi-Dirac distribution in $\nu$ phase space, one can study the evolution of the two systems according to the momentum-integrated Boltzmann equations:
\begin{eqnarray}
\label{eq:SMBoltzsys}
(\rho_{\gamma}^{\prime}+\rho_{e^{\pm}}^{\prime})\, \frac{dT_{\gamma}}{dt} & = &  - 4 H \, \rho_{\gamma} - 3 H (\rho_{e^{\pm}} + p_{e^{\pm}})+ \delta C_{e^{\pm}}   \ , \\ \nonumber
\rho_{\nu, \rm tot}^{\prime} \, \frac{dT_{\nu}}{dt} & = & - 4 \, H \, \rho_{\nu, \rm tot} + \delta C_{\nu} \ ,
\end{eqnarray}
with $^\prime \equiv d/dT$, $p$ the pressure density (equal to $\rho/3$ for a relativistic species), $\delta C$ the (momentum-integrated) collision term, and where we have conveniently traded energy densities for temperatures in light of Eq.~\eqref{eq:rhotot}. Due to energy-momentum conservation, the sum over all $\delta C$s must vanish, so that one recovers the continuity equation for the total energy density of the Universe:
\begin{equation}
\label{eq:TmunuConserv}
\frac{d \rho_{\rm tot}}{dt} + 3 H (\rho_{\rm tot} + p_{\rm tot}) = 0 \ .
\end{equation}
In the SM, where Eq.~\eqref{eq:SMBoltzsys} holds, such a constraint implies: $\delta C_{\nu} = -\delta C_{e^{\pm}}$. The collision term $\delta C_{\nu}$ has been evaluated in~\cite{Dolgov:2002wy} under Maxwell-Boltzmann approximation, nicely refined in~\cite{Escudero19,Escudero20} taking into account relativistic corrections as well as finite mass effects from $m_{e} \neq 0$, and more recently re-computed independently in~\cite{Chu:2022xuh}. Including finite temperature QED corrections to the electromagnetic plasma~\cite{Bennett:2019ewm}, one can solve the system of coupled differential equations in Eq.~\eqref{eq:SMBoltzsys}, to find $T_{\gamma}(t)$, $T_{\nu}(t)$, and, as a byproduct, $T_{\nu}(T_{\gamma})$.\footnote{In the current version of \texttt{PRyMordial} we adopt the computation of $\delta C_{\nu}$ as well as the next-to-leading (NLO) QED corrections to the electromagnetic pressure of the plasma directly from the numerical results tabulated in \texttt{NUDEC\_BSM}~\cite{Escudero20}.} Such a treatment naturally includes non-instantaneous decoupling effects, and allows one to perform a numerically fast, but accurate prediction of the effective number of relativistic degrees of freedom from first principles, yielding (in the SM) at $T_{\gamma} \ll$ MeV:
\begin{equation}
\label{eq:Neff}
N_{\rm eff} \equiv \frac{8}{7} \left(\frac{11}{4} \right)^{4/3}\left(\frac{\rho_{\rm rad}-\rho_{\gamma}}{\rho_{\gamma}}\right) = 3.044 \ ,
\end{equation}
while also opening up novel explorations of BSM physics in the Early Universe~\cite{Escudero20,Giovanetti21,Chu:2022xuh}.\footnote{
Eq.~\eqref{eq:SMBoltzsys} can be easily generalized to include new sectors.
This contrasts with typical existing BBN codes which compute the thermodynamic background by interpolating the
tabulated result of the (numerically intensive) integro-differential Boltzmann equation, solved for the neutrino phase-space density in the SM.}

Based on these results, one can also easily evaluate the relic density of neutrinos (neglecting phase space spectral distortions). From the CMB we know the photon temperature today is $T_{\gamma,0} = 0.2348$~meV; plugging this value into the solution of Eq.~\eqref{eq:SMBoltzsys} yields the temperature $T_{\nu,0} = 0.1682$~meV, corresponding to the cosmological abundance of SM neutrinos: 
\begin{eqnarray}
\label{eq:OmegaNuh2}
\Omega^{\rm(rel)}_{\nu} h^2 & = &  \left(\frac{7 \pi^{2}}{120} \, T_{\nu,0}^4 \right) \Big{/}\left(\frac{3}{8 \pi} \frac{M_{\rm Pl}^2 H_{0}^2}{h^2} \right) = 5.70 \times 10^{-6} \ , \ \\ 
\Omega^{\rm(nr)}_{\nu} h^2 & = & \left(\frac{3}{2} \frac{\zeta(3)}{\pi^2} T_{\nu,0}^3 \, \sum_{i} m_{{\nu}_{i}} \right) \Bigg{/}\left(\frac{3}{8\pi} \frac{M_{\rm Pl}^2 H_{0}^2}{h^2} \right) = \sum_{i}     \frac{m_{{\nu}_{i}}}{93.03 \, {\rm eV}} \ , \nonumber
\end{eqnarray}
which reproduces the relic neutrino abundance computed, e.g., in Ref.~\cite{Mangano:2005cc} to the per mil level.

In order to obtain $T_\gamma(t)$ and $T_\nu(t)$ from Eq.~\eqref{eq:SMBoltzsys}, we have made use both of Eq.~\eqref{eq:Fried} together with Eq.~\eqref{eq:rhotot}. At this point, to complete the study of the thermodynamic background, we must extract the scale factor $a$ as a function of time $t$ and temperature $T_{\gamma}$. This can be accomplished by applying (again) the notion of local thermodynamic equilibrium, which allows one to introduce the entropy density for each species $i$ as: $s_{i} = (\rho_{i} + p_{i} - \mu_{i} \, n_{i})/T_{i}$, where $n_{i}$ is the number density of the species with associated  chemical potential $\mu_{i}$.

For negligible chemical potentials, the total entropy density of the Universe $s_{tot}$ per comoving volume must be conserved as a consequence of energy-momentum conservation, Eq.~\eqref{eq:TmunuConserv}. Then, during radiation domination $s_{tot}$ roughly scales as $T_{\gamma}^3$, underlying the approximate relation $a \propto 1/T_{\gamma}$.
Nevertheless, even under the assumption of $\mu_{i}/T_{i} \ll 1 $, the entropy of each species is generally not separately conserved due to heat exchanges related to the interactions with other species. The Boltzmann equation for $s_{i}$ generally follows~(see, e.g., the discussion in Refs.~\cite{Grohs16,Pitrou:2018cgg}):
\begin{equation}
\label{eq:SMentropies}
\frac{d s_{i}}{dt} + 3 H s_{i}  = \frac{\delta C_{i}}{T_{i}} -\frac{\mu_{i}}{T_{i}} \left(\frac{d n_{i}}{dt} + 3 H n_{i} \right)\ ,
\end{equation}
where the first collision term (divided by the temperature) is the one appearing in the Boltzmann equation for the density $\rho_{i}$, while the second collision term has been rewritten using the Boltzmann equation for the number density $n_{i}$.\footnote{Notice that in absence of interactions for the species $i$, entropy conservation can be guaranteed either by a negligible chemical potential, $\mu_{i} \ll T_{i}$ or by number density conservation per comoving volume, $d(n_{i} a^3)/dt = 0$.}  In the SM, in the limit\footnote{$\mu_{e}/T_{\gamma} \ll 1 $ is justified in the SM by $\eta_{B}\sim \mathcal{O}(10^{-10})$ and the condition of electric charge neutrality in the Early Universe.} $\mu_{e}/T_{\gamma} \ll 1 $, we use Eq.~\eqref{eq:SMentropies} for the electromagnetic bath to pin down the relation between $a$ and $T_{\gamma}$; with $\bar{s}_{\rm pl} \equiv (s_{\gamma} + s_{e^{\pm}})/T^3_{\gamma}$, we get:
\begin{equation}
    \label{eq:aofT}
    \frac{1}{(T_{\gamma}a)^3}\frac{d\left(\bar{s}_{\rm pl} T_{\gamma}^3 a^3\right)}{d \ln a} = - \frac{\delta C_{\nu}}{H T^4_{\gamma}} \equiv - \mathcal{N}_{\nu}\ \ \Leftrightarrow \ \ a(T_{\gamma}) = a_{0} \exp \left(- \int_{T_{\gamma,0}}^{T_{\gamma}} \, \frac{d T}{T} \frac{3  \bar{s}_{\rm pl} + T \, \bar{s}_{\rm pl}^{\,\prime}}{3 \bar{s}_{\rm pl}+ \mathcal{N}_{\nu}}\right) \ .
\end{equation}
Knowing all the thermodynamical quantities as a function of $T_{\gamma}$ in the integrand above, Eq.~\eqref{eq:aofT} allows one to extract $a(T_{\gamma})$ up to the scale-factor value of today, $a_{0}$, customarily defined as 1. Note that for $T_{\gamma} \lesssim m_{e}$ one has $\bar{s}_{\rm pl}^{\,\prime} = 0$, and taking the limit $\mathcal{N}_{\nu} \to 0$, the expected scaling set by $d(s_{\gamma} a^3)/dt = 0$ is easily recovered.
The solution in Eq.~\eqref{eq:aofT} precisely tracks the relation between the scale factor and $T_{\gamma}$ in the case of non-instantaneous decoupling of neutrinos. While in the SM these effects are tiny (since $\mathcal{N}_{\nu}/3 \ll \bar{s}_{\rm pl}$), they could become non-negligible in a BSM scenario.

It is worth noting that given $T_{\gamma}(t)$ from the solution of Eq.~\eqref{eq:SMBoltzsys} and $a(T_{\gamma})$ from Eq.~\eqref{eq:aofT}, one obtains $a(t)$ as a byproduct, which allows to assess the evolution of the number density of baryons in $t$ or $T_{\gamma}$ during the BBN era, since by definition: $n_{B} \propto 1/a^{3}$. 
%Baryons will be discussed next.

\subsection{Neutron Freeze Out beyond the Born Approximation}\label{sec:nTOp}

Shortly after hadrons form, neutrons and protons are non-relativistic species that do not contribute appreciably to the total energy budget stored in the thermal bath. Nevertheless, their abundance is eventually responsible for the tiny fraction of light primordial elements relative to hydrogen which are observable today in pristine astrophysical environments.

According to local thermodynamic equilibrium, the relative number density of nucleons is initially given by the Maxwell-Boltzmann distribution:
\begin{equation}
    \label{eq:nTOpratio}
    \left(\frac{n_{\rm n}}{n_{\rm p}}\right){\Big|}_{T_{\gamma} \, \gg {\rm \, MeV}} = \left(\frac{m_{\rm n}}{m_{\rm p}}\right)^{3/2}\exp\left(-\frac{\mathcal{Q}}{T_{\gamma}}-\frac{\mu_{\mathcal{Q}}}{T_{\nu}}\right) \,,
\end{equation}
where $\mathcal{Q} = m_{\rm n} -m_{\rm p}$, $\mu_{\mathcal{Q}} = \mu_{\rm n}-\mu_{\rm p}$, $m_{\rm n,p}$ and $\mu_{\rm n,p}$ are the mass and chemical potential of neutrons and protons. 
For clarity, we have used $T_{\nu} = T_{\gamma}$ (valid for temperatures well above MeV) in the $\mathcal{Q}$ term, but retain $T_{\nu}$ explicitly in the $\mu_{\mathcal{Q}}$ term.
Assuming $\mu_{\rm n} \simeq \mu_{\rm p}$ (e.g. a negligible contribution from lepton chemical potentials), Eq.~\eqref{eq:nTOpratio} implies that at equilibrium $n_{\rm n} \simeq n_{\rm p}$. Indeed, fast electroweak processes efficiently convert $n \leftrightarrow  p$:
\begin{eqnarray*}
     \Gamma_{\rm n \, \to \, \rm p} & \equiv & \Gamma({n \, e^{+} \,\to \, p \, \bar{\nu}}) + 
    \Gamma(n \, \bar{\nu} \, \to \,  p \, e^{-}) +
    \Gamma(n  \, \to \,  p \, e^{-} \, \bar{\nu}) 
    \gg H  \, , \\ 
    \Gamma_{\rm p \, \to \, \rm n} & \equiv & \Gamma(p \, e^{-} \,\to \, n \, \bar{\nu}) + 
    \Gamma(p \, \bar{\nu} \, \to \,  n \, e^{+}) +
    \Gamma(p \, e^{-} \, \bar{\nu} \, \to \, n)
    \gg H \, , 
\end{eqnarray*}
and govern the Boltzmann equations for the nucleon yields $Y_{\rm n,p} \equiv n_{\rm n,p} \, / \, n_{B} = n_{\rm n,p} \, / \, (n_{\rm n}+n_{\rm p})$:
\begin{eqnarray}
\label{eq:nTop_rates}
\frac{d Y_{\rm n}}{dt} & = & \Gamma_{\rm p \, \to \, \rm n} \, Y_{\rm p} -  \Gamma_{\rm n \, \to \, \rm p} \, Y_{\rm n} \ , \\
\frac{d Y_{\rm p}}{dt} & = &  \Gamma_{\rm n \, \to \, \rm p} \, Y_{\rm n} - \Gamma_{\rm \rm p \, \to \, \rm n} \, Y_{\rm p} \ , \nonumber
\end{eqnarray}
whose equilibrium solutions:  $Y_{\rm n} = 1-Y_{\rm p} = \Gamma_{\rm p \, \to \, \rm n}/(\Gamma_{\rm p \, \to \, \rm n} + \Gamma_{\rm n \, \to \,\rm p}) \simeq 1/2$, are in agreement with Eq.~\eqref{eq:nTOpratio}. These reactions guarantee chemical equilibrium among the involved species, implying $\mu_{\mathcal{Q}} \simeq -\mu_{\nu}$.  Eq.~\eqref{eq:nTOpratio} thus demonstrates that a primordial non-zero lepton asymmetry in the neutrino sector~\cite{March-Russell:1999hpw,Kawasaki:2022hvx} can impact the initial conditions for BBN by altering the neutron-to-proton ratio, with notable cosmological consequences~\cite{Burns:2022hkq,Escudero:2022okz}. 

At temperatures close to neutrino decoupling, $n \leftrightarrow p $ conversion falls out of equilibrium, freezing out the neutron-to-proton ratio to $\sim 1/6$ (in the SM), up to finite neutron lifetime effects~\cite{Kolb:1990vq,Rubakov:2017xzr}. The weak rates for  neutron freeze out require the evaluation of an involved multi-dimensional phase-space integral: e.g. for $n \, e^{+} \to p \, \bar{\nu}$ (and similarly for the others)~\cite{Lopez:1997ki}:
\begin{equation}
\label{eq:GammaWeak}
Y_{\rm n} \, \Gamma({n \, e^{+} \,\to \, p \, \bar{\nu}})  = \frac{16 \pi^4}{n_{B}} \int d \Pi_{\rm n} d \Pi_{e} d \Pi_{\rm p} d \Pi_{\nu} \, \delta^{(4)}(P_{\rm n}+P_{e}-P_{\rm p}-P_{\nu}) \, |{\mathcal{M}}|^2 \, f_{\rm n} f_{e} (1-f_{\rm p})(1-f_{\nu}),
\end{equation}
where $ d \Pi_{i}$ and $P_{i}$ are the Lorentz-invariant phase-space element and 4-momentum of the particle $i$, $f_{i}$ is the relativistic thermal distribution of the species $i$ in the rest frame of the thermal bath, and $\mathcal{M}$ is the full matrix element of the process summed over initial and final spins. The latter can be computed from the weak effective theory for $\beta$ decay~\cite{Donoghue:1992dd}:
\begin{equation}
    \mathcal{L}_{\rm F} = - \frac{2G_{\rm F}}{\sqrt{2}} \, V_{\rm ud} \, \, \bar{\nu}(x) \, \gamma_{\mu}\,e_{L}(x) \, 
    \left\{\,\bar{n}(x) \gamma^{\mu}(1 - g_{\rm A} \, \gamma_{5})p(x) + \frac{\kappa}{2 m_{\rm N}} \partial_\nu \left[ \bar{n}(x) \, \sigma^{\mu \nu} \, p(x) \right] \,\right\} + h.c. \, ,
\end{equation}
where $G_{\rm F}$ is the Fermi constant~\cite{PDG:2022pth}, $V_{\rm ud}$ corresponds to the Cabibbo angle~\cite{UTfit:2022hsi}, $g_{\rm A}$ and $\kappa$ are the axial-current and weak-magnetism constant of the nucleon of mass $m_{\rm N}$~\cite{PhysRevD.88.073002}, and $\sigma_{\mu \nu} \equiv i \, (\gamma_{\mu}\gamma_{\nu}-\gamma_{\nu}\gamma_{\mu})/2$. The computation of $|\mathcal{M}|^2$ can be found in detail in Appendix~B of Ref.~\cite{Pitrou:2018cgg} (see also~\cite{Seckel:1993dc,Lopez:1997ki}).

While expressions like Eq.~\eqref{eq:GammaWeak} can be reduced to a five-dimensional integral in phase space by exploiting the symmetries of the problem, a dramatic simplification is obtained in the limit of infinite nucleon-mass at fixed $\mathcal{Q}$~\cite{Seckel:1993dc,Lopez:1997ki}. This is the so-called Born approximation, in which the kinetic energy of the `infinitely' heavy neutrons and protons may be neglected, leading to the simplification: $|\mathcal{M}|^2= 32 \, G_{\rm F}^2V_{\rm ud}^2(1+3g_{\rm A}^2)E_e E_\nu E_{\rm p} E_{\rm n}\,$. In that limit the $n \leftrightarrow p$ rates read:
\begin{eqnarray}
    \label{eq:GammaBorn}
    \Gamma_{\rm n \to p}^{\infty} & = & \widetilde{G}_{\rm F}^2 \int_{0}^{\infty} dE_e \,E_e \, \sqrt{E_e^2-m_e^2} \, (E_\nu^-)^2 \left[ f_{\nu}(E_\nu^-)f_{e}(-E_e)+
    f_{\nu}(-E_\nu^-)f_{e}(E_e)\right] \ , \\
    \Gamma_{\rm p \to n}^{\infty} & = & \widetilde{G}_{\rm F}^2 \int_{0}^{\infty} dE_e \,E_e \, \sqrt{E_e^2-m_e^2} \, (E_\nu^+)^2 \left[ f_{\nu}(E_\nu^+)f_{e}(-E_e)+
    f_{\nu}(-E_\nu^+)f_{e}(E_e)\right] \ , \nonumber
\end{eqnarray}
where $\widetilde{G}_{\rm F} \equiv G_{\rm F}V_{\rm ud}\sqrt{(1+3 g_{\rm A}^2)/(2\pi^3)}$ and $E_\nu^\pm = E_{e} \pm \mathcal{Q}$. The outcome of Eq.~\eqref{eq:GammaBorn} are rates that generally depend on both background temperatures and chemical potentials (i.e. $T_{\gamma},T_{\nu}$ and $\mu_{\nu}$). For $T_{\nu} = T_{\gamma}$ (and negligible chemical potentials) detailed balance follows as: $\Gamma_{\rm p \to n}^{\infty}/\Gamma_{\rm n \to p}^{\infty} = \exp(-\mathcal{Q}/T_{\gamma})$.
The dimensionful factor $\widetilde{G}_{\rm F}$ depends on $V_{\rm ud}$, $g_{\rm A}$, and $G_{\rm F}$, whose value is precisely determined by the muon lifetime. However, this factor is often more conveniently extracted from neutron decay in the vacuum, since in the SM: 
\begin{equation}
\label{eq:invtaun}
\tau_{\rm n}^{-1} = \widetilde{G}_{\rm F}^2 \,  m_{e}^5  \, \mathcal{F}_{\rm n} \ , 
\end{equation} 
where $\mathcal{F}_{\rm n}$ incorporates a phase-space statistical factor for the neutron decay at zero temperature~\cite{Wilkinson:1982hu} plus electroweak radiative corrections~\cite{Marciano:2005ec}. For a precise calculation of $\mathcal{F}_{\rm n}$, see the very recent reassessment in Ref.~\cite{Cirigliano:2023fnz} and references therein. This approach allows one to trade the combination $V_{ud}^2(1+3g_A^2)$ for the measured $\tau_{n}.$\footnote{Any treatment must confront both the \textit{neutron lifetime puzzle}, i.e. the tension between ``bottle''~\cite{UCNt:2021pcg} and ``beam''~\cite{Yue:2013qrc} measurements of $\tau_{\rm n}$, see, e.g.,~\cite{Chowdhury:2022ahn}; and the \textit{Cabibbo angle anomaly}~\cite{Cirigliano:2022yyo}, i.e. the extraction of $V_{\rm ud}$ from super-allowed $\beta$ decays and $V_{\rm us}$ from semi-leptonic decays versus unitarity in the Cabibbo-Kobayashi-Maskawa matrix~\cite{UTfit:2022hsi}.} Using Eq.~\eqref{eq:invtaun}, in \texttt{PRyMordial} one can choose to adopt either a normalization of the weak rates based on the determination of the neutron lifetime, or one involving the knowledge of the modified Fermi constant $\widetilde{G}_{\rm F}$.

In the SM the Born approximation predicts a neutron freeze-out temperature of slightly below 1~MeV. At smaller temperatures, the neutron-to-proton ratio is still affected by $\beta$ decay until the Universe cools down sufficiently enough to preclude photo-dissociation of deuterium: for a binding energy $B_{\rm D} = 2.2$~MeV, this happens at temperatures around $B_{\rm D}/\log(1/\eta_{B})\sim\,$0.1~MeV~\cite{Kolb:1990vq,Rubakov:2017xzr}. At that point, virtually all of the neutrons experience two-body nuclear reactions, ultimately resulting in their binding in helium-4, the most stable light element.  As a result, the uncertainty on the Born-level theory prediction for helium-4 is only a few \%~(see Table~5 in~\cite{Pitrou:2018cgg}). 

That said, the present percent-level inference of primordial helium-4 and deuterium~\cite{PDG:2022pth} and the sub-percent target of future observational campaigns~\cite{Grohs:2019cae} demand the following refinements to Eq.~\eqref{eq:GammaBorn}:
\begin{itemize}
\item QED radiative corrections (in the vacuum) to the $n \leftrightarrow p$ amplitudes of order $\mathcal{O}(\alpha_{\rm em})$  via virtual- and real-photon emission~\cite{Sirlin:1967zza,Abers:1968zz,Dicus:1982bz,Ivanov:2012qe} must be computed;
\item Finite nucleon-mass effects and non-zero weak magnetism, which induce relative shifts in the weak rates of $\Delta \Gamma / \Gamma \sim \mathcal{O}(10^{-2})$ ~\cite{Seckel:1993dc,Lopez:1997ki}, must be taken into account;
\item Finite-temperature effects~\cite{Dicus:1982bz,Brown:2000cp} must be evaluated for sub-percent accuracy.
\end{itemize}
\texttt{PRyMordial} implements all of these corrections, following the treatment in \texttt{PRIMAT} (see Appendix~B of~\cite{Pitrou:2018cgg}), where particular care was taken to attempt to combine several existing state-of-the-art recipes for electroweak rates beyond the Born approximation. 

It is worth noticing that in the context of the SM, the corrections to the Born rates due to the incomplete neutrino decoupling are only marginal~\cite{Grohs:2015tfy,Froustey:2019owm}. Nevertheless, NP could dramatically alter $T_{\nu}(T_{\gamma})$, $a(T_{\gamma})$ and $a(t)$, and the departure from the standard value for the weak rates can impact the final BBN abundances in a non-negligible way~\cite{Sabti:2019mhn,Giovanetti21}. As a result, the approach undertaken in~\autoref{sec:thermodyn} is particularly useful not only for the study of neutrino decoupling, but also for a careful assessment of the neutron-to-proton ratio in BSM scenarios.

\subsection{BBN Thermonuclear Reactions}
\label{sec:NuclearRates}

Local thermodynamical equilibrium implies that at temperatures above neutron decoupling, a nuclear species $i$ of atomic number $Z_{i}$, mass number $A_{i}$, spin $s_{i}$, and binding energy $B_{i}$ follows a Boltzmann distribution with internal degrees of freedom: $g_{i} = 2 s_{i}+1$; mass: $m_{i} = Z_{i} m_{\rm p} + (A_{i}-Z_{i}) m_{\rm n }- B_{i}\,$; and chemical potential: $\mu_{i}= Z_{i} \mu_{\rm p} + (A_{i}-Z_{i}) \mu_{\rm n }$. In terms of the yield $Y_{i} \equiv n_{i}/n_{B}$, this equilibrium distribution reads:
\begin{equation}
\label{eq:Ynucl}
Y_{i}{\big|}_{T_{\gamma}\gtrsim \rm MeV} = g_{i} \, 2^{(3 A_{i}-5)/2} \, \pi^{(1- A_{i})/2} \, \left(\zeta(3)\, \eta_{B}\right)^{A_{i}-1} \left( \frac{m_{i} \, T_{\gamma}^{A_{i}-1}}{m_{\rm p}^{Z_i} m_{\rm n}^{A_i-Z_i}} \right)^{3/2} Y_{\rm p}^{Z_i}Y_{\rm n}^{A_i-Z_i} \exp\left( B_{i}/T_{\gamma}\right) \, ,
\end{equation}
where we made use of the relation: $n_{B}/\eta_{B} = 3 \, \zeta(3)\,T^3_{\gamma}/(2\pi^2)$. This expression holds for the nucleons ($A_{\rm N} = 1$, $B_{\rm N} = 0 $) themselves, and is consistent with Eq.~\eqref{eq:nTOpratio}. Importantly, it offers another handle on the estimate for the start of nucleosynthesis as the time in which the relative abundance of neutrons after freeze out becomes comparable to deuterium as dictated by Eq.~\eqref{eq:Ynucl}, and pointing again to a temperature of about 0.1~MeV. 

Starting from the initial conditions, abundances are determined by a network of Boltzmann equations that generalize  Eq.~\eqref{eq:nTop_rates} (see, e.g., Refs.~ \cite{Fowler:1967ty,Wagoner69}) to include the relevant nuclei:
\begin{equation}
    \label{eq:YiBoltz}
    \frac{dY_{i}}{dt} =  \sum_{R} \mathcal{S}_{i,R}  \left[\, \Gamma^{(R)}_{\dots \, \to \, i \, \dots} \times \prod_{j} \left( \frac{Y_{j}^{\mathcal{S}_{j,R}}}{\mathcal{S}_{j,R}!} \right)
    - \Gamma^{(R)}_{i \, \dots \to \, \dots} \times \prod_{k} \left( \frac{Y_{k}^{\mathcal{S}_{k,R}}}{\mathcal{S}_{k,R}!} \right) \, \right] \, ,
\end{equation}
where the sum $R$ is performed over all reactions involving the nuclear species $i$; $\mathcal{S}_{i,R}$ is the stoichiometric coefficient for the species $i$ in the nuclear reaction $R$; and the products $j$ and $k$ run over all of the initial and final states of the reaction with nuclear rate $\Gamma^{(R)}_{ \dots \to i \, \dots}$ or $\Gamma^{(R)}_{i \, \dots \to \dots}$.

Given the range of energies characterizing the BBN era, the nuclear reaction rates of interest can be measured in the laboratory, and are often tabulated as $\widetilde{\Gamma}_{ i  \dots l\to j \dots m} \equiv N_{A}^{\mathcal{S}_{i} \dots \mathcal{S}_{l}-1} \langle \sigma_{ i  \dots l\to j \dots m} \, v \rangle$~\cite{Angulo:1999zz}, where $N_{A}$ is Avogadro's number (typically expressed in units of mol$^{-1}$), and the velocity averaged cross section is obtained by weighting the appropriate cross section by the Maxwell-Boltzmann velocity distribution for the non-relativistic species (see e.g. Ref.~\cite{Serpico:2004gx} for a detailed description). By definition, for a given number-density rate $\langle \sigma^{(R)}_{i \dots \to \dots} v \rangle$, the corresponding abundance rate $\Gamma^{(R)}_{i \dots \to \dots}$ is:
\begin{equation}
    \Gamma_{ i  \dots l\to j \dots m} = n_{B}^{\mathcal{S}_{i} \dots \mathcal{S}_{l}-1} \langle \sigma_{i  \dots l\to j \dots m} \, v \rangle =  (n_{B}/N_{A})^{\mathcal{S}_{i} \dots \mathcal{S}_{l}-1} \,  \widetilde{\Gamma}_{ i  \dots l\to j \dots m} \ .
\end{equation}

A priori, Eq.~\eqref{eq:YiBoltz} includes the rates of both forward and reverse reactions in the evolution of the abundance of the nuclear species $i$. Nevertheless detailed balance implies
\begin{equation}
   \left(\frac{Y_{j}^{\mathcal{S}_{j}} \dots Y_{m}^{\mathcal{S}_{m}}}{Y_{i}^{\mathcal{S}_{i}} \dots Y_{l}^{\mathcal{S}_{l}}}\right){\Bigg|}_{T_{\gamma}\gtrsim \rm MeV} = \  \frac{\langle \sigma_{i  \dots l\to j \dots m} v \rangle / (\mathcal{S}_{i}! \dots \mathcal{S}_{l}!)}{\langle \sigma_{j  \dots m\to i \dots l} v \rangle /(\mathcal{S}_{j}! \dots \mathcal{S}_{m}!)} \ ,
\end{equation}
since local thermodynamical equilibrium ensures that the forward and reverse reactions should balance.
Thus, it is easy to evaluate the reverse reaction rates given the forward ones.  
It is customary to parameterize the relationship as:
\begin{equation}
    \label{eq:reverseGamma}
    \langle \sigma_{j  \dots m\to i \dots l} v \rangle = \alpha_R \ T_{9}^{\beta_R} \,\exp(\gamma_R / T_{9}) \,  \langle \sigma_{i  \dots l\to j \dots m} v \rangle \ \, , \ \textrm{with:} \ \, T_{9} \equiv T_{\gamma}/ (10^9 \, \textrm{K})  \, ,
\end{equation}
where the constants $\alpha_R$, $\beta_R$, and $\gamma_R$ for a given process $R$ from e.g. the up-to-date nuclear database of Ref.~\cite{Kondev:2021lzi} via Eq.~\eqref{eq:Ynucl}.

\texttt{PRyMordial} solves the general system of equations Eq.~\eqref{eq:YiBoltz} following the strategy of Ref.~\cite{Pitrou:2018cgg} which conveniently breaks nucleosynthesis into three steps: 
\begin{itemize}
    \item[\textit{1)}] We analyze $n \leftrightarrow p$ conversion by solving Eq.~\eqref{eq:nTop_rates} from an initial temperature of $\mathcal{O}(10)$ MeV (and initial conditions from Eq.~\eqref{eq:nTOpratio}) down to standard neutron freeze out, around MeV;
    \item[\textit{2)}] We use the values of $Y_{\rm n,p}$ obtained from \textit{1)} together with Eq.~\eqref{eq:Ynucl} and evolve with a network comprised of the 18 key thermonuclear rates for the abundance of $n,p$ together with all of the nuclides up to $A=8$ and $Z=5$\footnote{In the current version of \texttt{PRyMordial} we include up to boron-8 in the nuclear chains, which is sufficient for an accurate prediction of lithium-7, likely the heaviest nuclide of interest when confronting BBN with observations~\cite{Fields:2022mpw}. For this purpose, the largest implemented set of thermonuclear rates comprises 63 reactions, see \autoref{app:ThermoNuclRates}. } down to the temperature where deuterium photo-dissociation becomes inefficient, around 0.1~MeV;
    \item[\textit{3)}] We further evolve the network with the full set of thermonuclear processes and with initial conditions given by the nuclide yields obtained in step \textit{2)}, evolving the abundances of the aforementioned nuclides down to $\mathcal{O}$(keV) (i.e., well below $e^{\pm}$ annihilation), when BBN is over.
\end{itemize}
The output of Step \textit{3)} is the abundances of the light-element originating from BBN. To compare with data, it is customary to quote helium-4 in terms of the primordial mass fraction:\footnote{Notice that this definition differs at the sub-percent level from the helium mass fraction adopted in the context of the CMB \cite{Planck20}: $Y_{P}^{\rm CMB} \equiv (m_{\rm ^4 He}/4) \, Y_{P}/[(m_{\rm ^4 He}/4) Y_{P} + m_{\rm H} \, (1-Y_{P})] $, with $m_{\rm H, ^4 He}$ the atomic mass of hydrogen and helium. } 
\begin{equation}
\label{eq:YP}
Y_{P} \equiv 4 \times Y_{^4 \rm He}  \simeq \rho_{^4 \rm He}/\rho_{B}  \, .
\end{equation}
The other primordial elements under the lamppost of astrophysical observations are deuterium, helium-3 and lithium-7~(see, e.g., \cite{Grohs:2023voo} for a recent report on the status of these measurements), which are usually quoted in terms of the relative number densities with respect to hydrogen: 
\begin{equation}
\label{eq:YioYH}
i / {\rm H} \equiv Y_{i}/Y_{ \rm p} = n_{i}/n_{ \rm H} \  \ , \  {\rm where}~i = {\rm D},\, ^3 {\rm He},\, ^7 {\rm Li} \, . 
\end{equation}
Notice that the final yield of primordial helium-3 receives a contribution from unstable species such as tritium; likewise,  the final amount of lithium-7 includes the decay of beryllium-7.

The literature contains several publicly accessible compilations of the thermonuclear rates relevant for BBN. It is important to note that there are several different parameterizations of these rates adopted in BBN studies, and they differ not only with respect to the theoretical approach, but also with respect to the measured nuclear reaction data included in fitting them. To highlight a few of the more important approaches:
\begin{itemize}
    \item The NACRE~II database~\cite{Xu:2013fha} collects an extended evaluation of reaction rates of charged-particle induced reactions on target nuclides with mass number $A<16$, adopting the so-called potential model~\cite{Angulo:1999zz} to describe nuclear cross sections in the energy range of interest.
    \item \texttt{PRIMAT} tabulates an extensive catalogue (comprising more than 400 reactions), characterized by several nuclear cross sections evaluated via refined statistical analyses within $R$-matrix theory~\cite{Descouvemont04,Longland10,Iliadis16,Gomez17} or computed using dedicated numerical tools, e.g., the TALYS code~\cite{Coc:2011az}.
    \item \texttt{PArthENoPE} implements semi-analytic expressions resulting from polynomial fits to nuclear data including theory modeling of screening and thermal effects~\cite{Serpico:2004gx,Pisanti:2020efz}; data-oriented analyses relevant for BBN rates can be also found in the work of Refs.~\cite{Cyburt:2004cq,Fields:2019pfx}.
\end{itemize}

If one limits the scope to precise predictions of the helium-4 and deuterium abundances, the relevant portion of the nuclear network simplifies considerably, contracting to $\mathcal{O}(10)$ processes~\cite{Iliadis:2020jtc}. 
Thus, \texttt{PRyMordial} offers the option of restricting the BBN analysis to a small network of 12 key reactions~\cite{Coc:2010zz}, implemented according to two different sets of thermonuclear rates: the first is largely based on the NACRE~II compilation, whereas the second is based on the tabulated rates in \texttt{PRIMAT}.
These two sets differ marginally in their predictions for
helium-4, but lead to relevant differences in the prediction for deuterium, as discussed at length in Ref.~\cite{Pitrou:2021vqr}, after the important measurement carried out by the LUNA collaboration~\cite{Mossa:2020gjc}.\footnote{This fact has been  more quantitatively acknowledged in Ref.~\cite{Burns:2022hkq} which used a beta version of \texttt{PRyMordial}.} For the most precise prediction of lithium-7, \texttt{PRyMordial} offers the possibility to solve a nuclear network including the 51 additional reactions listed in \autoref{app:ThermoNuclRates}, by adopting part of the network in Ref.~\cite{Coc:2011az} included in the \texttt{PRIMAT} database.

\texttt{PRyMordial} handles uncertainties on the tabulated thermonuclear rates $\widetilde{\Gamma}^{(R)}$ by providing (for each forward\footnote{The corresponding reverse reactions are obtained via Eq.~\eqref{eq:reverseGamma} from the interpolated forward rates.} nuclear reaction) a set of median values, $ \langle \widetilde{\Gamma}^{(R)} \rangle $ together with an uncertainty factor $ \Delta \widetilde{\Gamma}^{(R)} $, corresponding to a sample of temperatures. 
Following the method outlined in Refs.~\cite{Longland:2010gs,Coc:2014oia}, to perform a MC analysis with \texttt{PRyMordial} one should treat the provided thermonuclear rates as log-normal distributed, implying that for each nuclear process $R$ a random realization of the thermonuclear rate will be:
\begin{equation}
     \log \widetilde{\Gamma}^{(R)} = \log \,\langle \widetilde{\Gamma}^{(R)} \rangle +  p^{(R)} \log \Delta \widetilde{\Gamma}^{(R)} \ ,
\end{equation} 
where $p^{(R)}$ is a temperature-independent coefficient following a normal distribution~\cite{Sallaska:2013xqa}.
Hence, in order to properly take into account the uncertainties of the thermonuclear rates in a MC analysis of BBN, one should independently vary the nuisance parameters $p^{(R)}$ for all the reactions $R$ included in the study, see, e.g., the work carried out in Ref.~\cite{Burns:2022hkq} and the MC examples presented in \autoref{sec:HowPRyM}.

\section{How to Use \href{https://github.com/vallima/PRyMordial}{\texttt{PRyMordial}}} \label{sec:HowPRyM}

In this section we provide some example code that demonstrates the use of {\tt PRyMordial}. We start by detailing the modules of the code including their inputs and key parameters. We show how to implement a state-of-the-art analysis of the BBN era within the SM. Finally, we provide a concise description on how to use the code for the study of NP, and discuss how to implement and analyze generic BSM scenarios. 

\subsection{Structure of the Code and \texttt{Hello, World!}}

\texttt{PRyMordial} is a numerical tool dedicated to efficiently and accurately evaluate in the SM and beyond all the key observables related to the BBN era, discussed in \autoref{sec:PhysPRyM}, namely:
\begin{itemize}
\item The number of effective relativistic degrees of freedom, $N_{\rm eff}$, Eq.~\eqref{eq:Neff}$\,$;
\item The cosmic neutrino abundance today, $\Omega_{\nu} h^2$, Eq.~\eqref{eq:OmegaNuh2}$\,$;
\item The helium-4 mass fraction (both for BBN and CMB), $Y_{P}$, Eq.~\eqref{eq:YP}$\,$;
\item The relative number density of deuterium, helium-3 and lithium-7, Eq.~\eqref{eq:YioYH}$\,$.
\end{itemize}

\begin{figure}[t!]
\centering
\includegraphics[scale=0.6]{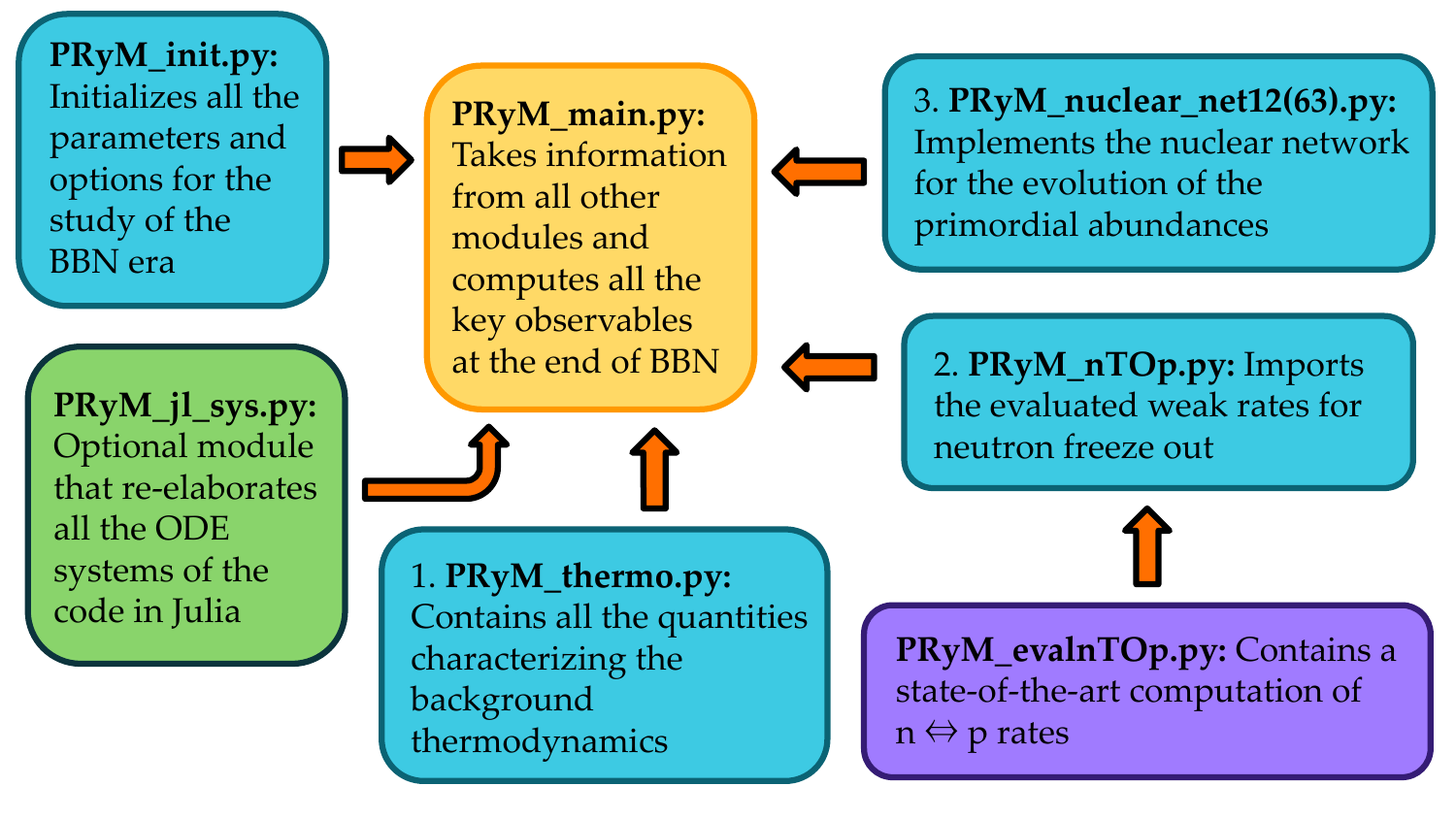}
\caption{\texttt{PRyMordial} {\it in a nutshell: Schematic of the modules making it up and their inter-relations.}}
\label{fig:PRyMnutshell}
\end{figure}

In contrast to other BBN codes available, \texttt{PRyMordial} begins by computing the thermal background from first principles. As a byproduct of the determination of $N_{\rm eff}$ and $\Omega_{\nu}h^2$, the relationship between time, scale factor and temperature of relativistic species is determined precisely, including effects from non-instantaneous decoupling within and beyond the Standard Model. 

Next, \texttt{PRyMordial} evaluates the weak rates for neutron freeze out via a state-of-the-art implementation that includes nucleon finite-mass effects, one-loop QED corrections and finite-temperature effects. While the latter are typically negligible within current observational precision and can be conveniently stored between runs, the remainder are generally recomputed for each iteration of a generic BBN analysis.

Finally, \texttt{PRyMordial} solves a network of nuclide reactions for their yields within three different physical regimes: \textit{i)} a high-temperature era in which one can restrict the study to nucleons with an initial temperature of $\mathcal{O}(10)~$MeV and a final temperature close to neutrino decoupling; \textit{ii)} a mid-temperature era from $\mathcal{O}(1)~$MeV down to $\mathcal{O}(0.1)~$MeV, during which photo-dissociation of nuclear bound states is relevant; \textit{iii)} and a low temperature era starting at $\mathcal{O}(0.1)~$MeV during which \texttt{PRyMordial} follows all of the nuclear species of interest, which ends at a temperature well below $e^{\pm}$ heating of the thermal bath, i.e. down to $\mathcal{O}(1)~$keV. Local thermal equilibrium sets the initial nuclide abundances and detailed balance determines all of the reverse reactions included in the chosen set of nuclear reactions.  These three regimes are matched such that the solution for each one provides the initial conditions for the successive period.

\texttt{PRyMordial} is a Python package with optional dependencies  which allow more advanced users to speed up execution by exploiting the Julia programming language. The recommended libraries and general requirements are tabulated in~\autoref{app:InstallPRyM}. 
As highlighted in~\autoref{fig:PRyMnutshell}, \texttt{PRyMordial} is organized in five primary modules:
\begin{itemize}
\item \texttt{PRyM\_init.py} is an initialization module where physical constants and Boolean flags for user-controlled options are defined; in particular, three main blocks for input parameters are found: 
\begin{itemize}
 \item[$\star$] Fundamental constants, masses (in natural units), initialized according to the PDG~\cite{PDG:2022pth};\footnote{For the electroweak sector we adopt $\{\alpha_{\rm em}$, $G_{\rm F}$, $M_{\rm Z}\}$ as inputs and derive the rest via tree-level relations.}
 \item[$\star$] Additional parameters needed for the evaluation of the $n \leftrightarrow p$ rates beyond the Born level;
 \item[$\star$] Cosmological inputs including the CMB temperature and the abundance of baryonic matter~\cite{Planck20}.
\end{itemize}
Boolean flags allow the user to switch on/off the following options:
\begin{itemize}
\item[$\circ$] \texttt{verbose\_flag}: Allows the user to run the code with all of the internal messages enabled;
\item[$\circ$] \texttt{numba\_flag}: If \texttt{True}, speeds up some numerical integrations, if the \texttt{Numba} library is installed;
\item[$\circ$] \texttt{numdiff\_flag}: If \texttt{True}, performs numerical derivatives using \texttt{Numdifftools} library;
\item[$\circ$] \texttt{aTid\_flag}: Controls the inclusion of incomplete-decoupling effects in the determination of the scale factor as a function of time and temperature;
\item[$\circ$] \texttt{compute\_bckg\_flag}: If \texttt{True}, recomputes the thermodynamical background as presented in \autoref{sec:thermodyn} (via \texttt{save\_bckg\_flag} the outcome can be stored in a file for future runs); 
\item[$\circ$] \texttt{NP\_thermo\_flag}: If \texttt{True}, includes the contribution(s) of new (interacting) species to the dynamics of the thermal bath (by default, one must also provide a NP temperature);
\item[$\circ$] \texttt{NP\_nu\_flag}: If \texttt{True}, includes new species thermalized with the neutrino bath;
\item[$\circ$] \texttt{NP\_e\_flag}: If \texttt{True}, includes new species thermalized with the plasma;
\item[$\circ$] \texttt{compute\_nTOp\_flag}: If \texttt{True}, recomputes weak rates beyond Born as discussed in \autoref{sec:nTOp} (via \texttt{save\_nTOp\_flag} the outcome can be stored in a file for future runs);
\item[$\circ$] \texttt{nTOpBorn\_flag}: If \texttt{True}, adopts the Born approximation for the neutron freeze out;
\item[$\circ$] \texttt{compute\_nTOp\_thermal\_flag}: If \texttt{True}, recomputes thermal corrections to $n \leftrightarrow p$ rates via \texttt{Vegas} (since this is numerically intensive, we recommend \texttt{save\_nTOp\_thermal\_flag = True});
\item[$\circ$] \texttt{tau\_n\_flag}: If \texttt{True}, uses the neutron lifetime to normalize the weak rates, see \autoref{sec:nTOp};
\item[$\circ$] \texttt{NP\_nTOp\_flag}: If \texttt{True}, includes NP affecting $n \leftrightarrow p$ rates in units of the Born rates;
\item[$\circ$] \texttt{smallnet\_flag}: If \texttt{True}, restricts the nuclear network to the set of 12 key nuclear processes collected in~\autoref{table:12} of \autoref{app:ThermoNuclRates};
\item[$\circ$] \texttt{nacreii\_flag}: If \texttt{True}, the key nuclear rates adopted in \texttt{PRyMordial} will be mostly based on NACRE~II compilation rather than those of \texttt{PRIMAT}, see \autoref{sec:NuclearRates};
\item[$\circ$] \texttt{NP\_nuclear\_flag}: If \texttt{True}, shifts the nuclear rates due to NP in units of the standard ones;
\item[$\circ$] \texttt{julia\_flag}: If \texttt{True}, solves all of the systems of ordinary differential equations using routines in the {SciML} kit~\cite{rackauckas2017differentialequations} developed for the Julia programming language; the optional dependencies described in \autoref{app:InstallPRyM} are then required.
\end{itemize}
This module also loads the tabulated nuclear rates (as well as the coefficients of Eq.~\eqref{eq:reverseGamma}).

\item \texttt{PRyM\_thermo.py} is the module where all of the thermodynamical quantities for the species contributing to the expansion of the Universe during radiation domination are defined, together with all the collision terms that enter in Eq.~\eqref{eq:SMBoltzsys} and Eq.~\eqref{eq:SMentropies}. 

\item \texttt{PRyM\_nTOp.py} is the module which imports the weak rates for $n \leftrightarrow p$ conversion described in \autoref{sec:nTOp}, either relying on the additional module \texttt{PRyM\_evalnTOp.py} -- where the actual computation of the rates is performed from scratch --  or by loading pre-stored rates from a file. 

\item \texttt{PRyM\_nuclear\_net12.py} and \texttt{PRyM\_nuclear\_net63.py} are the modules which set up the systems of ordinary differential equations -- see Eq.~\eqref{eq:YiBoltz} -- involving the nuclear rates loaded by \texttt{PRyM\_init.py}. The Boolean flag \texttt{smallnet\_flag} controls whether \texttt{PRyMordial} sets up and solves the smaller network of 12 key reactions or the full set of 63 nuclear processes.

\item \texttt{PRyM\_main.py} is the main module, which calls the other modules to solve for the thermodynamical background,  compute $N_{\rm eff}$ and the cosmic neutrino abundance, and solve for the nuclide yields.  It contains the Python class \texttt{PRyMclass()}, designated to return all the cosmological observables implemented in the package.
\end{itemize}
\begin{itemize}
\item \texttt{PRyM\_jl\_sys.py} is an optional module which allows the user to solve all of the systems of ordinary differential equations in \texttt{PRyM\_main.py} by taking advantage of the numerically efficient routines that are part of the \href{https://sciml.ai}{SciML} kit~\cite{rackauckas2017differentialequations} developed in Julia.  In some cases, this significantly speed up the execution time of the code (to a degree depending on both the adopted precision of the computation as well as the specific choice of differential-equation solver).
\end{itemize}

After downloading \texttt{PRyMordial}, the code can be used immediately. To run a \texttt{Hello, World!}-style example, the user would enter the package folder, start an interactive Python session, and type:
\usemintedstyle{emacs}
\begin{minted}{python}
# Hello, World! of PRyMordial
import PRyM.PRyM_main as PRyMmain
res = PRyMmain.PRyMclass().PRyMresults()
\end{minted}
which executes a BBN computation and fills the array \texttt{res} with the values of:
\begin{equation}
\left[~N_{\rm eff}, \Omega_{\nu}h^2 \times 10^6~{\rm (rel)},\sum m_{\nu}/\Omega_{\nu}h^2 [{\rm eV}], Y_P^{\rm (CMB)}, Y_P^{\rm (BBN)}, D/H \times 10^5, ^3{\rm He/H} \times 10^5 , ^7{\rm Li/H} \times 10^{10}~\right] \nonumber 
\end{equation}
\noindent Located in the same folder are: 
\begin{itemize}
    \item a folder \texttt{PRyM} in which all of the modules described above reside;
    \item a folder \texttt{PRyMrates} in which all the essential thermal, weak and nuclear rates are present, and where new evaluations of them can be stored;
    \item a script named \texttt{runPryM\_julia.py} that provides a simple example for the user as to how to use the package, with execution-time benchmarking in both standard and Julia modes.
\end{itemize}  
In the following subsections we present more sophisticated examples illustrating some of \texttt{PRyMordial}'s capabilities.

\subsection{SM examples: the PDG Plot and Monte Carlo Analysis}

\begin{figure}[t]
\centering
\includegraphics[scale=1.5]{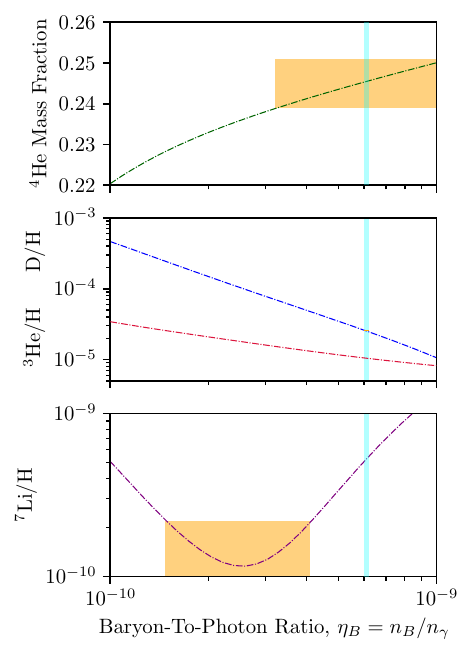}
\caption{{\it Primordial abundances of helium-4, deuterium, helium-3, and lithium-7 as predicted by }\texttt{PRyMordial} {\it within the SM, as a function of the cosmic baryon density. Central predictions are shown without theory uncertainties (i.e. using the nominal nuclear rates for the largest set implemented in the package with the NACRE II compilation for the key processes) and at the central values of all of the inputs. Measurements of light-element abundances (orange) as well as the CMB constraint on the baryon-to-photon ratio (cyan) follow from Figure 24.1 of the PDG \cite{PDG:2022pth}.}}
\label{fig:PDGplot}
\end{figure}

In an interactive session in Python, any default value in \texttt{PRyM\_init.py} can be changed using the syntax:
\usemintedstyle{emacs}
\begin{minted}{python}
import PRyM.PRyM_init as PRyMini
# New assignment x for parameter X
PRyMini.X = x
\end{minted}
This includes the Boolean flags listed in the previous subsection. Hence -- to perform a run with: \textit{i)} the computation of the thermal background from scratch, including non-instantaneous decoupling effects; \textit{ii)} the ab-initio evaluation of the weak rates for neutron freeze out; and \textit{iii)} the inclusion of key nuclear processes based on the tabulated rates of the NACRE~II compilation -- one should type:
\usemintedstyle{emacs}
\begin{minted}{python}
import PRyM.PRyM_init as PRyMini
# Include incomplete decoupling in a(T)
aTid_flag = True
# Recompute the background from scratch
PRyMini.compute_bckg_flag = True
# Save the background in PRyMrates/thermo
PRyMini.save_bckg_flag = True
# Recompute n <--> p rates from scratch
PRyMini.compute_nTOp_flag = True
# Save n <--> p rates in PRyMrates/nTOp
PRyMini.save_nTOp_flag = True
# Include only key rates in nuclear network
PRyMini.smallnet_flag = True
# NACRE II compilation for key rates
PRyMini.nacreii_flag = True
# Compute PRyMordial observables
import PRyM.PRyM_main as PRyMmain
res = PRyMmain.PRyMclass().PRyMresults()
\end{minted}
The array \texttt{res} is assigned the same values as in the \texttt{Hello, World!} example, above.
This code also stores the results for the thermal background and $n \leftrightarrow p$ rates for future runs. Consequently, a subsequent call with the same setup can be made faster:
\usemintedstyle{emacs}
\begin{minted}{python}
import PRyM.PRyM_init as PRyMini
# No need to recompute background since stored
PRyMini.compute_bckg_flag = False
PRyMini.save_bckg_flag = False
# No need to recompute n <--> p  rates as well
PRyMini.compute_nTOp_flag = False
PRyMini.save_nTOp_flag = False
# Compute PRyMordial observables: now faster!
import PRyM.PRyM_main as PRyMmain
res = PRyMmain.PRyMclass().PRyMresults()
\end{minted}

While it may be necessary in general to recompute the thermal background and/or the rates for neutron freeze out, there are cases for which storing the outcome of these computations can be computationally advantageous. 
An example is the classic PDG review BBN plot of the primordial abundances as a function of the baryon-to-photon ratio $\eta_{B}$~\cite{PDG:2022pth}.
Once thermal background and weak rates have been stored, the behaviour of the abundances in the PDG Figure~24.1 can be reproduced with \texttt{PRyMordial}:
\usemintedstyle{emacs}
\begin{minted}{python}
# PDG plot
npoints = 50
import numpy as np
etabvec = np.logspace(-10,-9,npoints)
# Initialization of array of observables
YP_vec, DoH_vec, He3oH_vec, Li7oH_vec = np.zeros((4,npoints))
for i in range(npoints):
    # Update value of baryon-to-photon ratio and store new obs
    PRyMini.eta0b = etabvec[i]
    YP_vec[i], DoH_vec[i], He3oH_vec[i], Li7oH_vec[i] =
    PRyMmain.PRyMclass().PRyMresults()[4:8]
\end{minted}
The outcome of this code is illustrated in \autoref{fig:PDGplot}, which adopts the largest nuclear network for the most accurate prediction of the relative abundance of lithium-7.
It is worth noting that the BBN prediction for deuterium matches observations of quasar absorption systems, and is also in line with the cosmological abundance of baryons independently determined from the CMB (without a BBN prior). 
As pointed out in Ref.~\cite{Pitrou:2021vqr} and further scrutinized in Ref.~\cite{Burns:2022hkq}, this test of concordance would fail if the \texttt{PRIMAT} rates were to be adopted, i.e. \texttt{nacreii\_flag = False}.

\begin{figure}[t!]
\centering
\includegraphics[scale=0.4]{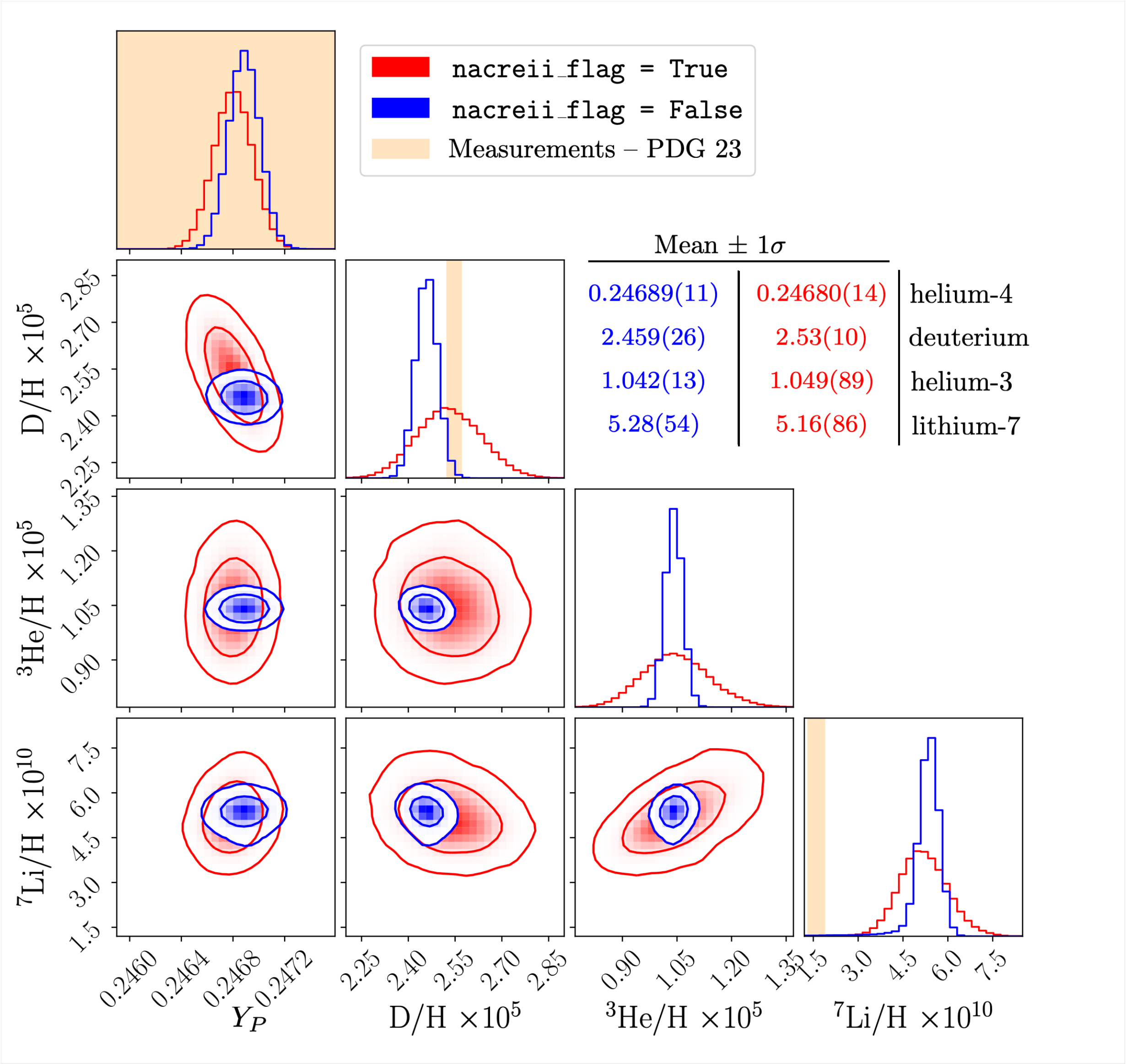}
\caption{{\it 1D probability distributions (and 2D joint 68\% and 95\% probability regions) for the light primordial abundances predicted in the SM with }\texttt{PRyMordial}{\it. Predictions are obtained using a Gaussian prior for the neutron lifetime $\tau_{n} = 878.4 \pm 0.5$ s (comprising the eight best measurements from ultra-cold neutron experiments combined in Ref.~\cite{PDG:2022pth}), and the cosmic baryon density, $\Omega_{B}h^2 = 0.02230 \pm 0.00020$ (from Table 5 of Ref.~\cite{Planck20} for the analysis with an uninformative $Y_{P}$ prior). The large network of nuclear reactions has been used, implying an additional 63 nuisance parameters varied with a log-normal distribution. Two different sets of key nuclear rates have been considered on the basis of the Boolean flag }\texttt{nacreii\_flag}{\it , and the statistics of the marginalized distributions for each case is presented.  %The change in the posterior distribution of deuterium is of particular significance in relation to the modeling of the key nuclear rates.
}}
\label{fig:MC_SM_BBN}
\end{figure}

To perform a Monte Carlo analysis of the SM predictions taking into account uncertainties (similar to the one presented in Ref.~\cite{Burns:2022hkq}):
\usemintedstyle{emacs}
\begin{minted}{python}
# SM MC run
num_it = 10000 # number of iterations
import numpy as np
Yp_vec, YDoH_vec, YHe3oH_vec, YLi7oH_vec = np.zeros((4,num_it))
# Import PRyM modules 
import PRyM.PRyM_init as PRyMini
import PRyM.PRyM_main as PRyMmain
# Baryon eta from Planck 18 (no BBN prior)
mean_eta0b = PRyMini.eta0b
std_Omegabh2 = 2*1.e-4
std_eta0b = PRyMini.Omegabh2_to_eta0b*std_Omegabh2
# Neutron lifetime from PDG 2023
mean_tau_n = PRyMini.tau_n
std_tau_n = 0.5

# Compute primordial abundances at each iteration
def ComputeAbundances(i):
    # Settings to speed up the SM MC 
    PRyMini.recompute_bckg = False
    PRyMini.recompute_nTOp_rates = False
    # Large network for nuclear rates
    PRyMini.smallnet_flag = False
    # Gaussian prior on baryon-to-photon ratio
    PRyMini.eta0b = np.random.normal(mean_eta0b,std_eta0b)
    # Gaussian prior on neutron lifetime
    PRyMini.tau_n = np.random.normal(mean_tau_n,std_tau_n)
    # Log-normal prior on nuclear rates
    PRyMini.p_npdg,PRyMini.p_dpHe3g,PRyMini.p_ddHe3n,
    ... # for the sake of brevity not listing all 63 process 
    PRyMini.p_ppndp, PRyMini.p_Li7taann = np.random.normal(0,1,PRyMini.num_reactions)
    # NACRE II compilation for key rates
    PRyMini.nacreii_flag = True
    PRyMini.ReloadKeyRates()
    return PRyMmain.PRyMclass().PRyMresults()[4:8]

# Parallelizing w/ joblib + multiprocessing
from joblib import Parallel, delayed
import multiprocessing
num_cpu = int(multiprocessing.cpu_count())
FinalAbundances = Parallel(n_jobs = num_cpu)(delayed(ComputeAbundances)((i)) 
for i in range(num_it))
Yp_vec,YDoH_vec,YHe3oH_vec,YLi7oH_vec = np.array(FinalAbundances).transpose()
\end{minted}
The output maps out the probability distributions, shown in \autoref{fig:MC_SM_BBN}, where the light elements at the end of the BBN era are predicted within the SM via a MC analysis that involves: \textit{i)} a cosmological prior on the cosmic baryon abundance; \textit{ii)} a particle-physics measurement prior on the neutron lifetime; and \textit{iii)} a dedicated treatment of the uncertainties in the rates of the nuclear processes. \autoref{fig:MC_SM_BBN} displays the ``deuterium anomaly'' present for the \texttt{PRIMAT} compilation of the key nuclear rates, and further shows that it is completely washed out when one employs the NACRE II database\footnote{The results in \autoref{fig:MC_SM_BBN} slightly differ from Ref.~\cite{Burns:2022hkq} due to an update on the Gaussian prior for the neutron lifetime and the different choice for the cosmological baryon abundance adopted in that study.}.

\autoref{fig:MC_SM_BBN} suggests that the ``primordial lithium problem'' stands out as statistically significant, regardless of the approach undertaken for the nuclear network.  However, the up-to-date analysis of the lithium problem in Ref.~\cite{Fields:2022mpw} points out that the predicted primordial abundance of lithium-7 could be depleted via stellar (and cosmic-ray) nucleosynthesis. Given this argument, the observational inference of \autoref{fig:PDGplot} and \autoref{fig:MC_SM_BBN}, in which the observations lie below the theoretical prediction for primordial lithium-7, are consistent with a resolution for this long-standing puzzle.

\subsection{NP examples: New Interacting Sectors and BBN}

\texttt{PRyMordial} allows the user to perform state-of-the-art analyses for Physics beyond the SM in the Early Universe. A few options already built-in to the current release include: 
\begin{itemize}
\item additional relativistic degrees of freedom contributing to the expansion rate of the Universe in the form of a shift of $N_{\rm eff}$, see Eq.~\eqref{eq:Neff};
\item a non-zero chemical potential for neutrinos, influencing both the cosmological expansion rate as well as the equilibrium distributions in the weak processes for neutron-to-proton conversion;
\item Boolean flags specific to the study of new species interacting with the plasma and/or neutrino bath, as well as flags implementing a new entire sector with temperature $T_{\rm NP} \neq T_{\gamma,\nu}\,$;
\item a Boolean flag and a dedicated parameter encoding NP effects as a phenomenological modification of $n \leftrightarrow p$ conversion rates (in units of the Born rates);
\item a set of parameters that allow one to similarly investigate NP effects in the nuclear processes as a simple shift in terms of the median rate of each process.
\end{itemize}
The first two have been extensively investigated in Ref~\cite{Burns:2022hkq}, and thus we focus here on the others. The following is code demonstrating how to implement an electrophilic species in thermal equilibrium with the SM bath during BBN:
\usemintedstyle{emacs}
\begin{minted}{python}
import PRyM.PRyM_init as PRyMini
# Electrophilic
PRyMini.NP_e_flag = True
import numpy as np
from scipy.integrate import quad
# Scalar with mass mX = 5 MeV
gX = 1; mX = 5.;
def rho_NP(T_NP):
    if T_NP < mX/30.: return 0.
    else:
        res_int = quad(lambda E: E**2*(E**2-(mX/T_NP)**2)**0.5
        /(np.exp(E)-1.) ,mX/T_NP,100.,epsrel=1e-9,epsabs=1e-12)[0]
        return gX/(2*np.pi**2)*T_NP**4*res_int
def p_NP(T_NP):
    if T_NP < mX/30.: return 0.
    else:
        res_int = quad(lambda E: (E**2-(mX/T_NP)**2)**1.5
        /(np.exp(E)-1.) ,mX/T_NP,100.,epsabs=1e-9,epsrel=1e-12)[0]
        return gX/(6*np.pi**2)*T_NP**4*res_int
def drho_NP_dT(T_NP):
    if T_NP < mX/30.: return 0.
    else:
        res_int = quad(lambda E: 0.25*E**3*(E**2-(mX/T_NP)**2)**0.5*
        np.sinh(E/2.0)**-2 ,mX/T_NP,100,epsabs=1e-9,epsrel=1e-12)[0]
        return gX/(2*np.pi**2)*T_NP**3*res_int
import PRyM.PRyM_main as PRyMmain
res = PRyMmain.PRyMclass(rho_NP,p_NP,drho_NP_dT).PRyMresults()
\end{minted}
One can similarly evaluate a neutrinophilic species thermalized with the SM bath by replacing the Boolean flag at the top of the script with: \texttt{PRyMini.NP\_nu\_flag = True}. 

\begin{figure}[th!]
\centering
\includegraphics[scale=1.2]{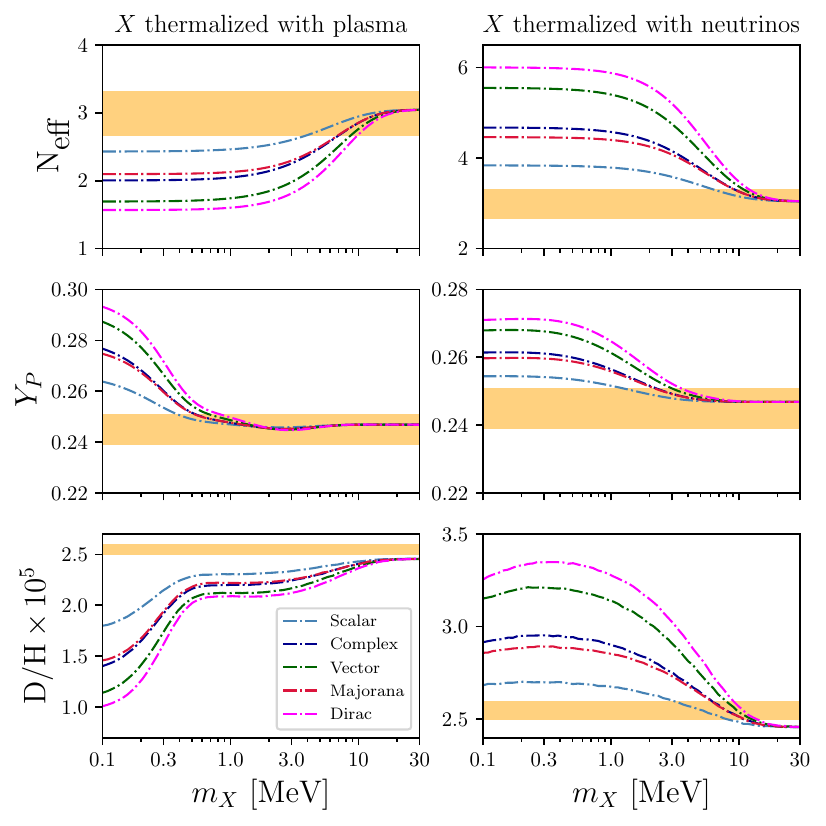}
\caption{{\it Investigation of the cosmological impact at the end of the BBN era from a new relativistic species $X$ with degrees of freedom corresponding to a real / complex scalar (light / dark-blue lines), a real massive vector (magenta), or a Majorana / Dirac fermion (red / green); $X$ is assumed to be in thermal equilibrium with either the electron-positron-photon plasma (left panels) or with the SM neutrino thermal bath (right panels). The orange bands represent the observational constraints at the 2$\sigma$ level from Refs.~\cite{Planck20,PDG:2022pth}. Predictions with }\texttt{PRyMordial}{\it ~are obtained at nominal inputs and rates.}}
\label{fig:NP_thermo}
\end{figure}

In \autoref{fig:NP_thermo} we present the results for NP scenarios of this type, reproducing the qualitative features already well-discussed, e.g., in Ref.~\cite{Sabti:2019mhn}.  In particular, we observe three primary NP effects: \textit{i)} a change in the cosmological expansion rate, affecting the time-temperature relation; \textit{ii)} an impact on the evolution of the neutrino-to-photon temperature ratio, relevant for both neutrino and neutron decoupling; and \textit{iii)} additional entropy released in the plasma, altering the number of baryons per a given baryon-to-photon ratio. 
Note that in \autoref{fig:NP_thermo} we use the set of nuclear reactions from \texttt{PRIMAT} (\texttt{nacreii\_flag = False}) and as a result a neutrinophilic species around $\sim 10~$MeV in mass appears to be favored by current observations of primordial $D/H$ while remaining compatible with the other cosmological NP probes based on helium-4 and $N_{\rm eff}$.

In contrast to the previous scripts, this code calls \texttt{PRyMclass()} with three functions (of temperature) as arguments:
 the contribution to the energy density, its derivative, and the pressure of the new species added to the bath. More generally, one can include a new interacting sector with its own temperature $T_{\rm NP}$ and non-trivial collision term $\delta C_{\rm NP}$ along the lines of the recent work in Ref.~\cite{Chu:2022xuh}. In \texttt{PRyMordial} one may study such ``dark sectors'' consistently by generalizing the set of equations in Eq.~\eqref{eq:SMBoltzsys} to follow $T_{\rm NP}$ together with $T_{\gamma, \nu}$, and solving for the entropy density involved in Eq~\eqref{eq:aofT} taking into account the effect of the NP. To do this, one switches on the Boolean flag \texttt{NP\_thermo\_flag} and codes all of the relevant contributions to the energy density, its derivative (which can optionally be evaluated numerically via \texttt{Numdifftools}), pressure and collision term for the NP sector, and passes them to \texttt{PRyMresults}.

One can also study NP resulting in changes to the weak rates for neutron freeze out and/or any of the implemented thermonuclear rates. To modify the weak rates, one sets the Boolean flag \texttt{NP\_nTOp\_flag = True} and change the parameter \texttt{NP\_delta\_nTOp} from its default of zero. Analogously, for the nuclear rates one switches on the flag \texttt{NP\_nuclear\_flag} and modifies the value of \texttt{NP\_delta\_R} with \texttt{R} being the reaction of interest. 
 
As an example, we consider NP which results in a small change to the $n \leftrightarrow p$ conversion rates.  We perform a Bayesian fit to $Y_{P}$ and $D/H$ (as quoted by the PDG~\cite{PDG:2022pth}) and allowing $\tau_{\rm n}$, $\Omega_{B}h^2$, and the other key nuclear rates to vary within their uncertainties (in line with the SM MC analysis of the previous subsection):
\usemintedstyle{emacs}
\begin{minted}{python}
# Bayesian analysis w/ PRyMordial and emcee 
import emcee
# BBN measurements from PDG 2023
YP_ave = 0.245; YP_std = 0.003;
DoH_ave = 2.547; DoH_std = 0.025;
# Mean and standard deviation on neutron lifetime [s]
mean_tau_n = PRyMini.tau_n
std_tau_n = 0.5
# Mean and standard deviation on cosmic baryonic abundance
mean_Omegabh2 = PRyMini.Omegabh2
std_Omegabh2 =  2*1.e-4
# Test statistic for the fit
def log_L(theta):
    delta_nTOp = theta
    PRyMini.NP_delta_nTOp = delta_nTOp
    # Gaussian extraction of neutron lifetime
    PRyMini.tau_n = np.random.normal(mean_tau_n,std_tau_n)
    # Gaussian extraction of cosmic baryonic abundance
    PRyMini.Omegabh2 = np.random.normal(mean_Omegabh2,std_Omegabh2)
    # IMPORTANT: Assign etab after updating Omegab (or directly vary etab)
    PRyMini.eta0b =  PRyMini.Omegabh2_to_eta0b*PRyMini.Omegabh2 
    # Gaussian weights for log-normal nuclear rates
    PRyMini.p_npdg,PRyMini.p_dpHe3g,PRyMini.p_ddHe3n,PRyMini.p_ddtp, PRyMini.p_tpag,
    PRyMini.p_tdan,PRyMini.p_taLi7g,PRyMini.p_He3ntp,PRyMini.p_He3dap, PRyMini.p_He3aBe7g,
    PRyMini.p_Be7nLi7p,PRyMini.p_Li7paa = np.random.normal(0,1,12)
    YPth, DoHth = PRyMmain.PRyMclass().PRyMresults()[4:6]
    m2LogL = (YPth-YP_ave)**2/(YP_std**2) + (DoHth-DoH_ave)**2/(DoH_std**2)
    return -0.5*m2LogL
def log_prior(theta):
    delta_nTOp = theta
    if -0.3 < delta_nTOp < 0.3:
        return 0.0
    return -np.inf
def log_prob(theta):
    lp = log_prior(theta)
    if not np.isfinite(lp):
        return -np.inf
    ll = log_L(theta)
    return lp + ll

if __name__ == '__main__' :
    # Total number of steps x walker
    nsteps = 2100
    # Guess on burn-in steps
    discsteps = int(nsteps/3.)
    nwalkers = 6
    ndim = 1
    pos = np.array([0.]) + [1e-2]*np.random.randn(nwalkers,ndim)
    def RunMCMCserial(i):
        import PRyM.PRyM_init as PRyMini
        PRyMini.smallnet_flag = True
        PRyMini.nacreii_flag = True
        PRyMini.ReloadKeyRates()
        PRyMini.NP_nTOp_flag = True
        sampler = emcee.EnsembleSampler(nwalkers, ndim, log_prob)
        sampler.run_mcmc(pos, nsteps, progress = True)
        all_samples = sampler.get_chain(discard=discsteps,flat=True)
        return all_samples
    from joblib import Parallel, delayed
    import multiprocessing
    num_cpu = int(multiprocessing.cpu_count())
    start = time.time()
    FinalRes = Parallel(n_jobs = num_cpu)(delayed(RunMCMCserial)((i)) 
    for i in range(num_cpu))
    my_samples_1,my_samples_2,my_samples_3,my_samples_4, 
    my_samples_5,my_samples_6,my_samples_7,my_samples_8 = FinalRes
    # Collecting the samples all together
    final_samples = np.concatenate((my_samples_1,my_samples_2,my_samples_3,my_samples_4,
    my_samples_5,my_samples_6,my_samples_7,my_samples_8))
\end{minted}
This code can be simply generalized to modify any of the other nuclear reactions.

\begin{figure}[t!]
\centering
\includegraphics[scale=0.5]{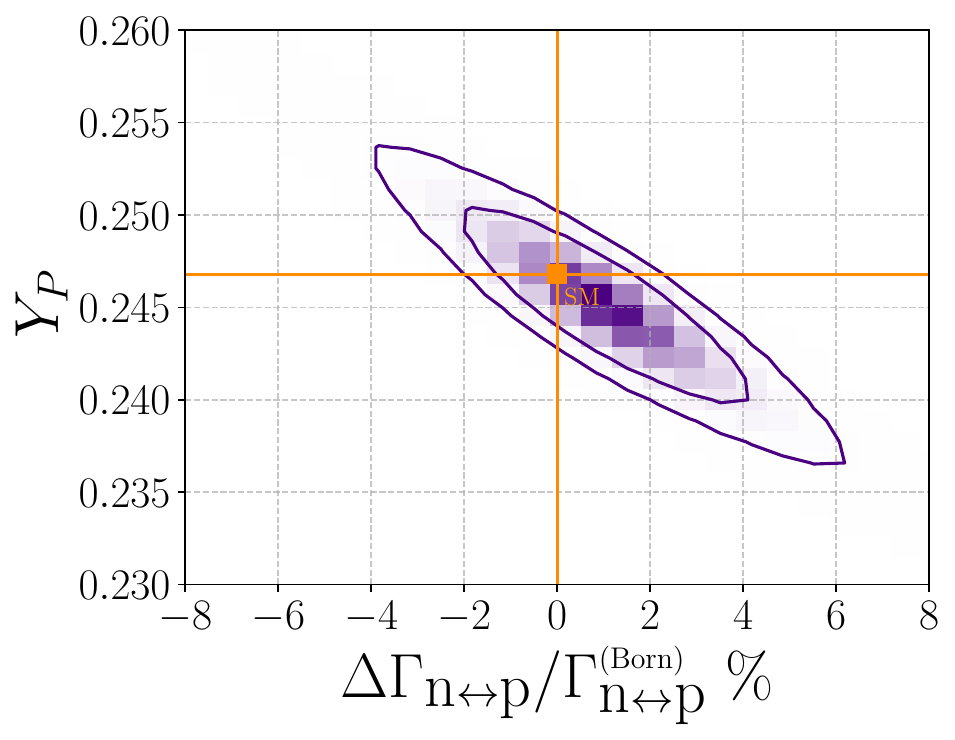}
\includegraphics[scale=0.5]{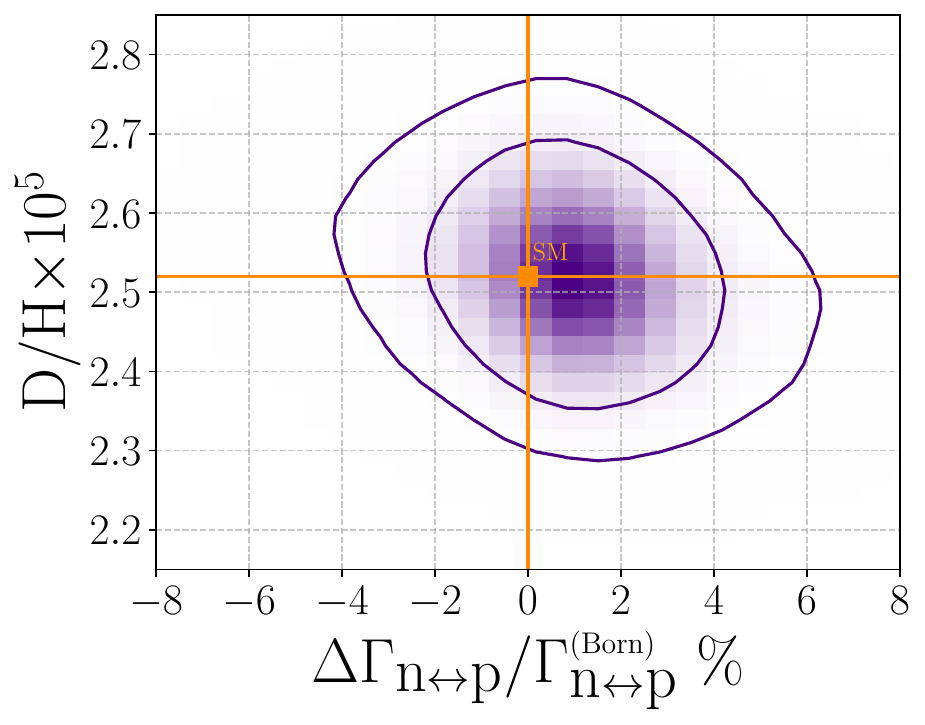}
\caption{{\it Constraint on a relative change of the weak $n \leftrightarrow p$ conversion rates from NP, based on a Bayesian fit performed with }\texttt{PRyMordial} {\it with the use of the} \texttt{emcee}~\cite{2013PASP..125..306F}{\it package.  Gaussian priors on the neutron lifetime and the cosmic baryon abundance are assumed (as for \autoref{fig:MC_SM_BBN}) and flags} \texttt{smallnet\_flag} \textit{and} \texttt{nacreii\_flag} {\it are both switched on. Helium-4, deuterium measurements correspond to the recommended values from the PDG~\cite{PDG:2022pth}.}}
\label{fig:NP_nTOp}
\end{figure}

\autoref{fig:NP_nTOp} shows the resulting 2D joint (68\% and 95\%) probability regions for \texttt{NP\_delta\_nTOp} correlated with the measurements of primordial helium-4 and deuterium. To perform the statistical analysis, we adopt the \texttt{emcee} package~\cite{2013PASP..125..306F}.  For the sake of computational efficiency, we restrict the analysis to the network of 12 key reactions (with \texttt{nacreii\_flag = True}), as is sufficient given the focus on helium-4 and deuterium.  \autoref{fig:NP_nTOp} indicates that BBN is consistent with NP in the $n \leftrightarrow p$ conversion rates at the level of at most a few percent relative to the standard Born rates. The tight correlation with $Y_{P}$ illustrates the importance of neutron freeze out in determining the primordial helium-4 abundance. 

\section{Outlook}\label{sec:Concl}

In this work we have presented \href{https://github.com/vallima/PRyMordial}{\texttt{PRyMordial}}: A new tool to explore the physics of BBN in great detail, with an unprecedented eye toward applications for physics beyond the SM. The package also allows for fast, user-friendly precision analyses of the BBN era within the SM of Particle Physics, reaching the same level of accuracy as the state-of-the-art codes already publicly available.

In \autoref{sec:PhysPRyM} we have provided in some detail a review of the BBN era, highlighting the physics implemented in the code. The main novelties in \texttt{PRyMordial} are that it is:
\begin{itemize}
    \item A package entirely written in Python, easy to install, run and modify, efficient in the evaluation of the key quantities for the study of BBN; moreover, an  optional dependence on Julia allows the user to make the code run even faster;
    \item A computation of the thermal background based on the Boltzmann equations governing the evolution of the relativistic species present at that time.  This allows for an accurate prediction of $N_{\rm eff}$ from first principles and opens up new avenues for the study of BSM Physics;
    \item A fast and accurate evaluation of the weak rates including QED, nucleon-finite mass and thermal corrections for a prediction of the neutron-to-proton ratio that confronts the precision of current and next-generation measurements;
    \item A BBN code that easily allows exploration of uncertainties and changes in all of the input parameters and most importantly, includes by default different treatments for the nuclear rates in order to give to the user a better handle on the overall theoretical systematics.
\end{itemize}

In \autoref{sec:HowPRyM} we describe the structure of the code and provide several examples of its usage within the Standard Model and for a few interesting scenarios of NP.

There are many directions that can be pursued in the future to make \texttt{PRyMordial} an even more compelling and flexible tool for the community.
 One important aspect we plan to expand upon is the characterization of the thermal background. At the moment, only a single common temperature for neutrinos is considered and no evolution equation for primordial chemical potentials is given by default. All of these can be easily implemented along the lines of Ref.~\cite{Escudero20}.
 
Also relevant for precision studies would be an approach to efficiently include effects from phase-space spectral distortions of relativistic species. In this regard, we plan to further enrich the physics in \texttt{PRyMordial} with a dedicated framework for neutrino decoupling that includes effects from oscillations at non-zero lepton chemical potentials, see Ref.~\cite{Froustey:2022sla}. 
 
 It would be a very interesting (though formidable) task to further improve the current next-to-leading order computation of neutron freeze out in the Early Universe, filling in the gaps of some of the approximations undertaken in the literature (see Appendix~B of ~\cite{Pitrou18} as well as the improvements brought by the recent effective-field-theory study at zero temperature of Ref.~\cite{Cirigliano:2023fnz}). We would also eventually like to include higher-order QED corrections such as the ones available in Refs~\cite{Bennett:2019ewm} and~\cite{Pitrou:2019pqh}, as well as the NLO QED corrections to $e^{+}e^{-} \leftrightarrow \nu \bar{\nu}$ matrix elements recently inspected in Ref.~\cite{Cielo:2023bqp}. 
 
 Finally, in the future we would like to enlarge the  nuclear network beyond the 63 nuclear reactions currently implemented, which encode all of the processes involving nuclides up to boron-8 in atomic and mass number (needed for an accurate prediction of lithium-7 in the Standard Model). 

With the public release of \texttt{PRyMordial} we hope to provide to the community an important new tool to address fundamental questions about the Early Universe, whose study remains central to further progress in our understanding of Nature. In the wise words of a giant of our time~\cite{Weinberg:1977ji}: 
\begin{quote} \textit{
``[Human beings] are not content to comfort themselves with tales of gods and giants, or to confine their thoughts to the daily affairs of life; they also build telescopes and satellites and accelerators, and sit at their desks for endless hours working out the meaning of the data they gather. The effort to understand the universe is one of the very few things that lifts human life a little above the level of farce, and gives it some of the grace of tragedy.''}
\end{quote}

\vspace*{0.5cm}
\noindent
{\bf Note about referencing:} \texttt{PRyMordial} makes use of previous work in the literature.  When using it, please make certain to appropriately reference the original literature as well as \texttt{PRyMordial} itself.

\begin{appendix}

\section{How to Install \href{https://github.com/vallima/PRyMordial}{\texttt{PRyMordial}}}\label{app:InstallPRyM}

\texttt{PRyMordial} is publicly released on \faGithub \href{https://github.com/vallima/PRyMordial}{\,GitHub}. Once in the desired directory, from your terminal type:
\usemintedstyle{fruity}
\begin{minted}[bgcolor=black,mathescape]{r}
git clone https://github.com/vallima/PRyMordial
\end{minted}
The code requires a modern distribution of Python (Python~3 recommended) in order to properly run, and features only a couple of standard libraries as mandatory dependencies:
\begin{itemize}
    \item \href{https://numpy.org}{NumPy} (mandatory) -- pip install numpy;
    \item \href{https://scipy.org}{SciPy} (mandatory) -- pip install scipy;
    \item \href{https://scipy.org}{Vegas} (mandatory) -- pip install vegas;
    \item \href{https://scipy.org}{Numba} (recommended) -- pip install numba;
    \item \href{https://numdifftools.readthedocs.io/en/master/index.html}{Numdifftools} (recommended) -- pip install numdifftools;
    \item \href{https://pyjulia.readthedocs.io/en/latest/installation.html#step-1-install-julia}{PyJulia} (optional) -- pip install julia; \item \href{https://github.com/SciML/diffeqpy}{diffeqpy} (optional) -- pip install diffeqpy.
\end{itemize}
Indeed, the code can easily avoid dependencies on Numba and Numdifftools. Nevertheless, giving up on Numba will slightly slow down a few routines in \texttt{PRyM\_thermo.py} which involve SciPy integration. Also, the installation of Vegas library is required only in the case where thermal corrections to the weak rates governing neutron freeze out have to be recomputed. This is usually not the case, since those are already tiny effects in the SM and can be reasonably neglected in studies of NP during BBN.

The optional dependencies above require the Julia programming language to be installed. It can be downloaded at \href{https://julialang.org}{https://julialang.org}. 
Once Julia is installed, it is recommended to create a soft link from the terminal typing something like:

\usemintedstyle{fruity}
\begin{minted}[bgcolor=black,mathescape]{r}
ln -s path-where-/bin/julia-is /usr/local/bin/julia
\end{minted}
Then, launch Julia and install \texttt{DifferentialEquations.jl} of the \href{https://sciml.ai}{SciML} kit (with Sundials wrapper):
\usemintedstyle{emacs}
\begin{minted}{julia}
using Pkg; 
Pkg.add("DifferentialEquations");
Pkg.add("Sundials");
\end{minted}
After a successful installation of the package, one needs to open a Python shell and type:
\usemintedstyle{emacs}
\begin{minted}{python}
import julia
julia.install()
import diffeqpy
diffeqpy.install()
\end{minted}
At this point the user will be able to exploit the SciML routines developed in Julia to solve the nuclear-reaction network in \texttt{PRyMordial}, speeding up the execution of time by a factor of two or more, and with the possibility of cherry-picking from a large collection of differential-equation solvers built-in in the package, see the documentation at \href{https://docs.sciml.ai/DiffEqDocs/stable/solvers/ode_solve}{https://docs.sciml.ai/DiffEqDocs/stable/solvers/ode\_solve}. 

To use the SciML routines, the user must set the flag \texttt{PRyM\_init.flag\_julia = True}. In some systems, the very first call of \texttt{PRyM\_main.PRyMresults()} might need to be in Python and therefore requires initially \texttt{PRyM\_init.flag\_julia = False}. Also, the first call in Julia will inevitably be slow, since it will compile \texttt{PRyM\_jl\_sys.py}. As a concise example of the dedicated script \texttt{runPRyM\_julia.py} coming with the present release, here below is how things should work in the Julia mode:

\usemintedstyle{emacs}
\begin{minted}{python}
import PRyM.PRyM_init as PRyMini
import PRyM.PRyM_main as PRyMmain
# Initialization call in Python:
PRyMini.julia_flag = False
res = PRyMmain.PRyMclass().PRyMresults()
# First call in Julia will be slow:
PRyMini.julia_flag = True
res = PRyMmain.PRyMclass().PRyMresults()
# From here on, any call will be fast!
\end{minted}

\section{Nuclear Processes in \href{https://github.com/vallima/PRyMordial}{\texttt{PRyMordial}}}
\label{app:ThermoNuclRates}

In this appendix we collect the 12 key reactions necessary to accurately predict helium-4 and deuterium, see \autoref{table:12}, as well as the 51 additional reactions comprising the full set recommended for a more robust prediction of lithium-7, \autoref{table:12_63}. For the general aspects of the evaluation of the nuclear rates in the Early Universe as well as the theoretical and statistical details behind the compilation of the nuclear rates present in \texttt{PRyMordial}, we refer the interested reader to Ref.~\cite{Serpico:2004gx} and Refs.~\cite{Longland:2010gs,Coc:2014oia}.

\begin{table}[th!]
\centering
\begin{tabular}{|c|c|c|c|c|c|c|}
\hline
\textbf{Nuclear Reaction} & Ref. & Ref. &  \textbf{Nuclear Reaction} & Ref. & Ref. \\ 
\hline
 n+p $\to$ D+$\gamma$ &  \cellcolor{blue!25}  \cite{Ando06}  & \cellcolor{red!25} \cite{Ando06} &  D+p $\to$ $^{3}$He+$\gamma$ & \cellcolor{blue!25} \cite{Mossa:2020gjc} & \cellcolor{red!25}  \cite{Mossa:2020gjc} \\ 
 D+D $\to$ $^{3}$He+n & \cellcolor{blue!25} \cite{Iliadis16} & \cellcolor{red!25} \cite{Xu13} &  D+D $\to$ $^{3}$H+p & \cellcolor{blue!25} \cite{Iliadis16} &  \cellcolor{red!25} \cite{Xu13} \\ 
$^{3}$H+p $\to$ $^{4}$He+$\gamma$ & \cellcolor{blue!25} \cite{Serpico:2004gx} & \cellcolor{red!25} \cite{Serpico:2004gx} & $^{3}$H+D $\to$ $^{4}$He+n & \cellcolor{blue!25} \cite{Descouvemont04} & \cellcolor{red!25} \cite{Xu13} \\ 
$^{3}$H+$^{4}$He $\to$ $^{7}$Li+$\gamma$ & \cellcolor{blue!25} \cite{Descouvemont04} & \cellcolor{red!25} \cite{Xu13} & $^{3}$He+n $\to$ $^{3}$H+p & \cellcolor{blue!25} \cite{Descouvemont04} & \cellcolor{red!25} \cite{Cyburt:2004cq} \\ 
$^{3}$He+D $\to$ $^{4}$He+p & \cellcolor{blue!25}  \cite{Descouvemont04} & \cellcolor{red!25}  \cite{Xu13} & $^{3}$He+$^{4}$He $\to$ $^{7}$Be+$\gamma$ & \cellcolor{blue!25} \cite{Iliadis16} & \cellcolor{red!25} \cite{Xu13}  \\ 
$^{7}$Be+n $\to$ $^{7}$Li+p & \cellcolor{blue!25} \cite{Descouvemont04} & \cellcolor{red!25} \cite{Fields:2019pfx} & $^{7}$Li+p $\to$ $^{4}$He+$^{4}$He & \cellcolor{blue!25} \cite{Descouvemont04} & \cellcolor{red!25} \cite{Xu13} \\ 
\hline
\end{tabular}
\caption{{\it The key nuclear reactions adopted in }\texttt{PRyMordial}{\it, with corresponding references. The red (blue) column refers to the option }\texttt{nacreii\_flag = True (False)}{\it, see \autoref{sec:NuclearRates} for further details. Notice that the compilation of the blue column is present also in the code }\texttt{PRIMAT}{\it~\cite{Pitrou:2018cgg}.}}
\label{table:12}
\end{table}

\begin{table}
\centering
\begin{tabular}{|c|c|c|c|c|}
\hline
\textbf{Nuclear Reaction} & Ref. &  \textbf{Nuclear Reaction} & Ref. \\ 
\hline
 $^{7}$Li+p $\to$ $^{4}$He+$^{4}$He+$\gamma$   &  \cite{Xu13}  & $^{7}$Be+n $\to$ $^{4}$He+$^{4}$He &   \cite{Barbagallo16}  \\ 
 $^{7}$Be+D $\to$ $^{4}$He+$^{4}$He+p   &  \cite{Caughlan88} &  D+$^{4}$He $\to$ $^{6}$Li+$\gamma$    &   \cite{Hammache10}   \\ 
    $^{6}$Li+p $\to$ $^{7}$Be+$\gamma$    &   \cite{Xu13} &  $^{6}$Li+p $\to$ $^{3}$He+$^{4}$He & \cite{Xu13}    \\ 
 $^{8}$B+n $\to$ $^{4}$He+$^{4}$He+p &     \cite{Goriely08} &  $^{6}$Li+$^{3}$He $\to$ $^{4}$He+$^{4}$He+p   &  \cite{Goriely08} \\ 
    $^{6}$Li+$^{3}$H $\to$ $^{4}$He+$^{4}$He+n  &   \cite{Goriely08} &   $^{6}$Li+$^{3}$H $\to$ $^{8}$Li+p    &   \cite{Goriely08}    \\ 
    $^{7}$Li+$^{3}$He $\to$ $^{6}$Li+$^{4}$He      &   \cite{Goriely08}  &  $^{8}$Li+$^{3}$He $\to$ $^{7}$Li+$^{4}$He   &   \cite{Goriely08}   \\ 
$^{7}$Be+$^{3}$H $\to$ $^{6}$Li+$^{4}$He &   \cite{Goriely08}  &  $^{8}$B+$^{3}$H $\to$ $^{7}$Be+$^{4}$He &     \cite{Goriely08}   \\ 
  $^{8}$B+n $\to$ $^{6}$Li+$^{3}$He  &     \cite{Goriely08}  &    $^{8}$B+n $\to$ $^{7}$Be+D   &    \cite{Goriely08}  \\ 
    $^{6}$Li+$^{3}$H $\to$ $^{7}$Li+D    &     \cite{Goriely08} &   $^{6}$Li+$^{3}$He $\to$ $^{7}$Be+D     &    \cite{Goriely08}    \\ 
     $^{7}$Li+$^{3}$He $\to$ $^{4}$He+$^{4}$He+D        &   \cite{Goriely08}  &  $^{8}$Li+$^{3}$He $\to$ $^{4}$He+$^{4}$He+$^{3}$H      &    \cite{Goriely08}   \\ 
$^{7}$Be+$^{3}$H $\to$ $^{4}$He+$^{4}$He+D  &   \cite{Goriely08}  & $^{7}$Be+$^{3}$H $\to$ $^{7}$Li+$^{3}$He&   \cite{Goriely08}   \\  
$^{8}$B+D $\to$ $^{7}$Be+$^{3}$He &  \cite{Goriely08} & $^{8}$B+$^{3}$H $\to$ $^{4}$He+$^{4}$He+$^{3}$He &  \cite{Goriely08}    \\ 
$^{7}$Be+$^{3}$He $\to$ p+p+$^{4}$He+$^{4}$He &   \cite{Goriely08}  & D+D $\to$ $^{4}$He+$\gamma$ &   \cite{Xu13}   \\ 
 $^{3}$He+$^{3}$He $\to$ $^{4}$He+p+p &  \cite{Xu13}  & $^{7}$Be+p $\to$ $^{8}$B+$\gamma$ &  \cite{Xu13}   \\ 
$^{7}$Li+D $\to$ $^{4}$He+$^{4}$He+n &  \cite{Coc11}  &  D+n $\to$ $^{3}$H+$\gamma$ &     \cite{Nagai06}  \\ 
$^{3}$H+$^{3}$H $\to$ $^{4}$He+n+n &    \cite{Nagai06}  & $^{3}$He+n $\to$ $^{4}$He+$\gamma$ &   \cite{Wagoner69}   \\ 
$^{3}$He+$^{3}$H $\to$ $^{4}$He+D &    \cite{Caughlan88}  & $^{3}$He+$^{3}$H $\to$ $^{4}$He+n+p &     \cite{Caughlan88}  \\ 
 $^{7}$Li+$^{3}$H $\to$ $^{4}$He+$^{4}$He+n+n & \cite{Caughlan88}, \cite{Malaney89}  & $^{7}$Li+$^{3}$He $\to$ $^{4}$He+$^{4}$He+n+p &  \cite{Caughlan88}, \cite{Malaney89}    \\ 
$^{8}$Li+D $\to$ $^{7}$Li+$^{3}$H &  \cite{Hashimoto09} & $^{7}$Be+$^{3}$H $\to$ $^{4}$He+$^{4}$He+n+p &   \cite{Caughlan88}, \cite{Malaney89}     \\ 
$^{7}$Be+$^{3}$He $\to$ $^{4}$He+$^{4}$He+p+p &  \cite{Caughlan88},  \cite{Malaney89}   & $^{6}$Li+n $\to$ $^{3}$H+$^{4}$He &   \cite{Caughlan88}   \\ 
$^{3}$He+$^{3}$H $\to$ $^{6}$Li+$\gamma$ &  \cite{Fukugita90}  & $^{4}$He+n+p $\to$ $^{6}$Li+$\gamma$ &    \cite{Caughlan88}  \\ 
$^{6}$Li+n $\to$ $^{7}$Li+$\gamma$ &   \cite{Malaney89} & $^{6}$Li+D $\to$ $^{7}$Li+p &   \cite{Malaney89}   \\ 
$^{6}$Li+D $\to$ $^{7}$Be+n &   \cite{Malaney89}  & $^{7}$Li+n $\to$ $^{8}$Li+$\gamma$ &  \cite{Malaney89}, \cite{Heil89}   \\ 
$^{7}$Li+D $\to$ $^{8}$Li+p &  \cite{Malaney89}   & $^{8}$Li+p $\to$ $^{4}$He+$^{4}$He+n &    \cite{Mendes18}   \\ 
$^{4}$He+n+n $\to$ $^{6}$He+$\gamma$ &   \cite{Efros96}    & p+p+n $\to$ D+p &   \cite{Caughlan88}  \\ 
$^{7}$Li+$^{3}$H $\to$ $^{4}$He+$^{4}$He+n+n &   \cite{Caughlan88}, \cite{Malaney89}  & &  \\ 
\hline
\end{tabular}
\caption{{\it Nuclear processes beyond the key ones implemented in the package }\texttt{PRyMordial}{,\it with related references. Those processes are particularly needed for a precise prediction of the primordial abundance of lithium-7. Notice that the compilation above is part of the larger one present in the code }\texttt{PRIMAT}{\it~\cite{Pitrou:2018cgg}.}}
\label{table:12_63}
\end{table}

\end{appendix}

\newpage
\acknowledgments We are grateful to Cara Giovanetti and Federico Bianchini for providing us valuable feedback for the present release after $\beta$-testing \texttt{PRyMordial}. We acknowledge Kevork Abazajian, Kim Berghaus, Federico Bianchini, Miguel Escudero, Rouven Essig, Cara Giovanetti, Seyda Ipek, Mariangela Lisanti, Hongwan Liu, Jessie Shelton for discussion. We are in debt to all of the authors of \texttt{AlterBBN}, \texttt{NUDEC\_BSM}, \texttt{PArthENoPE}, and \texttt{PRIMAT} for making their codes publicly accessible: The present work and the release of \texttt{PRyMordial} greatly benefited from the open-source community. 

M.V. is supported in part by the Simons Foundation under the Simons Bridge for Postdoctoral Fellowships at SCGP and YITP, award number 815892. T.M.P.T. is supported in part by the U.S. National Science Foundation under Grant PHY-2210283.
This work was performed in part at Aspen Center for Physics, which is supported by National Science Foundation grant PHY-2210452.

\bibliography{main}

\begin{thebibliography}{127}
\expandafter\ifx\csname natexlab\endcsname\relax\def\natexlab#1{#1}\fi
\expandafter\ifx\csname bibnamefont\endcsname\relax
  \def\bibnamefont#1{#1}\fi
\expandafter\ifx\csname bibfnamefont\endcsname\relax
  \def\bibfnamefont#1{#1}\fi
\expandafter\ifx\csname citenamefont\endcsname\relax
  \def\citenamefont#1{#1}\fi
\expandafter\ifx\csname url\endcsname\relax
  \def\url#1{\texttt{#1}}\fi
\expandafter\ifx\csname urlprefix\endcsname\relax\def\urlprefix{URL }\fi
\providecommand{\bibinfo}[2]{#2}
\providecommand{\eprint}[2][]{\url{#2}}

\bibitem[{\citenamefont{Weinberg}(1977)}]{Weinberg:1977ji}
\bibinfo{author}{\bibfnamefont{S.}~\bibnamefont{Weinberg}},
  \emph{\bibinfo{title}{{The First Three Minutes. A Modern View of the Origin
  of the Universe}}} (\bibinfo{publisher}{Basic Books}, \bibinfo{year}{1977}),
  ISBN \bibinfo{isbn}{978-0-465-02437-7}.

\bibitem[{\citenamefont{Alpher et~al.}(1948)\citenamefont{Alpher, Bethe, and
  Gamow}}]{PhysRev.73.803}
\bibinfo{author}{\bibfnamefont{R.~A.} \bibnamefont{Alpher}},
  \bibinfo{author}{\bibfnamefont{H.}~\bibnamefont{Bethe}}, \bibnamefont{and}
  \bibinfo{author}{\bibfnamefont{G.}~\bibnamefont{Gamow}},
  \bibinfo{journal}{Phys. Rev.} \textbf{\bibinfo{volume}{73}},
  \bibinfo{pages}{803} (\bibinfo{year}{1948}),
  \urlprefix\url{https://link.aps.org/doi/10.1103/PhysRev.73.803}.

\bibitem[{\citenamefont{{Penzias} and {Wilson}}(1965)}]{1965ApJ...142..419P}
\bibinfo{author}{\bibfnamefont{A.~A.} \bibnamefont{{Penzias}}}
  \bibnamefont{and} \bibinfo{author}{\bibfnamefont{R.~W.}
  \bibnamefont{{Wilson}}}, \bibinfo{journal}{\apj}
  \textbf{\bibinfo{volume}{142}}, \bibinfo{pages}{419} (\bibinfo{year}{1965}).

\bibitem[{\citenamefont{{Dicke} et~al.}(1965)\citenamefont{{Dicke}, {Peebles},
  {Roll}, and {Wilkinson}}}]{1965ApJ...142..414D}
\bibinfo{author}{\bibfnamefont{R.~H.} \bibnamefont{{Dicke}}},
  \bibinfo{author}{\bibfnamefont{P.~J.~E.} \bibnamefont{{Peebles}}},
  \bibinfo{author}{\bibfnamefont{P.~G.} \bibnamefont{{Roll}}},
  \bibnamefont{and} \bibinfo{author}{\bibfnamefont{D.~T.}
  \bibnamefont{{Wilkinson}}}, \bibinfo{journal}{\apj}
  \textbf{\bibinfo{volume}{142}}, \bibinfo{pages}{414} (\bibinfo{year}{1965}).

\bibitem[{\citenamefont{Schramm and
  Wagoner}(1977)}]{doi:10.1146/annurev.ns.27.120177.000345}
\bibinfo{author}{\bibfnamefont{D.~N.} \bibnamefont{Schramm}} \bibnamefont{and}
  \bibinfo{author}{\bibfnamefont{R.~V.} \bibnamefont{Wagoner}},
  \bibinfo{journal}{Annual Review of Nuclear Science}
  \textbf{\bibinfo{volume}{27}}, \bibinfo{pages}{37} (\bibinfo{year}{1977}),
  \urlprefix\url{https://doi.org/10.1146/annurev.ns.27.120177.000345}.

\bibitem[{\citenamefont{Sarkar}(1996)}]{Sarkar:1995dd}
\bibinfo{author}{\bibfnamefont{S.}~\bibnamefont{Sarkar}},
  \bibinfo{journal}{Rept. Prog. Phys.} \textbf{\bibinfo{volume}{59}},
  \bibinfo{pages}{1493} (\bibinfo{year}{1996}),
  \eprint{https://arxiv.org/abs/hep-ph/9602260}.

\bibitem[{\citenamefont{Olive et~al.}(2000)\citenamefont{Olive, Steigman, and
  Walker}}]{Olive:1999ij}
\bibinfo{author}{\bibfnamefont{K.~A.} \bibnamefont{Olive}},
  \bibinfo{author}{\bibfnamefont{G.}~\bibnamefont{Steigman}}, \bibnamefont{and}
  \bibinfo{author}{\bibfnamefont{T.~P.} \bibnamefont{Walker}},
  \bibinfo{journal}{Phys. Rept.} \textbf{\bibinfo{volume}{333}},
  \bibinfo{pages}{389} (\bibinfo{year}{2000}),
  \eprint{https://arxiv.org/abs/astro-ph/9905320}.

\bibitem[{\citenamefont{Steigman}(2007)}]{Steigman:2007xt}
\bibinfo{author}{\bibfnamefont{G.}~\bibnamefont{Steigman}},
  \bibinfo{journal}{Ann. Rev. Nucl. Part. Sci.} \textbf{\bibinfo{volume}{57}},
  \bibinfo{pages}{463} (\bibinfo{year}{2007}),
  \eprint{https://arxiv.org/abs/0712.1100}.

\bibitem[{\citenamefont{Pospelov and Pradler}(2010)}]{Pospelov:2010hj}
\bibinfo{author}{\bibfnamefont{M.}~\bibnamefont{Pospelov}} \bibnamefont{and}
  \bibinfo{author}{\bibfnamefont{J.}~\bibnamefont{Pradler}},
  \bibinfo{journal}{Ann. Rev. Nucl. Part. Sci.} \textbf{\bibinfo{volume}{60}},
  \bibinfo{pages}{539} (\bibinfo{year}{2010}),
  \eprint{https://arxiv.org/abs/1011.1054}.

\bibitem[{\citenamefont{Cyburt et~al.}(2016)\citenamefont{Cyburt, Fields,
  Olive, and Yeh}}]{Cyburt:2015mya}
\bibinfo{author}{\bibfnamefont{R.~H.} \bibnamefont{Cyburt}},
  \bibinfo{author}{\bibfnamefont{B.~D.} \bibnamefont{Fields}},
  \bibinfo{author}{\bibfnamefont{K.~A.} \bibnamefont{Olive}}, \bibnamefont{and}
  \bibinfo{author}{\bibfnamefont{T.-H.} \bibnamefont{Yeh}},
  \bibinfo{journal}{Rev. Mod. Phys.} \textbf{\bibinfo{volume}{88}},
  \bibinfo{pages}{015004} (\bibinfo{year}{2016}),
  \eprint{https://arxiv.org/abs/1505.01076}.

\bibitem[{\citenamefont{Grohs and Fuller}(2023)}]{Grohs:2023voo}
\bibinfo{author}{\bibfnamefont{E.}~\bibnamefont{Grohs}} \bibnamefont{and}
  \bibinfo{author}{\bibfnamefont{G.~M.} \bibnamefont{Fuller}}
  (\bibinfo{year}{2023}), \eprint{https://arxiv.org/abs/2301.12299}.

\bibitem[{\citenamefont{Esteban et~al.}(2020)\citenamefont{Esteban,
  Gonzalez-Garcia, Maltoni, Schwetz, and Zhou}}]{Esteban:2020cvm}
\bibinfo{author}{\bibfnamefont{I.}~\bibnamefont{Esteban}},
  \bibinfo{author}{\bibfnamefont{M.~C.} \bibnamefont{Gonzalez-Garcia}},
  \bibinfo{author}{\bibfnamefont{M.}~\bibnamefont{Maltoni}},
  \bibinfo{author}{\bibfnamefont{T.}~\bibnamefont{Schwetz}}, \bibnamefont{and}
  \bibinfo{author}{\bibfnamefont{A.}~\bibnamefont{Zhou}},
  \bibinfo{journal}{JHEP} \textbf{\bibinfo{volume}{09}}, \bibinfo{pages}{178}
  (\bibinfo{year}{2020}), \eprint{https://arxiv.org/abs/2007.14792}.

\bibitem[{\citenamefont{Akita and Yamaguchi}(2022)}]{Akita:2022hlx}
\bibinfo{author}{\bibfnamefont{K.}~\bibnamefont{Akita}} \bibnamefont{and}
  \bibinfo{author}{\bibfnamefont{M.}~\bibnamefont{Yamaguchi}},
  \bibinfo{journal}{Universe} \textbf{\bibinfo{volume}{8}},
  \bibinfo{pages}{552} (\bibinfo{year}{2022}),
  \eprint{https://arxiv.org/abs/2210.10307}.

\bibitem[{\citenamefont{Bennett et~al.}(2021)\citenamefont{Bennett, Buldgen,
  De~Salas, Drewes, Gariazzo, Pastor, and Wong}}]{Bennett:2020zkv}
\bibinfo{author}{\bibfnamefont{J.~J.} \bibnamefont{Bennett}},
  \bibinfo{author}{\bibfnamefont{G.}~\bibnamefont{Buldgen}},
  \bibinfo{author}{\bibfnamefont{P.~F.} \bibnamefont{De~Salas}},
  \bibinfo{author}{\bibfnamefont{M.}~\bibnamefont{Drewes}},
  \bibinfo{author}{\bibfnamefont{S.}~\bibnamefont{Gariazzo}},
  \bibinfo{author}{\bibfnamefont{S.}~\bibnamefont{Pastor}}, \bibnamefont{and}
  \bibinfo{author}{\bibfnamefont{Y.~Y.~Y.} \bibnamefont{Wong}},
  \bibinfo{journal}{JCAP} \textbf{\bibinfo{volume}{04}}, \bibinfo{pages}{073}
  (\bibinfo{year}{2021}), \eprint{https://arxiv.org/abs/2012.02726}.

\bibitem[{\citenamefont{Froustey et~al.}(2020)\citenamefont{Froustey, Pitrou,
  and Volpe}}]{Froustey:2020mcq}
\bibinfo{author}{\bibfnamefont{J.}~\bibnamefont{Froustey}},
  \bibinfo{author}{\bibfnamefont{C.}~\bibnamefont{Pitrou}}, \bibnamefont{and}
  \bibinfo{author}{\bibfnamefont{M.~C.} \bibnamefont{Volpe}},
  \bibinfo{journal}{JCAP} \textbf{\bibinfo{volume}{12}}, \bibinfo{pages}{015}
  (\bibinfo{year}{2020}), \eprint{https://arxiv.org/abs/2008.01074}.

\bibitem[{\citenamefont{Akita and Yamaguchi}(2020)}]{Akita20}
\bibinfo{author}{\bibfnamefont{K.}~\bibnamefont{Akita}} \bibnamefont{and}
  \bibinfo{author}{\bibfnamefont{M.}~\bibnamefont{Yamaguchi}},
  \bibinfo{journal}{JCAP} \textbf{\bibinfo{volume}{08}}, \bibinfo{pages}{012}
  (\bibinfo{year}{2020}), \eprint{https://arxiv.org/abs/2005.07047}.

\bibitem[{\citenamefont{{Planck Collaboration}
  et~al.}(2016)\citenamefont{{Planck Collaboration}, {Ade, P. A. R.}, {Aghanim,
  N.}, {Arnaud, M.}, {Ashdown, M.}, {Aumont, J.}, {Baccigalupi, C.}, {Banday,
  A. J.}, {Barreiro, R. B.}, {Bartlett, J. G.} et~al.}}]{Ade16}
\bibinfo{author}{\bibnamefont{{Planck Collaboration}}},
  \bibinfo{author}{\bibnamefont{{Ade, P. A. R.}}},
  \bibinfo{author}{\bibnamefont{{Aghanim, N.}}},
  \bibinfo{author}{\bibnamefont{{Arnaud, M.}}},
  \bibinfo{author}{\bibnamefont{{Ashdown, M.}}},
  \bibinfo{author}{\bibnamefont{{Aumont, J.}}},
  \bibinfo{author}{\bibnamefont{{Baccigalupi, C.}}},
  \bibinfo{author}{\bibnamefont{{Banday, A. J.}}},
  \bibinfo{author}{\bibnamefont{{Barreiro, R. B.}}},
  \bibinfo{author}{\bibnamefont{{Bartlett, J. G.}}}, \bibnamefont{et~al.},
  \bibinfo{journal}{A\&A} \textbf{\bibinfo{volume}{594}}, \bibinfo{pages}{A13}
  (\bibinfo{year}{2016}),
  \urlprefix\url{https://doi.org/10.1051/0004-6361/201525830}.

\bibitem[{\citenamefont{Canetti et~al.}(2012)\citenamefont{Canetti, Drewes, and
  Shaposhnikov}}]{Canetti:2012zc}
\bibinfo{author}{\bibfnamefont{L.}~\bibnamefont{Canetti}},
  \bibinfo{author}{\bibfnamefont{M.}~\bibnamefont{Drewes}}, \bibnamefont{and}
  \bibinfo{author}{\bibfnamefont{M.}~\bibnamefont{Shaposhnikov}},
  \bibinfo{journal}{New J. Phys.} \textbf{\bibinfo{volume}{14}},
  \bibinfo{pages}{095012} (\bibinfo{year}{2012}),
  \eprint{https://arxiv.org/abs/1204.4186}.

\bibitem[{\citenamefont{Serpico and Raffelt}(2005)}]{Serpico:2005bc}
\bibinfo{author}{\bibfnamefont{P.~D.} \bibnamefont{Serpico}} \bibnamefont{and}
  \bibinfo{author}{\bibfnamefont{G.~G.} \bibnamefont{Raffelt}},
  \bibinfo{journal}{Phys. Rev. D} \textbf{\bibinfo{volume}{71}},
  \bibinfo{pages}{127301} (\bibinfo{year}{2005}),
  \eprint{https://arxiv.org/abs/astro-ph/0506162}.

\bibitem[{\citenamefont{Riemer-S\o{}rensen and
  Jenssen}(2017)}]{RiemerSorensen:2017vx}
\bibinfo{author}{\bibfnamefont{S.}~\bibnamefont{Riemer-S\o{}rensen}}
  \bibnamefont{and} \bibinfo{author}{\bibfnamefont{E.~S.}
  \bibnamefont{Jenssen}}, \bibinfo{journal}{Universe}
  \textbf{\bibinfo{volume}{3}}, \bibinfo{pages}{44} (\bibinfo{year}{2017}),
  \eprint{https://arxiv.org/abs/1705.03653}.

\bibitem[{\citenamefont{Cooke et~al.}(2018)\citenamefont{Cooke, Pettini, and
  Steidel}}]{Cooke:2017cwo}
\bibinfo{author}{\bibfnamefont{R.~J.} \bibnamefont{Cooke}},
  \bibinfo{author}{\bibfnamefont{M.}~\bibnamefont{Pettini}}, \bibnamefont{and}
  \bibinfo{author}{\bibfnamefont{C.~C.} \bibnamefont{Steidel}},
  \bibinfo{journal}{Astrophys. J.} \textbf{\bibinfo{volume}{855}},
  \bibinfo{pages}{102} (\bibinfo{year}{2018}),
  \eprint{https://arxiv.org/abs/1710.11129}.

\bibitem[{\citenamefont{{Hsyu} et~al.}(2020)\citenamefont{{Hsyu}, {Cooke},
  {Prochaska}, and {Bolte}}}]{2020ApJ89677H}
\bibinfo{author}{\bibfnamefont{T.}~\bibnamefont{{Hsyu}}},
  \bibinfo{author}{\bibfnamefont{R.~J.} \bibnamefont{{Cooke}}},
  \bibinfo{author}{\bibfnamefont{J.~X.} \bibnamefont{{Prochaska}}},
  \bibnamefont{and} \bibinfo{author}{\bibfnamefont{M.}~\bibnamefont{{Bolte}}},
  \bibinfo{journal}{\apj} \textbf{\bibinfo{volume}{896}}, \bibinfo{eid}{77}
  (\bibinfo{year}{2020}), \eprint{https://arxiv.org/abs/2005.12290}.

\bibitem[{\citenamefont{Aver et~al.}(2021)\citenamefont{Aver, Berg, Olive,
  Pogge, Salzer, and Skillman}}]{Aver:2020fon}
\bibinfo{author}{\bibfnamefont{E.}~\bibnamefont{Aver}},
  \bibinfo{author}{\bibfnamefont{D.~A.} \bibnamefont{Berg}},
  \bibinfo{author}{\bibfnamefont{K.~A.} \bibnamefont{Olive}},
  \bibinfo{author}{\bibfnamefont{R.~W.} \bibnamefont{Pogge}},
  \bibinfo{author}{\bibfnamefont{J.~J.} \bibnamefont{Salzer}},
  \bibnamefont{and} \bibinfo{author}{\bibfnamefont{E.~D.}
  \bibnamefont{Skillman}}, \bibinfo{journal}{JCAP}
  \textbf{\bibinfo{volume}{03}}, \bibinfo{pages}{027} (\bibinfo{year}{2021}),
  \eprint{https://arxiv.org/abs/2010.04180}.

\bibitem[{\citenamefont{Aghanim et~al.}(2020)}]{Planck20}
\bibinfo{author}{\bibfnamefont{N.}~\bibnamefont{Aghanim}} \bibnamefont{et~al.}
  (\bibinfo{collaboration}{Planck}), \bibinfo{journal}{Astron. Astrophys.}
  \textbf{\bibinfo{volume}{641}}, \bibinfo{pages}{A6} (\bibinfo{year}{2020}),
  \bibinfo{note}{[Erratum: Astron.Astrophys. 652, C4 (2021)]},
  \eprint{https://arxiv.org/abs/1807.06209}.

\bibitem[{\citenamefont{Yeh et~al.}(2022)\citenamefont{Yeh, Shelton, Olive, and
  Fields}}]{Yeh:2022heq}
\bibinfo{author}{\bibfnamefont{T.-H.} \bibnamefont{Yeh}},
  \bibinfo{author}{\bibfnamefont{J.}~\bibnamefont{Shelton}},
  \bibinfo{author}{\bibfnamefont{K.~A.} \bibnamefont{Olive}}, \bibnamefont{and}
  \bibinfo{author}{\bibfnamefont{B.~D.} \bibnamefont{Fields}},
  \bibinfo{journal}{JCAP} \textbf{\bibinfo{volume}{10}}, \bibinfo{pages}{046}
  (\bibinfo{year}{2022}), \eprint{https://arxiv.org/abs/2207.13133}.

\bibitem[{\citenamefont{Boehm et~al.}(2013)\citenamefont{Boehm, Dolan, and
  McCabe}}]{Boehm13}
\bibinfo{author}{\bibfnamefont{C.}~\bibnamefont{Boehm}},
  \bibinfo{author}{\bibfnamefont{M.~J.} \bibnamefont{Dolan}}, \bibnamefont{and}
  \bibinfo{author}{\bibfnamefont{C.}~\bibnamefont{McCabe}},
  \bibinfo{journal}{JCAP} \textbf{\bibinfo{volume}{08}}, \bibinfo{pages}{041}
  (\bibinfo{year}{2013}), \eprint{https://arxiv.org/abs/1303.6270}.

\bibitem[{\citenamefont{Hardy et~al.}(2015)\citenamefont{Hardy, Lasenby,
  March-Russell, and West}}]{Hardy:2014mqa}
\bibinfo{author}{\bibfnamefont{E.}~\bibnamefont{Hardy}},
  \bibinfo{author}{\bibfnamefont{R.}~\bibnamefont{Lasenby}},
  \bibinfo{author}{\bibfnamefont{J.}~\bibnamefont{March-Russell}},
  \bibnamefont{and} \bibinfo{author}{\bibfnamefont{S.~M.} \bibnamefont{West}},
  \bibinfo{journal}{JHEP} \textbf{\bibinfo{volume}{06}}, \bibinfo{pages}{011}
  (\bibinfo{year}{2015}), \eprint{https://arxiv.org/abs/1411.3739}.

\bibitem[{\citenamefont{Alvey et~al.}(2020)\citenamefont{Alvey, Sabti,
  Escudero, and Fairbairn}}]{Alvey:2019ctk}
\bibinfo{author}{\bibfnamefont{J.}~\bibnamefont{Alvey}},
  \bibinfo{author}{\bibfnamefont{N.}~\bibnamefont{Sabti}},
  \bibinfo{author}{\bibfnamefont{M.}~\bibnamefont{Escudero}}, \bibnamefont{and}
  \bibinfo{author}{\bibfnamefont{M.}~\bibnamefont{Fairbairn}},
  \bibinfo{journal}{Eur. Phys. J. C} \textbf{\bibinfo{volume}{80}},
  \bibinfo{pages}{148} (\bibinfo{year}{2020}),
  \eprint{https://arxiv.org/abs/1910.10730}.

\bibitem[{\citenamefont{Sibiryakov et~al.}(2020)\citenamefont{Sibiryakov,
  S\o{}rensen, and Yu}}]{Sibiryakov:2020eir}
\bibinfo{author}{\bibfnamefont{S.}~\bibnamefont{Sibiryakov}},
  \bibinfo{author}{\bibfnamefont{P.}~\bibnamefont{S\o{}rensen}},
  \bibnamefont{and} \bibinfo{author}{\bibfnamefont{T.-T.} \bibnamefont{Yu}},
  \bibinfo{journal}{JHEP} \textbf{\bibinfo{volume}{12}}, \bibinfo{pages}{075}
  (\bibinfo{year}{2020}), \eprint{https://arxiv.org/abs/2006.04820}.

\bibitem[{\citenamefont{Mahbubani et~al.}(2021)\citenamefont{Mahbubani, Redi,
  and Tesi}}]{Mahbubani:2020knq}
\bibinfo{author}{\bibfnamefont{R.}~\bibnamefont{Mahbubani}},
  \bibinfo{author}{\bibfnamefont{M.}~\bibnamefont{Redi}}, \bibnamefont{and}
  \bibinfo{author}{\bibfnamefont{A.}~\bibnamefont{Tesi}},
  \bibinfo{journal}{JCAP} \textbf{\bibinfo{volume}{02}}, \bibinfo{pages}{039}
  (\bibinfo{year}{2021}), \eprint{https://arxiv.org/abs/2007.07231}.

\bibitem[{\citenamefont{Sabti et~al.}(2020)\citenamefont{Sabti, Alvey,
  Escudero, Fairbairn, and Blas}}]{Sabti:2019mhn}
\bibinfo{author}{\bibfnamefont{N.}~\bibnamefont{Sabti}},
  \bibinfo{author}{\bibfnamefont{J.}~\bibnamefont{Alvey}},
  \bibinfo{author}{\bibfnamefont{M.}~\bibnamefont{Escudero}},
  \bibinfo{author}{\bibfnamefont{M.}~\bibnamefont{Fairbairn}},
  \bibnamefont{and} \bibinfo{author}{\bibfnamefont{D.}~\bibnamefont{Blas}},
  \bibinfo{journal}{JCAP} \textbf{\bibinfo{volume}{01}}, \bibinfo{pages}{004}
  (\bibinfo{year}{2020}), \eprint{https://arxiv.org/abs/1910.01649}.

\bibitem[{\citenamefont{Depta et~al.}(2021)\citenamefont{Depta, Hufnagel, and
  Schmidt-Hoberg}}]{Depta:2020zbh}
\bibinfo{author}{\bibfnamefont{P.~F.} \bibnamefont{Depta}},
  \bibinfo{author}{\bibfnamefont{M.}~\bibnamefont{Hufnagel}}, \bibnamefont{and}
  \bibinfo{author}{\bibfnamefont{K.}~\bibnamefont{Schmidt-Hoberg}},
  \bibinfo{journal}{JCAP} \textbf{\bibinfo{volume}{04}}, \bibinfo{pages}{011}
  (\bibinfo{year}{2021}), \eprint{https://arxiv.org/abs/2011.06519}.

\bibitem[{\citenamefont{Giovanetti et~al.}(2022)\citenamefont{Giovanetti,
  Lisanti, Liu, and Ruderman}}]{Giovanetti21}
\bibinfo{author}{\bibfnamefont{C.}~\bibnamefont{Giovanetti}},
  \bibinfo{author}{\bibfnamefont{M.}~\bibnamefont{Lisanti}},
  \bibinfo{author}{\bibfnamefont{H.}~\bibnamefont{Liu}}, \bibnamefont{and}
  \bibinfo{author}{\bibfnamefont{J.~T.} \bibnamefont{Ruderman}},
  \bibinfo{journal}{Phys. Rev. Lett.} \textbf{\bibinfo{volume}{129}},
  \bibinfo{pages}{021302} (\bibinfo{year}{2022}),
  \eprint{https://arxiv.org/abs/2109.03246}.

\bibitem[{\citenamefont{Chu et~al.}(2022)\citenamefont{Chu, Kuo, and
  Pradler}}]{Chu:2022xuh}
\bibinfo{author}{\bibfnamefont{X.}~\bibnamefont{Chu}},
  \bibinfo{author}{\bibfnamefont{J.-L.} \bibnamefont{Kuo}}, \bibnamefont{and}
  \bibinfo{author}{\bibfnamefont{J.}~\bibnamefont{Pradler}},
  \bibinfo{journal}{Phys. Rev. D} \textbf{\bibinfo{volume}{106}},
  \bibinfo{pages}{055022} (\bibinfo{year}{2022}),
  \eprint{https://arxiv.org/abs/2205.05714}.

\bibitem[{\citenamefont{Burns et~al.}(2023)\citenamefont{Burns, Tait, and
  Valli}}]{Burns:2022hkq}
\bibinfo{author}{\bibfnamefont{A.-K.} \bibnamefont{Burns}},
  \bibinfo{author}{\bibfnamefont{T.~M.~P.} \bibnamefont{Tait}},
  \bibnamefont{and} \bibinfo{author}{\bibfnamefont{M.}~\bibnamefont{Valli}},
  \bibinfo{journal}{Phys. Rev. Lett.} \textbf{\bibinfo{volume}{130}},
  \bibinfo{pages}{131001} (\bibinfo{year}{2023}),
  \eprint{https://arxiv.org/abs/2206.00693}.

\bibitem[{\citenamefont{{Abazajian} et~al.}(2019)\citenamefont{{Abazajian},
  {Addison}, {Adshead}, {Ahmed}, {Allen}, {Alonso}, {Alvarez}, {Anderson},
  {Arnold}, {Baccigalupi} et~al.}}]{2019arXiv190704473A}
\bibinfo{author}{\bibfnamefont{K.}~\bibnamefont{{Abazajian}}},
  \bibinfo{author}{\bibfnamefont{G.}~\bibnamefont{{Addison}}},
  \bibinfo{author}{\bibfnamefont{P.}~\bibnamefont{{Adshead}}},
  \bibinfo{author}{\bibfnamefont{Z.}~\bibnamefont{{Ahmed}}},
  \bibinfo{author}{\bibfnamefont{S.~W.} \bibnamefont{{Allen}}},
  \bibinfo{author}{\bibfnamefont{D.}~\bibnamefont{{Alonso}}},
  \bibinfo{author}{\bibfnamefont{M.}~\bibnamefont{{Alvarez}}},
  \bibinfo{author}{\bibfnamefont{A.}~\bibnamefont{{Anderson}}},
  \bibinfo{author}{\bibfnamefont{K.~S.} \bibnamefont{{Arnold}}},
  \bibinfo{author}{\bibfnamefont{C.}~\bibnamefont{{Baccigalupi}}},
  \bibnamefont{et~al.}, \bibinfo{journal}{arXiv e-prints}
  \bibinfo{eid}{arXiv:1907.04473} (\bibinfo{year}{2019}),
  \eprint{https://arxiv.org/abs/1907.04473}.

\bibitem[{\citenamefont{{Lee} et~al.}(2019)\citenamefont{{Lee}, {Abitbol},
  {Adachi}, {Ade}, {Aguirre}, {Ahmed}, {Aiola}, {Ali}, {Alonso}, {Alvarez}
  et~al.}}]{2019BAAS51g147L}
\bibinfo{author}{\bibfnamefont{A.}~\bibnamefont{{Lee}}},
  \bibinfo{author}{\bibfnamefont{M.~H.} \bibnamefont{{Abitbol}}},
  \bibinfo{author}{\bibfnamefont{S.}~\bibnamefont{{Adachi}}},
  \bibinfo{author}{\bibfnamefont{P.}~\bibnamefont{{Ade}}},
  \bibinfo{author}{\bibfnamefont{J.}~\bibnamefont{{Aguirre}}},
  \bibinfo{author}{\bibfnamefont{Z.}~\bibnamefont{{Ahmed}}},
  \bibinfo{author}{\bibfnamefont{S.}~\bibnamefont{{Aiola}}},
  \bibinfo{author}{\bibfnamefont{A.}~\bibnamefont{{Ali}}},
  \bibinfo{author}{\bibfnamefont{D.}~\bibnamefont{{Alonso}}},
  \bibinfo{author}{\bibfnamefont{M.~A.} \bibnamefont{{Alvarez}}},
  \bibnamefont{et~al.}, in \emph{\bibinfo{booktitle}{Bulletin of the American
  Astronomical Society}} (\bibinfo{year}{2019}), vol.~\bibinfo{volume}{51}, p.
  \bibinfo{pages}{147}, \eprint{https://arxiv.org/abs/1907.08284}.

\bibitem[{\citenamefont{{Sehgal} et~al.}(2019)\citenamefont{{Sehgal}, {Aiola},
  {Akrami}, {Basu}, {Boylan-Kolchin}, {Bryan}, {Clesse}, {Cyr-Racine}, {Di
  Mascolo}, {Dicker} et~al.}}]{2019BAAS51g6S}
\bibinfo{author}{\bibfnamefont{N.}~\bibnamefont{{Sehgal}}},
  \bibinfo{author}{\bibfnamefont{S.}~\bibnamefont{{Aiola}}},
  \bibinfo{author}{\bibfnamefont{Y.}~\bibnamefont{{Akrami}}},
  \bibinfo{author}{\bibfnamefont{K.}~\bibnamefont{{Basu}}},
  \bibinfo{author}{\bibfnamefont{M.}~\bibnamefont{{Boylan-Kolchin}}},
  \bibinfo{author}{\bibfnamefont{S.}~\bibnamefont{{Bryan}}},
  \bibinfo{author}{\bibfnamefont{S.}~\bibnamefont{{Clesse}}},
  \bibinfo{author}{\bibfnamefont{F.-Y.} \bibnamefont{{Cyr-Racine}}},
  \bibinfo{author}{\bibfnamefont{L.}~\bibnamefont{{Di Mascolo}}},
  \bibinfo{author}{\bibfnamefont{S.}~\bibnamefont{{Dicker}}},
  \bibnamefont{et~al.}, in \emph{\bibinfo{booktitle}{Bulletin of the American
  Astronomical Society}} (\bibinfo{year}{2019}), vol.~\bibinfo{volume}{51},
  p.~\bibinfo{pages}{6}, \eprint{https://arxiv.org/abs/1906.10134}.

\bibitem[{\citenamefont{Grohs et~al.}(2019)\citenamefont{Grohs, Bond, Cooke,
  Fuller, Meyers, and Paris}}]{Grohs:2019cae}
\bibinfo{author}{\bibfnamefont{E.~B.} \bibnamefont{Grohs}},
  \bibinfo{author}{\bibfnamefont{J.~R.} \bibnamefont{Bond}},
  \bibinfo{author}{\bibfnamefont{R.~J.} \bibnamefont{Cooke}},
  \bibinfo{author}{\bibfnamefont{G.~M.} \bibnamefont{Fuller}},
  \bibinfo{author}{\bibfnamefont{J.}~\bibnamefont{Meyers}}, \bibnamefont{and}
  \bibinfo{author}{\bibfnamefont{M.~W.} \bibnamefont{Paris}}
  (\bibinfo{year}{2019}), \eprint{https://arxiv.org/abs/1903.09187}.

\bibitem[{\citenamefont{Lagu\"e and Meyers}(2020)}]{Lague:2019yvs}
\bibinfo{author}{\bibfnamefont{A.}~\bibnamefont{Lagu\"e}} \bibnamefont{and}
  \bibinfo{author}{\bibfnamefont{J.}~\bibnamefont{Meyers}},
  \bibinfo{journal}{Phys. Rev. D} \textbf{\bibinfo{volume}{101}},
  \bibinfo{pages}{043509} (\bibinfo{year}{2020}),
  \eprint{https://arxiv.org/abs/1908.05291}.

\bibitem[{\citenamefont{Wagoner et~al.}(1967)\citenamefont{Wagoner, Fowler, and
  Hoyle}}]{Wagoner:1966pv}
\bibinfo{author}{\bibfnamefont{R.~V.} \bibnamefont{Wagoner}},
  \bibinfo{author}{\bibfnamefont{W.~A.} \bibnamefont{Fowler}},
  \bibnamefont{and} \bibinfo{author}{\bibfnamefont{F.}~\bibnamefont{Hoyle}},
  \bibinfo{journal}{Astrophys. J.} \textbf{\bibinfo{volume}{148}},
  \bibinfo{pages}{3} (\bibinfo{year}{1967}).

\bibitem[{\citenamefont{Workman et~al.}(2022)}]{PDG:2022pth}
\bibinfo{author}{\bibfnamefont{R.~L.} \bibnamefont{Workman}}
  \bibnamefont{et~al.} (\bibinfo{collaboration}{Particle Data Group}),
  \bibinfo{journal}{PTEP} \textbf{\bibinfo{volume}{2022}},
  \bibinfo{pages}{083C01} (\bibinfo{year}{2022}).

\bibitem[{\citenamefont{{Kawano}}(1992)}]{1992STIN9225163K}
\bibinfo{author}{\bibfnamefont{L.}~\bibnamefont{{Kawano}}},
  \emph{\bibinfo{title}{{Let's go: Early universe 2. Primordial nucleosynthesis
  the computer way}}} (\bibinfo{year}{1992}).

\bibitem[{\citenamefont{Pisanti et~al.}(2008)\citenamefont{Pisanti, Cirillo,
  Esposito, Iocco, Mangano, Miele, and Serpico}}]{Pisanti:2007hk}
\bibinfo{author}{\bibfnamefont{O.}~\bibnamefont{Pisanti}},
  \bibinfo{author}{\bibfnamefont{A.}~\bibnamefont{Cirillo}},
  \bibinfo{author}{\bibfnamefont{S.}~\bibnamefont{Esposito}},
  \bibinfo{author}{\bibfnamefont{F.}~\bibnamefont{Iocco}},
  \bibinfo{author}{\bibfnamefont{G.}~\bibnamefont{Mangano}},
  \bibinfo{author}{\bibfnamefont{G.}~\bibnamefont{Miele}}, \bibnamefont{and}
  \bibinfo{author}{\bibfnamefont{P.~D.} \bibnamefont{Serpico}},
  \bibinfo{journal}{Comput. Phys. Commun.} \textbf{\bibinfo{volume}{178}},
  \bibinfo{pages}{956} (\bibinfo{year}{2008}),
  \eprint{https://arxiv.org/abs/0705.0290}.

\bibitem[{\citenamefont{Consiglio et~al.}(2018)\citenamefont{Consiglio,
  de~Salas, Mangano, Miele, Pastor, and Pisanti}}]{Consiglio:2017pot}
\bibinfo{author}{\bibfnamefont{R.}~\bibnamefont{Consiglio}},
  \bibinfo{author}{\bibfnamefont{P.~F.} \bibnamefont{de~Salas}},
  \bibinfo{author}{\bibfnamefont{G.}~\bibnamefont{Mangano}},
  \bibinfo{author}{\bibfnamefont{G.}~\bibnamefont{Miele}},
  \bibinfo{author}{\bibfnamefont{S.}~\bibnamefont{Pastor}}, \bibnamefont{and}
  \bibinfo{author}{\bibfnamefont{O.}~\bibnamefont{Pisanti}},
  \bibinfo{journal}{Comput. Phys. Commun.} \textbf{\bibinfo{volume}{233}},
  \bibinfo{pages}{237} (\bibinfo{year}{2018}),
  \eprint{https://arxiv.org/abs/1712.04378}.

\bibitem[{\citenamefont{Gariazzo et~al.}(2022)\citenamefont{Gariazzo, F.~de
  Salas, Pisanti, and Consiglio}}]{Gariazzo:2021iiu}
\bibinfo{author}{\bibfnamefont{S.}~\bibnamefont{Gariazzo}},
  \bibinfo{author}{\bibfnamefont{P.}~\bibnamefont{F.~de Salas}},
  \bibinfo{author}{\bibfnamefont{O.}~\bibnamefont{Pisanti}}, \bibnamefont{and}
  \bibinfo{author}{\bibfnamefont{R.}~\bibnamefont{Consiglio}},
  \bibinfo{journal}{Comput. Phys. Commun.} \textbf{\bibinfo{volume}{271}},
  \bibinfo{pages}{108205} (\bibinfo{year}{2022}),
  \eprint{https://arxiv.org/abs/2103.05027}.

\bibitem[{\citenamefont{Pitrou et~al.}(2018{\natexlab{a}})\citenamefont{Pitrou,
  Coc, Uzan, and Vangioni}}]{Pitrou:2018cgg}
\bibinfo{author}{\bibfnamefont{C.}~\bibnamefont{Pitrou}},
  \bibinfo{author}{\bibfnamefont{A.}~\bibnamefont{Coc}},
  \bibinfo{author}{\bibfnamefont{J.-P.} \bibnamefont{Uzan}}, \bibnamefont{and}
  \bibinfo{author}{\bibfnamefont{E.}~\bibnamefont{Vangioni}},
  \bibinfo{journal}{Phys. Rept.} \textbf{\bibinfo{volume}{754}},
  \bibinfo{pages}{1} (\bibinfo{year}{2018}{\natexlab{a}}),
  \eprint{https://arxiv.org/abs/1801.08023}.

\bibitem[{\citenamefont{Pitrou et~al.}(2021{\natexlab{a}})\citenamefont{Pitrou,
  Coc, Uzan, and Vangioni}}]{Pitrou:2020etk}
\bibinfo{author}{\bibfnamefont{C.}~\bibnamefont{Pitrou}},
  \bibinfo{author}{\bibfnamefont{A.}~\bibnamefont{Coc}},
  \bibinfo{author}{\bibfnamefont{J.-P.} \bibnamefont{Uzan}}, \bibnamefont{and}
  \bibinfo{author}{\bibfnamefont{E.}~\bibnamefont{Vangioni}},
  \bibinfo{journal}{Mon. Not. Roy. Astron. Soc.}
  \textbf{\bibinfo{volume}{502}}, \bibinfo{pages}{2474}
  (\bibinfo{year}{2021}{\natexlab{a}}),
  \eprint{https://arxiv.org/abs/2011.11320}.

\bibitem[{\citenamefont{Arbey}(2012)}]{Arbey:2011nf}
\bibinfo{author}{\bibfnamefont{A.}~\bibnamefont{Arbey}},
  \bibinfo{journal}{Comput. Phys. Commun.} \textbf{\bibinfo{volume}{183}},
  \bibinfo{pages}{1822} (\bibinfo{year}{2012}),
  \eprint{https://arxiv.org/abs/1106.1363}.

\bibitem[{\citenamefont{Arbey et~al.}(2020)\citenamefont{Arbey, Auffinger,
  Hickerson, and Jenssen}}]{Arbey:2018zfh}
\bibinfo{author}{\bibfnamefont{A.}~\bibnamefont{Arbey}},
  \bibinfo{author}{\bibfnamefont{J.}~\bibnamefont{Auffinger}},
  \bibinfo{author}{\bibfnamefont{K.~P.} \bibnamefont{Hickerson}},
  \bibnamefont{and} \bibinfo{author}{\bibfnamefont{E.~S.}
  \bibnamefont{Jenssen}}, \bibinfo{journal}{Comput. Phys. Commun.}
  \textbf{\bibinfo{volume}{248}}, \bibinfo{pages}{106982}
  (\bibinfo{year}{2020}), \eprint{https://arxiv.org/abs/1806.11095}.

\bibitem[{\citenamefont{Madhavacheril et~al.}(2023)}]{ACT:2023kun}
\bibinfo{author}{\bibfnamefont{M.~S.} \bibnamefont{Madhavacheril}}
  \bibnamefont{et~al.} (\bibinfo{collaboration}{ACT}) (\bibinfo{year}{2023}),
  \eprint{https://arxiv.org/abs/2304.05203}.

\bibitem[{\citenamefont{Escudero}(2019)}]{Escudero19}
\bibinfo{author}{\bibfnamefont{M.}~\bibnamefont{Escudero}},
  \bibinfo{journal}{JCAP} \textbf{\bibinfo{volume}{02}}, \bibinfo{pages}{007}
  (\bibinfo{year}{2019}), \eprint{https://arxiv.org/abs/1812.05605}.

\bibitem[{\citenamefont{Escudero~Abenza}(2020)}]{Escudero20}
\bibinfo{author}{\bibfnamefont{M.}~\bibnamefont{Escudero~Abenza}},
  \bibinfo{journal}{JCAP} \textbf{\bibinfo{volume}{05}}, \bibinfo{pages}{048}
  (\bibinfo{year}{2020}), \eprint{https://arxiv.org/abs/2001.04466}.

\bibitem[{\citenamefont{Pitrou et~al.}(2021{\natexlab{b}})\citenamefont{Pitrou,
  Coc, Uzan, and Vangioni}}]{Pitrou:2021vqr}
\bibinfo{author}{\bibfnamefont{C.}~\bibnamefont{Pitrou}},
  \bibinfo{author}{\bibfnamefont{A.}~\bibnamefont{Coc}},
  \bibinfo{author}{\bibfnamefont{J.-P.} \bibnamefont{Uzan}}, \bibnamefont{and}
  \bibinfo{author}{\bibfnamefont{E.}~\bibnamefont{Vangioni}},
  \bibinfo{journal}{Nature Rev. Phys.} \textbf{\bibinfo{volume}{3}},
  \bibinfo{pages}{231} (\bibinfo{year}{2021}{\natexlab{b}}),
  \eprint{https://arxiv.org/abs/2104.11148}.

\bibitem[{\citenamefont{Sabti et~al.}(2021)\citenamefont{Sabti, Alvey,
  Escudero, Fairbairn, and Blas}}]{Sabti:2021reh}
\bibinfo{author}{\bibfnamefont{N.}~\bibnamefont{Sabti}},
  \bibinfo{author}{\bibfnamefont{J.}~\bibnamefont{Alvey}},
  \bibinfo{author}{\bibfnamefont{M.}~\bibnamefont{Escudero}},
  \bibinfo{author}{\bibfnamefont{M.}~\bibnamefont{Fairbairn}},
  \bibnamefont{and} \bibinfo{author}{\bibfnamefont{D.}~\bibnamefont{Blas}},
  \bibinfo{journal}{JCAP} \textbf{\bibinfo{volume}{08}}, \bibinfo{pages}{A01}
  (\bibinfo{year}{2021}), \eprint{https://arxiv.org/abs/2107.11232}.

\bibitem[{\citenamefont{{Foreman-Mackey}
  et~al.}(2013)\citenamefont{{Foreman-Mackey}, {Hogg}, {Lang}, and
  {Goodman}}}]{2013PASP..125..306F}
\bibinfo{author}{\bibfnamefont{D.}~\bibnamefont{{Foreman-Mackey}}},
  \bibinfo{author}{\bibfnamefont{D.~W.} \bibnamefont{{Hogg}}},
  \bibinfo{author}{\bibfnamefont{D.}~\bibnamefont{{Lang}}}, \bibnamefont{and}
  \bibinfo{author}{\bibfnamefont{J.}~\bibnamefont{{Goodman}}},
  \bibinfo{journal}{\pasp} \textbf{\bibinfo{volume}{125}}, \bibinfo{pages}{306}
  (\bibinfo{year}{2013}), \eprint{https://arxiv.org/abs/1202.3665}.

\bibitem[{\citenamefont{Schulz et~al.}(2021)\citenamefont{Schulz, Beaujean,
  Caldwell, Grunwald, Hafych, Kr{\"o}ninger, Cagnina, R{\"o}hrig, and
  Shtembari}}]{Schulz:2021BAT}
\bibinfo{author}{\bibfnamefont{O.}~\bibnamefont{Schulz}},
  \bibinfo{author}{\bibfnamefont{F.}~\bibnamefont{Beaujean}},
  \bibinfo{author}{\bibfnamefont{A.}~\bibnamefont{Caldwell}},
  \bibinfo{author}{\bibfnamefont{C.}~\bibnamefont{Grunwald}},
  \bibinfo{author}{\bibfnamefont{V.}~\bibnamefont{Hafych}},
  \bibinfo{author}{\bibfnamefont{K.}~\bibnamefont{Kr{\"o}ninger}},
  \bibinfo{author}{\bibfnamefont{S.~L.} \bibnamefont{Cagnina}},
  \bibinfo{author}{\bibfnamefont{L.}~\bibnamefont{R{\"o}hrig}},
  \bibnamefont{and}
  \bibinfo{author}{\bibfnamefont{L.}~\bibnamefont{Shtembari}},
  \bibinfo{journal}{SN Computer Science} \textbf{\bibinfo{volume}{2}},
  \bibinfo{pages}{210} (\bibinfo{year}{2021}), ISSN \bibinfo{issn}{2661-8907},
  \urlprefix\url{https://doi.org/10.1007/s42979-021-00626-4}.

\bibitem[{\citenamefont{Rackauckas and
  Nie}(2017)}]{rackauckas2017differentialequations}
\bibinfo{author}{\bibfnamefont{C.}~\bibnamefont{Rackauckas}} \bibnamefont{and}
  \bibinfo{author}{\bibfnamefont{Q.}~\bibnamefont{Nie}},
  \bibinfo{journal}{Journal of Open Research Software}
  \textbf{\bibinfo{volume}{5}}, \bibinfo{pages}{15} (\bibinfo{year}{2017}).

\bibitem[{\citenamefont{Kolb and Turner}(1990)}]{Kolb:1990vq}
\bibinfo{author}{\bibfnamefont{E.~W.} \bibnamefont{Kolb}} \bibnamefont{and}
  \bibinfo{author}{\bibfnamefont{M.~S.} \bibnamefont{Turner}},
  \emph{\bibinfo{title}{{The Early Universe}}}, vol.~\bibinfo{volume}{69}
  (\bibinfo{publisher}{CRC Press}, \bibinfo{year}{1990}), ISBN
  \bibinfo{isbn}{978-0-201-62674-2}.

\bibitem[{\citenamefont{Rubakov and Gorbunov}(2017)}]{Rubakov:2017xzr}
\bibinfo{author}{\bibfnamefont{V.~A.} \bibnamefont{Rubakov}} \bibnamefont{and}
  \bibinfo{author}{\bibfnamefont{D.~S.} \bibnamefont{Gorbunov}},
  \emph{\bibinfo{title}{{Introduction to the Theory of the Early Universe}:
  {Hot big bang theory}}} (\bibinfo{publisher}{World Scientific},
  \bibinfo{address}{Singapore}, \bibinfo{year}{2017}), ISBN
  \bibinfo{isbn}{978-981-320-987-9, 978-981-320-988-6, 978-981-322-005-8}.

\bibitem[{\citenamefont{Borsanyi et~al.}(2016)}]{Borsanyi:2016ksw}
\bibinfo{author}{\bibfnamefont{S.}~\bibnamefont{Borsanyi}}
  \bibnamefont{et~al.}, \bibinfo{journal}{Nature}
  \textbf{\bibinfo{volume}{539}}, \bibinfo{pages}{69} (\bibinfo{year}{2016}),
  \eprint{https://arxiv.org/abs/1606.07494}.

\bibitem[{\citenamefont{Dolgov et~al.}(2002)\citenamefont{Dolgov, Hansen,
  Pastor, Petcov, Raffelt, and Semikoz}}]{Dolgov:2002ab}
\bibinfo{author}{\bibfnamefont{A.~D.} \bibnamefont{Dolgov}},
  \bibinfo{author}{\bibfnamefont{S.~H.} \bibnamefont{Hansen}},
  \bibinfo{author}{\bibfnamefont{S.}~\bibnamefont{Pastor}},
  \bibinfo{author}{\bibfnamefont{S.~T.} \bibnamefont{Petcov}},
  \bibinfo{author}{\bibfnamefont{G.~G.} \bibnamefont{Raffelt}},
  \bibnamefont{and} \bibinfo{author}{\bibfnamefont{D.~V.}
  \bibnamefont{Semikoz}}, \bibinfo{journal}{Nucl. Phys. B}
  \textbf{\bibinfo{volume}{632}}, \bibinfo{pages}{363} (\bibinfo{year}{2002}),
  \eprint{https://arxiv.org/abs/hep-ph/0201287}.

\bibitem[{\citenamefont{Dolgov}(2002)}]{Dolgov:2002wy}
\bibinfo{author}{\bibfnamefont{A.~D.} \bibnamefont{Dolgov}},
  \bibinfo{journal}{Phys. Rept.} \textbf{\bibinfo{volume}{370}},
  \bibinfo{pages}{333} (\bibinfo{year}{2002}),
  \eprint{https://arxiv.org/abs/hep-ph/0202122}.

\bibitem[{\citenamefont{Bennett et~al.}(2020)\citenamefont{Bennett, Buldgen,
  Drewes, and Wong}}]{Bennett:2019ewm}
\bibinfo{author}{\bibfnamefont{J.~J.} \bibnamefont{Bennett}},
  \bibinfo{author}{\bibfnamefont{G.}~\bibnamefont{Buldgen}},
  \bibinfo{author}{\bibfnamefont{M.}~\bibnamefont{Drewes}}, \bibnamefont{and}
  \bibinfo{author}{\bibfnamefont{Y.~Y.~Y.} \bibnamefont{Wong}},
  \bibinfo{journal}{JCAP} \textbf{\bibinfo{volume}{03}}, \bibinfo{pages}{003}
  (\bibinfo{year}{2020}), \bibinfo{note}{[Addendum: JCAP 03, A01 (2021)]},
  \eprint{https://arxiv.org/abs/1911.04504}.

\bibitem[{\citenamefont{Mangano et~al.}(2005)\citenamefont{Mangano, Miele,
  Pastor, Pinto, Pisanti, and Serpico}}]{Mangano:2005cc}
\bibinfo{author}{\bibfnamefont{G.}~\bibnamefont{Mangano}},
  \bibinfo{author}{\bibfnamefont{G.}~\bibnamefont{Miele}},
  \bibinfo{author}{\bibfnamefont{S.}~\bibnamefont{Pastor}},
  \bibinfo{author}{\bibfnamefont{T.}~\bibnamefont{Pinto}},
  \bibinfo{author}{\bibfnamefont{O.}~\bibnamefont{Pisanti}}, \bibnamefont{and}
  \bibinfo{author}{\bibfnamefont{P.~D.} \bibnamefont{Serpico}},
  \bibinfo{journal}{Nucl. Phys. B} \textbf{\bibinfo{volume}{729}},
  \bibinfo{pages}{221} (\bibinfo{year}{2005}),
  \eprint{https://arxiv.org/abs/hep-ph/0506164}.

\bibitem[{\citenamefont{Grohs et~al.}(2016{\natexlab{a}})\citenamefont{Grohs,
  Fuller, Kishimoto, Paris, and Vlasenko}}]{Grohs16}
\bibinfo{author}{\bibfnamefont{E.}~\bibnamefont{Grohs}},
  \bibinfo{author}{\bibfnamefont{G.~M.} \bibnamefont{Fuller}},
  \bibinfo{author}{\bibfnamefont{C.~T.} \bibnamefont{Kishimoto}},
  \bibinfo{author}{\bibfnamefont{M.~W.} \bibnamefont{Paris}}, \bibnamefont{and}
  \bibinfo{author}{\bibfnamefont{A.}~\bibnamefont{Vlasenko}},
  \bibinfo{journal}{Phys. Rev. D} \textbf{\bibinfo{volume}{93}},
  \bibinfo{pages}{083522} (\bibinfo{year}{2016}{\natexlab{a}}),
  \urlprefix\url{https://link.aps.org/doi/10.1103/PhysRevD.93.083522}.

\bibitem[{\citenamefont{March-Russell et~al.}(1999)\citenamefont{March-Russell,
  Murayama, and Riotto}}]{March-Russell:1999hpw}
\bibinfo{author}{\bibfnamefont{J.}~\bibnamefont{March-Russell}},
  \bibinfo{author}{\bibfnamefont{H.}~\bibnamefont{Murayama}}, \bibnamefont{and}
  \bibinfo{author}{\bibfnamefont{A.}~\bibnamefont{Riotto}},
  \bibinfo{journal}{JHEP} \textbf{\bibinfo{volume}{11}}, \bibinfo{pages}{015}
  (\bibinfo{year}{1999}), \eprint{https://arxiv.org/abs/hep-ph/9908396}.

\bibitem[{\citenamefont{Kawasaki and Murai}(2022)}]{Kawasaki:2022hvx}
\bibinfo{author}{\bibfnamefont{M.}~\bibnamefont{Kawasaki}} \bibnamefont{and}
  \bibinfo{author}{\bibfnamefont{K.}~\bibnamefont{Murai}},
  \bibinfo{journal}{JCAP} \textbf{\bibinfo{volume}{08}}, \bibinfo{pages}{041}
  (\bibinfo{year}{2022}), \eprint{https://arxiv.org/abs/2203.09713}.

\bibitem[{\citenamefont{Escudero et~al.}(2023)\citenamefont{Escudero, Ibarra,
  and Maura}}]{Escudero:2022okz}
\bibinfo{author}{\bibfnamefont{M.}~\bibnamefont{Escudero}},
  \bibinfo{author}{\bibfnamefont{A.}~\bibnamefont{Ibarra}}, \bibnamefont{and}
  \bibinfo{author}{\bibfnamefont{V.}~\bibnamefont{Maura}},
  \bibinfo{journal}{Phys. Rev. D} \textbf{\bibinfo{volume}{107}},
  \bibinfo{pages}{035024} (\bibinfo{year}{2023}),
  \eprint{https://arxiv.org/abs/2208.03201}.

\bibitem[{\citenamefont{Lopez et~al.}(1997)\citenamefont{Lopez, Turner, and
  Gyuk}}]{Lopez:1997ki}
\bibinfo{author}{\bibfnamefont{R.~E.} \bibnamefont{Lopez}},
  \bibinfo{author}{\bibfnamefont{M.~S.} \bibnamefont{Turner}},
  \bibnamefont{and} \bibinfo{author}{\bibfnamefont{G.}~\bibnamefont{Gyuk}},
  \bibinfo{journal}{Phys. Rev. D} \textbf{\bibinfo{volume}{56}},
  \bibinfo{pages}{3191} (\bibinfo{year}{1997}),
  \eprint{https://arxiv.org/abs/astro-ph/9703065}.

\bibitem[{\citenamefont{Donoghue et~al.}(2014)\citenamefont{Donoghue, Golowich,
  and Holstein}}]{Donoghue:1992dd}
\bibinfo{author}{\bibfnamefont{J.~F.} \bibnamefont{Donoghue}},
  \bibinfo{author}{\bibfnamefont{E.}~\bibnamefont{Golowich}}, \bibnamefont{and}
  \bibinfo{author}{\bibfnamefont{B.~R.} \bibnamefont{Holstein}},
  \emph{\bibinfo{title}{{Dynamics of the Standard Model : Second edition}}},
  vol.~\bibinfo{volume}{2} (\bibinfo{publisher}{Oxford University Press},
  \bibinfo{year}{2014}), ISBN \bibinfo{isbn}{978-1-00-929103-3,
  978-1-00-929100-2, 978-1-00-929101-9}.

\bibitem[{\citenamefont{Bona et~al.}(2023)}]{UTfit:2022hsi}
\bibinfo{author}{\bibfnamefont{M.}~\bibnamefont{Bona}} \bibnamefont{et~al.}
  (\bibinfo{collaboration}{UTfit}), \bibinfo{journal}{Rend. Lincei Sci. Fis.
  Nat.} \textbf{\bibinfo{volume}{34}}, \bibinfo{pages}{37}
  (\bibinfo{year}{2023}), \eprint{https://arxiv.org/abs/2212.03894}.

\bibitem[{\citenamefont{Ivanov et~al.}(2013{\natexlab{a}})\citenamefont{Ivanov,
  Pitschmann, and Troitskaya}}]{PhysRevD.88.073002}
\bibinfo{author}{\bibfnamefont{A.~N.} \bibnamefont{Ivanov}},
  \bibinfo{author}{\bibfnamefont{M.}~\bibnamefont{Pitschmann}},
  \bibnamefont{and} \bibinfo{author}{\bibfnamefont{N.~I.}
  \bibnamefont{Troitskaya}}, \bibinfo{journal}{Phys. Rev. D}
  \textbf{\bibinfo{volume}{88}}, \bibinfo{pages}{073002}
  (\bibinfo{year}{2013}{\natexlab{a}}),
  \urlprefix\url{https://link.aps.org/doi/10.1103/PhysRevD.88.073002}.

\bibitem[{\citenamefont{Seckel}(1993)}]{Seckel:1993dc}
\bibinfo{author}{\bibfnamefont{D.}~\bibnamefont{Seckel}}
  (\bibinfo{year}{1993}), \eprint{https://arxiv.org/abs/hep-ph/9305311}.

\bibitem[{\citenamefont{Wilkinson}(1982)}]{Wilkinson:1982hu}
\bibinfo{author}{\bibfnamefont{D.~H.} \bibnamefont{Wilkinson}},
  \bibinfo{journal}{Nucl. Phys. A} \textbf{\bibinfo{volume}{377}},
  \bibinfo{pages}{474} (\bibinfo{year}{1982}).

\bibitem[{\citenamefont{Marciano and Sirlin}(2006)}]{Marciano:2005ec}
\bibinfo{author}{\bibfnamefont{W.~J.} \bibnamefont{Marciano}} \bibnamefont{and}
  \bibinfo{author}{\bibfnamefont{A.}~\bibnamefont{Sirlin}},
  \bibinfo{journal}{Phys. Rev. Lett.} \textbf{\bibinfo{volume}{96}},
  \bibinfo{pages}{032002} (\bibinfo{year}{2006}),
  \eprint{https://arxiv.org/abs/hep-ph/0510099}.

\bibitem[{\citenamefont{Cirigliano
  et~al.}(2023{\natexlab{a}})\citenamefont{Cirigliano, Dekens, Mereghetti, and
  Tomalak}}]{Cirigliano:2023fnz}
\bibinfo{author}{\bibfnamefont{V.}~\bibnamefont{Cirigliano}},
  \bibinfo{author}{\bibfnamefont{W.}~\bibnamefont{Dekens}},
  \bibinfo{author}{\bibfnamefont{E.}~\bibnamefont{Mereghetti}},
  \bibnamefont{and} \bibinfo{author}{\bibfnamefont{O.}~\bibnamefont{Tomalak}}
  (\bibinfo{year}{2023}{\natexlab{a}}),
  \eprint{https://arxiv.org/abs/2306.03138}.

\bibitem[{\citenamefont{Gonzalez et~al.}(2021)}]{UCNt:2021pcg}
\bibinfo{author}{\bibfnamefont{F.~M.} \bibnamefont{Gonzalez}}
  \bibnamefont{et~al.} (\bibinfo{collaboration}{UCN\ensuremath{\tau}}),
  \bibinfo{journal}{Phys. Rev. Lett.} \textbf{\bibinfo{volume}{127}},
  \bibinfo{pages}{162501} (\bibinfo{year}{2021}),
  \eprint{https://arxiv.org/abs/2106.10375}.

\bibitem[{\citenamefont{Yue et~al.}(2013)\citenamefont{Yue, Dewey, Gilliam,
  Greene, Laptev, Nico, Snow, and Wietfeldt}}]{Yue:2013qrc}
\bibinfo{author}{\bibfnamefont{A.~T.} \bibnamefont{Yue}},
  \bibinfo{author}{\bibfnamefont{M.~S.} \bibnamefont{Dewey}},
  \bibinfo{author}{\bibfnamefont{D.~M.} \bibnamefont{Gilliam}},
  \bibinfo{author}{\bibfnamefont{G.~L.} \bibnamefont{Greene}},
  \bibinfo{author}{\bibfnamefont{A.~B.} \bibnamefont{Laptev}},
  \bibinfo{author}{\bibfnamefont{J.~S.} \bibnamefont{Nico}},
  \bibinfo{author}{\bibfnamefont{W.~M.} \bibnamefont{Snow}}, \bibnamefont{and}
  \bibinfo{author}{\bibfnamefont{F.~E.} \bibnamefont{Wietfeldt}},
  \bibinfo{journal}{Phys. Rev. Lett.} \textbf{\bibinfo{volume}{111}},
  \bibinfo{pages}{222501} (\bibinfo{year}{2013}),
  \eprint{https://arxiv.org/abs/1309.2623}.

\bibitem[{\citenamefont{Chowdhury and Ipek}(2022)}]{Chowdhury:2022ahn}
\bibinfo{author}{\bibfnamefont{T.}~\bibnamefont{Chowdhury}} \bibnamefont{and}
  \bibinfo{author}{\bibfnamefont{S.}~\bibnamefont{Ipek}}
  (\bibinfo{year}{2022}), \eprint{https://arxiv.org/abs/2210.12031}.

\bibitem[{\citenamefont{Cirigliano
  et~al.}(2023{\natexlab{b}})\citenamefont{Cirigliano, Crivellin, Hoferichter,
  and Moulson}}]{Cirigliano:2022yyo}
\bibinfo{author}{\bibfnamefont{V.}~\bibnamefont{Cirigliano}},
  \bibinfo{author}{\bibfnamefont{A.}~\bibnamefont{Crivellin}},
  \bibinfo{author}{\bibfnamefont{M.}~\bibnamefont{Hoferichter}},
  \bibnamefont{and} \bibinfo{author}{\bibfnamefont{M.}~\bibnamefont{Moulson}},
  \bibinfo{journal}{Phys. Lett. B} \textbf{\bibinfo{volume}{838}},
  \bibinfo{pages}{137748} (\bibinfo{year}{2023}{\natexlab{b}}),
  \eprint{https://arxiv.org/abs/2208.11707}.

\bibitem[{\citenamefont{Sirlin}(1967)}]{Sirlin:1967zza}
\bibinfo{author}{\bibfnamefont{A.}~\bibnamefont{Sirlin}},
  \bibinfo{journal}{Phys. Rev.} \textbf{\bibinfo{volume}{164}},
  \bibinfo{pages}{1767} (\bibinfo{year}{1967}).

\bibitem[{\citenamefont{Abers et~al.}(1968)\citenamefont{Abers, Dicus, Norton,
  and Quinn}}]{Abers:1968zz}
\bibinfo{author}{\bibfnamefont{E.~S.} \bibnamefont{Abers}},
  \bibinfo{author}{\bibfnamefont{D.~A.} \bibnamefont{Dicus}},
  \bibinfo{author}{\bibfnamefont{R.~E.} \bibnamefont{Norton}},
  \bibnamefont{and} \bibinfo{author}{\bibfnamefont{H.~R.} \bibnamefont{Quinn}},
  \bibinfo{journal}{Phys. Rev.} \textbf{\bibinfo{volume}{167}},
  \bibinfo{pages}{1461} (\bibinfo{year}{1968}).

\bibitem[{\citenamefont{Dicus et~al.}(1982)\citenamefont{Dicus, Kolb, Gleeson,
  Sudarshan, Teplitz, and Turner}}]{Dicus:1982bz}
\bibinfo{author}{\bibfnamefont{D.~A.} \bibnamefont{Dicus}},
  \bibinfo{author}{\bibfnamefont{E.~W.} \bibnamefont{Kolb}},
  \bibinfo{author}{\bibfnamefont{A.~M.} \bibnamefont{Gleeson}},
  \bibinfo{author}{\bibfnamefont{E.~C.~G.} \bibnamefont{Sudarshan}},
  \bibinfo{author}{\bibfnamefont{V.~L.} \bibnamefont{Teplitz}},
  \bibnamefont{and} \bibinfo{author}{\bibfnamefont{M.~S.}
  \bibnamefont{Turner}}, \bibinfo{journal}{Phys. Rev. D}
  \textbf{\bibinfo{volume}{26}}, \bibinfo{pages}{2694} (\bibinfo{year}{1982}).

\bibitem[{\citenamefont{Ivanov et~al.}(2013{\natexlab{b}})\citenamefont{Ivanov,
  Pitschmann, and Troitskaya}}]{Ivanov:2012qe}
\bibinfo{author}{\bibfnamefont{A.~N.} \bibnamefont{Ivanov}},
  \bibinfo{author}{\bibfnamefont{M.}~\bibnamefont{Pitschmann}},
  \bibnamefont{and} \bibinfo{author}{\bibfnamefont{N.~I.}
  \bibnamefont{Troitskaya}}, \bibinfo{journal}{Phys. Rev. D}
  \textbf{\bibinfo{volume}{88}}, \bibinfo{pages}{073002}
  (\bibinfo{year}{2013}{\natexlab{b}}),
  \eprint{https://arxiv.org/abs/1212.0332}.

\bibitem[{\citenamefont{Brown and Sawyer}(2001)}]{Brown:2000cp}
\bibinfo{author}{\bibfnamefont{L.~S.} \bibnamefont{Brown}} \bibnamefont{and}
  \bibinfo{author}{\bibfnamefont{R.~F.} \bibnamefont{Sawyer}},
  \bibinfo{journal}{Phys. Rev. D} \textbf{\bibinfo{volume}{63}},
  \bibinfo{pages}{083503} (\bibinfo{year}{2001}),
  \eprint{https://arxiv.org/abs/astro-ph/0006370}.

\bibitem[{\citenamefont{Grohs et~al.}(2016{\natexlab{b}})\citenamefont{Grohs,
  Fuller, Kishimoto, Paris, and Vlasenko}}]{Grohs:2015tfy}
\bibinfo{author}{\bibfnamefont{E.}~\bibnamefont{Grohs}},
  \bibinfo{author}{\bibfnamefont{G.~M.} \bibnamefont{Fuller}},
  \bibinfo{author}{\bibfnamefont{C.~T.} \bibnamefont{Kishimoto}},
  \bibinfo{author}{\bibfnamefont{M.~W.} \bibnamefont{Paris}}, \bibnamefont{and}
  \bibinfo{author}{\bibfnamefont{A.}~\bibnamefont{Vlasenko}},
  \bibinfo{journal}{Phys. Rev. D} \textbf{\bibinfo{volume}{93}},
  \bibinfo{pages}{083522} (\bibinfo{year}{2016}{\natexlab{b}}),
  \eprint{https://arxiv.org/abs/1512.02205}.

\bibitem[{\citenamefont{Froustey and Pitrou}(2020)}]{Froustey:2019owm}
\bibinfo{author}{\bibfnamefont{J.}~\bibnamefont{Froustey}} \bibnamefont{and}
  \bibinfo{author}{\bibfnamefont{C.}~\bibnamefont{Pitrou}},
  \bibinfo{journal}{Phys. Rev. D} \textbf{\bibinfo{volume}{101}},
  \bibinfo{pages}{043524} (\bibinfo{year}{2020}),
  \eprint{https://arxiv.org/abs/1912.09378}.

\bibitem[{\citenamefont{Fowler et~al.}(1967)\citenamefont{Fowler, Caughlan, and
  Zimmerman}}]{Fowler:1967ty}
\bibinfo{author}{\bibfnamefont{W.~A.} \bibnamefont{Fowler}},
  \bibinfo{author}{\bibfnamefont{G.~R.} \bibnamefont{Caughlan}},
  \bibnamefont{and} \bibinfo{author}{\bibfnamefont{B.~A.}
  \bibnamefont{Zimmerman}}, \bibinfo{journal}{Ann. Rev. Astron. Astrophys.}
  \textbf{\bibinfo{volume}{5}}, \bibinfo{pages}{525} (\bibinfo{year}{1967}).

\bibitem[{\citenamefont{Wagoner}(1969)}]{Wagoner69}
\bibinfo{author}{\bibfnamefont{R.~V.} \bibnamefont{Wagoner}},
  \bibinfo{journal}{Astrophys. J., 18: Suppl. Ser., 162, 247-95(June 1969).}
  \textbf{\bibinfo{volume}{18}} (\bibinfo{year}{1969}), ISSN
  \bibinfo{issn}{0067--0049},
  \urlprefix\url{https://www.osti.gov/biblio/4772978}.

\bibitem[{\citenamefont{Angulo et~al.}(1999)}]{Angulo:1999zz}
\bibinfo{author}{\bibfnamefont{C.}~\bibnamefont{Angulo}} \bibnamefont{et~al.},
  \bibinfo{journal}{Nucl. Phys. A} \textbf{\bibinfo{volume}{656}},
  \bibinfo{pages}{3} (\bibinfo{year}{1999}).

\bibitem[{\citenamefont{Serpico et~al.}(2004)\citenamefont{Serpico, Esposito,
  Iocco, Mangano, Miele, and Pisanti}}]{Serpico:2004gx}
\bibinfo{author}{\bibfnamefont{P.~D.} \bibnamefont{Serpico}},
  \bibinfo{author}{\bibfnamefont{S.}~\bibnamefont{Esposito}},
  \bibinfo{author}{\bibfnamefont{F.}~\bibnamefont{Iocco}},
  \bibinfo{author}{\bibfnamefont{G.}~\bibnamefont{Mangano}},
  \bibinfo{author}{\bibfnamefont{G.}~\bibnamefont{Miele}}, \bibnamefont{and}
  \bibinfo{author}{\bibfnamefont{O.}~\bibnamefont{Pisanti}},
  \bibinfo{journal}{JCAP} \textbf{\bibinfo{volume}{12}}, \bibinfo{pages}{010}
  (\bibinfo{year}{2004}), \eprint{https://arxiv.org/abs/astro-ph/0408076}.

\bibitem[{\citenamefont{Kondev et~al.}(2021)\citenamefont{Kondev, Wang, Huang,
  Naimi, and Audi}}]{Kondev:2021lzi}
\bibinfo{author}{\bibfnamefont{F.~G.} \bibnamefont{Kondev}},
  \bibinfo{author}{\bibfnamefont{M.}~\bibnamefont{Wang}},
  \bibinfo{author}{\bibfnamefont{W.~J.} \bibnamefont{Huang}},
  \bibinfo{author}{\bibfnamefont{S.}~\bibnamefont{Naimi}}, \bibnamefont{and}
  \bibinfo{author}{\bibfnamefont{G.}~\bibnamefont{Audi}},
  \bibinfo{journal}{Chin. Phys. C} \textbf{\bibinfo{volume}{45}},
  \bibinfo{pages}{030001} (\bibinfo{year}{2021}).

\bibitem[{\citenamefont{Fields and Olive}(2022)}]{Fields:2022mpw}
\bibinfo{author}{\bibfnamefont{B.~D.} \bibnamefont{Fields}} \bibnamefont{and}
  \bibinfo{author}{\bibfnamefont{K.~A.} \bibnamefont{Olive}},
  \bibinfo{journal}{JCAP} \textbf{\bibinfo{volume}{10}}, \bibinfo{pages}{078}
  (\bibinfo{year}{2022}), \eprint{https://arxiv.org/abs/2204.03167}.

\bibitem[{\citenamefont{Xu et~al.}(2013{\natexlab{a}})\citenamefont{Xu,
  Takahashi, Goriely, Arnould, Ohta, and Utsunomiya}}]{Xu:2013fha}
\bibinfo{author}{\bibfnamefont{Y.}~\bibnamefont{Xu}},
  \bibinfo{author}{\bibfnamefont{K.}~\bibnamefont{Takahashi}},
  \bibinfo{author}{\bibfnamefont{S.}~\bibnamefont{Goriely}},
  \bibinfo{author}{\bibfnamefont{M.}~\bibnamefont{Arnould}},
  \bibinfo{author}{\bibfnamefont{M.}~\bibnamefont{Ohta}}, \bibnamefont{and}
  \bibinfo{author}{\bibfnamefont{H.}~\bibnamefont{Utsunomiya}},
  \bibinfo{journal}{Nucl. Phys. A} \textbf{\bibinfo{volume}{918}},
  \bibinfo{pages}{61} (\bibinfo{year}{2013}{\natexlab{a}}),
  \eprint{https://arxiv.org/abs/1310.7099}.

\bibitem[{\citenamefont{Descouvemont et~al.}(2004)\citenamefont{Descouvemont,
  Adahchour, Angulo, Coc, and Vangioni-Flam}}]{Descouvemont04}
\bibinfo{author}{\bibfnamefont{P.}~\bibnamefont{Descouvemont}},
  \bibinfo{author}{\bibfnamefont{A.}~\bibnamefont{Adahchour}},
  \bibinfo{author}{\bibfnamefont{C.}~\bibnamefont{Angulo}},
  \bibinfo{author}{\bibfnamefont{A.}~\bibnamefont{Coc}}, \bibnamefont{and}
  \bibinfo{author}{\bibfnamefont{E.}~\bibnamefont{Vangioni-Flam}},
  \bibinfo{journal}{Atomic Data and Nuclear Data Tables}
  \textbf{\bibinfo{volume}{88}}, \bibinfo{pages}{203–236}
  (\bibinfo{year}{2004}), ISSN \bibinfo{issn}{0092-640X},
  \urlprefix\url{http://dx.doi.org/10.1016/j.adt.2004.08.001}.

\bibitem[{\citenamefont{Longland
  et~al.}(2010{\natexlab{a}})\citenamefont{Longland, Iliadis, Champagne,
  Newton, Ugalde, Coc, and Fitzgerald}}]{Longland10}
\bibinfo{author}{\bibfnamefont{R.}~\bibnamefont{Longland}},
  \bibinfo{author}{\bibfnamefont{C.}~\bibnamefont{Iliadis}},
  \bibinfo{author}{\bibfnamefont{A.}~\bibnamefont{Champagne}},
  \bibinfo{author}{\bibfnamefont{J.}~\bibnamefont{Newton}},
  \bibinfo{author}{\bibfnamefont{C.}~\bibnamefont{Ugalde}},
  \bibinfo{author}{\bibfnamefont{A.}~\bibnamefont{Coc}}, \bibnamefont{and}
  \bibinfo{author}{\bibfnamefont{R.}~\bibnamefont{Fitzgerald}},
  \bibinfo{journal}{Nuclear Physics A} \textbf{\bibinfo{volume}{841}},
  \bibinfo{pages}{1} (\bibinfo{year}{2010}{\natexlab{a}}), ISSN
  \bibinfo{issn}{0375-9474}, \bibinfo{note}{the 2010 Evaluation of Monte Carlo
  based Thermonuclear Reaction Rates},
  \urlprefix\url{https://www.sciencedirect.com/science/article/pii/S0375947410004185}.

\bibitem[{\citenamefont{Iliadis et~al.}(2016)\citenamefont{Iliadis, Anderson,
  Coc, Timmes, and Starrfield}}]{Iliadis16}
\bibinfo{author}{\bibfnamefont{C.}~\bibnamefont{Iliadis}},
  \bibinfo{author}{\bibfnamefont{K.~S.} \bibnamefont{Anderson}},
  \bibinfo{author}{\bibfnamefont{A.}~\bibnamefont{Coc}},
  \bibinfo{author}{\bibfnamefont{F.~X.} \bibnamefont{Timmes}},
  \bibnamefont{and}
  \bibinfo{author}{\bibfnamefont{S.}~\bibnamefont{Starrfield}},
  \bibinfo{journal}{The Astrophysical Journal} \textbf{\bibinfo{volume}{831}},
  \bibinfo{pages}{107} (\bibinfo{year}{2016}), ISSN \bibinfo{issn}{1538-4357},
  \urlprefix\url{http://dx.doi.org/10.3847/0004-637X/831/1/107}.

\bibitem[{\citenamefont{I{\~{n}}esta et~al.}(2017)\citenamefont{I{\~{n}}esta,
  Iliadis, and Coc}}]{Gomez17}
\bibinfo{author}{\bibfnamefont{{\'{A} }.~G.} \bibnamefont{I{\~{n}}esta}},
  \bibinfo{author}{\bibfnamefont{C.}~\bibnamefont{Iliadis}}, \bibnamefont{and}
  \bibinfo{author}{\bibfnamefont{A.}~\bibnamefont{Coc}}, \bibinfo{journal}{The
  Astrophysical Journal} \textbf{\bibinfo{volume}{849}}, \bibinfo{pages}{134}
  (\bibinfo{year}{2017}),
  \urlprefix\url{https://doi.org/10.3847%2F1538-4357%2Faa9025}.

\bibitem[{\citenamefont{Coc et~al.}(2012{\natexlab{a}})\citenamefont{Coc,
  Goriely, Xu, Saimpert, and Vangioni}}]{Coc:2011az}
\bibinfo{author}{\bibfnamefont{A.}~\bibnamefont{Coc}},
  \bibinfo{author}{\bibfnamefont{S.}~\bibnamefont{Goriely}},
  \bibinfo{author}{\bibfnamefont{Y.}~\bibnamefont{Xu}},
  \bibinfo{author}{\bibfnamefont{M.}~\bibnamefont{Saimpert}}, \bibnamefont{and}
  \bibinfo{author}{\bibfnamefont{E.}~\bibnamefont{Vangioni}},
  \bibinfo{journal}{Astrophys. J.} \textbf{\bibinfo{volume}{744}},
  \bibinfo{pages}{158} (\bibinfo{year}{2012}{\natexlab{a}}),
  \eprint{https://arxiv.org/abs/1107.1117}.

\bibitem[{\citenamefont{Pisanti et~al.}(2021)\citenamefont{Pisanti, Mangano,
  Miele, and Mazzella}}]{Pisanti:2020efz}
\bibinfo{author}{\bibfnamefont{O.}~\bibnamefont{Pisanti}},
  \bibinfo{author}{\bibfnamefont{G.}~\bibnamefont{Mangano}},
  \bibinfo{author}{\bibfnamefont{G.}~\bibnamefont{Miele}}, \bibnamefont{and}
  \bibinfo{author}{\bibfnamefont{P.}~\bibnamefont{Mazzella}},
  \bibinfo{journal}{JCAP} \textbf{\bibinfo{volume}{04}}, \bibinfo{pages}{020}
  (\bibinfo{year}{2021}), \eprint{https://arxiv.org/abs/2011.11537}.

\bibitem[{\citenamefont{Cyburt}(2004)}]{Cyburt:2004cq}
\bibinfo{author}{\bibfnamefont{R.~H.} \bibnamefont{Cyburt}},
  \bibinfo{journal}{Phys. Rev. D} \textbf{\bibinfo{volume}{70}},
  \bibinfo{pages}{023505} (\bibinfo{year}{2004}),
  \eprint{https://arxiv.org/abs/astro-ph/0401091}.

\bibitem[{\citenamefont{Fields et~al.}(2020)\citenamefont{Fields, Olive, Yeh,
  and Young}}]{Fields:2019pfx}
\bibinfo{author}{\bibfnamefont{B.~D.} \bibnamefont{Fields}},
  \bibinfo{author}{\bibfnamefont{K.~A.} \bibnamefont{Olive}},
  \bibinfo{author}{\bibfnamefont{T.-H.} \bibnamefont{Yeh}}, \bibnamefont{and}
  \bibinfo{author}{\bibfnamefont{C.}~\bibnamefont{Young}},
  \bibinfo{journal}{JCAP} \textbf{\bibinfo{volume}{03}}, \bibinfo{pages}{010}
  (\bibinfo{year}{2020}), \bibinfo{note}{[Erratum: JCAP 11, E02 (2020)]},
  \eprint{https://arxiv.org/abs/1912.01132}.

\bibitem[{\citenamefont{Iliadis and Coc}(2020)}]{Iliadis:2020jtc}
\bibinfo{author}{\bibfnamefont{C.}~\bibnamefont{Iliadis}} \bibnamefont{and}
  \bibinfo{author}{\bibfnamefont{A.}~\bibnamefont{Coc}},
  \bibinfo{journal}{Astrophys. J.} \textbf{\bibinfo{volume}{901}},
  \bibinfo{pages}{127} (\bibinfo{year}{2020}),
  \eprint{https://arxiv.org/abs/2008.12200}.

\bibitem[{\citenamefont{Coc and Vangioni}(2010)}]{Coc:2010zz}
\bibinfo{author}{\bibfnamefont{A.}~\bibnamefont{Coc}} \bibnamefont{and}
  \bibinfo{author}{\bibfnamefont{E.}~\bibnamefont{Vangioni}},
  \bibinfo{journal}{J. Phys. Conf. Ser.} \textbf{\bibinfo{volume}{202}},
  \bibinfo{pages}{012001} (\bibinfo{year}{2010}).

\bibitem[{\citenamefont{Mossa et~al.}(2020)}]{Mossa:2020gjc}
\bibinfo{author}{\bibfnamefont{V.}~\bibnamefont{Mossa}} \bibnamefont{et~al.},
  \bibinfo{journal}{Nature} \textbf{\bibinfo{volume}{587}},
  \bibinfo{pages}{210} (\bibinfo{year}{2020}).

\bibitem[{\citenamefont{Longland
  et~al.}(2010{\natexlab{b}})\citenamefont{Longland, Iliadis, Champagne,
  Newton, Ugalde, Coc, and Fitzgerald}}]{Longland:2010gs}
\bibinfo{author}{\bibfnamefont{R.}~\bibnamefont{Longland}},
  \bibinfo{author}{\bibfnamefont{C.}~\bibnamefont{Iliadis}},
  \bibinfo{author}{\bibfnamefont{A.}~\bibnamefont{Champagne}},
  \bibinfo{author}{\bibfnamefont{J.}~\bibnamefont{Newton}},
  \bibinfo{author}{\bibfnamefont{C.}~\bibnamefont{Ugalde}},
  \bibinfo{author}{\bibfnamefont{A.}~\bibnamefont{Coc}}, \bibnamefont{and}
  \bibinfo{author}{\bibfnamefont{R.}~\bibnamefont{Fitzgerald}},
  \bibinfo{journal}{Nucl. Phys. A} \textbf{\bibinfo{volume}{841}},
  \bibinfo{pages}{1} (\bibinfo{year}{2010}{\natexlab{b}}),
  \eprint{https://arxiv.org/abs/1004.4136}.

\bibitem[{\citenamefont{Coc et~al.}(2014)\citenamefont{Coc, Uzan, and
  Vangioni}}]{Coc:2014oia}
\bibinfo{author}{\bibfnamefont{A.}~\bibnamefont{Coc}},
  \bibinfo{author}{\bibfnamefont{J.-P.} \bibnamefont{Uzan}}, \bibnamefont{and}
  \bibinfo{author}{\bibfnamefont{E.}~\bibnamefont{Vangioni}},
  \bibinfo{journal}{JCAP} \textbf{\bibinfo{volume}{10}}, \bibinfo{pages}{050}
  (\bibinfo{year}{2014}), \eprint{https://arxiv.org/abs/1403.6694}.

\bibitem[{\citenamefont{Sallaska et~al.}(2013)\citenamefont{Sallaska, Iliadis,
  Champagne, Goriely, Starrfield, and Timmes}}]{Sallaska:2013xqa}
\bibinfo{author}{\bibfnamefont{A.~L.} \bibnamefont{Sallaska}},
  \bibinfo{author}{\bibfnamefont{C.}~\bibnamefont{Iliadis}},
  \bibinfo{author}{\bibfnamefont{A.~E.} \bibnamefont{Champagne}},
  \bibinfo{author}{\bibfnamefont{S.}~\bibnamefont{Goriely}},
  \bibinfo{author}{\bibfnamefont{S.}~\bibnamefont{Starrfield}},
  \bibnamefont{and} \bibinfo{author}{\bibfnamefont{F.~X.}
  \bibnamefont{Timmes}}, \bibinfo{journal}{Astrophys. J. Suppl.}
  \textbf{\bibinfo{volume}{207}}, \bibinfo{pages}{18} (\bibinfo{year}{2013}),
  \eprint{https://arxiv.org/abs/1304.7811}.

\bibitem[{\citenamefont{Froustey}(2022)}]{Froustey:2022sla}
\bibinfo{author}{\bibfnamefont{J.}~\bibnamefont{Froustey}}, Ph.D. thesis,
  \bibinfo{school}{Institut d'Astrophysique de Paris, France, Paris, Inst.
  Astrophys.} (\bibinfo{year}{2022}),
  \eprint{https://arxiv.org/abs/2209.06672}.

\bibitem[{\citenamefont{Pitrou et~al.}(2018{\natexlab{b}})\citenamefont{Pitrou,
  Coc, Uzan, and Vangioni}}]{Pitrou18}
\bibinfo{author}{\bibfnamefont{C.}~\bibnamefont{Pitrou}},
  \bibinfo{author}{\bibfnamefont{A.}~\bibnamefont{Coc}},
  \bibinfo{author}{\bibfnamefont{J.-P.} \bibnamefont{Uzan}}, \bibnamefont{and}
  \bibinfo{author}{\bibfnamefont{E.}~\bibnamefont{Vangioni}},
  \bibinfo{journal}{Physics Reports} \textbf{\bibinfo{volume}{754}},
  \bibinfo{pages}{1–66} (\bibinfo{year}{2018}{\natexlab{b}}), ISSN
  \bibinfo{issn}{0370-1573},
  \urlprefix\url{http://dx.doi.org/10.1016/j.physrep.2018.04.005}.

\bibitem[{\citenamefont{Pitrou and Pospelov}(2020)}]{Pitrou:2019pqh}
\bibinfo{author}{\bibfnamefont{C.}~\bibnamefont{Pitrou}} \bibnamefont{and}
  \bibinfo{author}{\bibfnamefont{M.}~\bibnamefont{Pospelov}},
  \bibinfo{journal}{Phys. Rev. C} \textbf{\bibinfo{volume}{102}},
  \bibinfo{pages}{015803} (\bibinfo{year}{2020}),
  \eprint{https://arxiv.org/abs/1904.07795}.

\bibitem[{\citenamefont{Cielo et~al.}(2023)\citenamefont{Cielo, Escudero,
  Mangano, and Pisanti}}]{Cielo:2023bqp}
\bibinfo{author}{\bibfnamefont{M.}~\bibnamefont{Cielo}},
  \bibinfo{author}{\bibfnamefont{M.}~\bibnamefont{Escudero}},
  \bibinfo{author}{\bibfnamefont{G.}~\bibnamefont{Mangano}}, \bibnamefont{and}
  \bibinfo{author}{\bibfnamefont{O.}~\bibnamefont{Pisanti}}
  (\bibinfo{year}{2023}), \eprint{https://arxiv.org/abs/2306.05460}.

\bibitem[{\citenamefont{Ando et~al.}(2006)\citenamefont{Ando, Cyburt, Hong, and
  Hyun}}]{Ando06}
\bibinfo{author}{\bibfnamefont{S.}~\bibnamefont{Ando}},
  \bibinfo{author}{\bibfnamefont{R.~H.} \bibnamefont{Cyburt}},
  \bibinfo{author}{\bibfnamefont{S.~W.} \bibnamefont{Hong}}, \bibnamefont{and}
  \bibinfo{author}{\bibfnamefont{C.~H.} \bibnamefont{Hyun}},
  \bibinfo{journal}{Phys. Rev. C} \textbf{\bibinfo{volume}{74}},
  \bibinfo{pages}{025809} (\bibinfo{year}{2006}),
  \eprint{https://arxiv.org/abs/nucl-th/0511074}.

\bibitem[{\citenamefont{Xu et~al.}(2013{\natexlab{b}})\citenamefont{Xu,
  Takahashi, Goriely, Arnould, Ohta, and Utsunomiya}}]{Xu13}
\bibinfo{author}{\bibfnamefont{Y.}~\bibnamefont{Xu}},
  \bibinfo{author}{\bibfnamefont{K.}~\bibnamefont{Takahashi}},
  \bibinfo{author}{\bibfnamefont{S.}~\bibnamefont{Goriely}},
  \bibinfo{author}{\bibfnamefont{M.}~\bibnamefont{Arnould}},
  \bibinfo{author}{\bibfnamefont{M.}~\bibnamefont{Ohta}}, \bibnamefont{and}
  \bibinfo{author}{\bibfnamefont{H.}~\bibnamefont{Utsunomiya}},
  \bibinfo{journal}{Nucl. Phys. A} \textbf{\bibinfo{volume}{918}},
  \bibinfo{pages}{61} (\bibinfo{year}{2013}{\natexlab{b}}),
  \eprint{https://arxiv.org/abs/1310.7099}.

\bibitem[{\citenamefont{Barbagallo et~al.}(2016)\citenamefont{Barbagallo,
  Musumarra, Cosentino, Maugeri, Heinitz, Mengoni, Dressler, Schumann,
  Käppeler, Colonna et~al.}}]{Barbagallo16}
\bibinfo{author}{\bibfnamefont{M.}~\bibnamefont{Barbagallo}},
  \bibinfo{author}{\bibfnamefont{A.}~\bibnamefont{Musumarra}},
  \bibinfo{author}{\bibfnamefont{L.}~\bibnamefont{Cosentino}},
  \bibinfo{author}{\bibfnamefont{E.}~\bibnamefont{Maugeri}},
  \bibinfo{author}{\bibfnamefont{S.}~\bibnamefont{Heinitz}},
  \bibinfo{author}{\bibfnamefont{A.}~\bibnamefont{Mengoni}},
  \bibinfo{author}{\bibfnamefont{R.}~\bibnamefont{Dressler}},
  \bibinfo{author}{\bibfnamefont{D.}~\bibnamefont{Schumann}},
  \bibinfo{author}{\bibfnamefont{F.}~\bibnamefont{Käppeler}},
  \bibinfo{author}{\bibfnamefont{N.}~\bibnamefont{Colonna}},
  \bibnamefont{et~al.}, \bibinfo{journal}{Physical Review Letters}
  \textbf{\bibinfo{volume}{117}} (\bibinfo{year}{2016}), ISSN
  \bibinfo{issn}{1079-7114},
  \urlprefix\url{http://dx.doi.org/10.1103/PhysRevLett.117.152701}.

\bibitem[{\citenamefont{Caughlan and Fowler}(1988)}]{Caughlan88}
\bibinfo{author}{\bibfnamefont{G.~R.} \bibnamefont{Caughlan}} \bibnamefont{and}
  \bibinfo{author}{\bibfnamefont{W.~A.} \bibnamefont{Fowler}},
  \bibinfo{journal}{Atomic Data and Nuclear Data Tables}
  \textbf{\bibinfo{volume}{40}}, \bibinfo{pages}{283} (\bibinfo{year}{1988}),
  ISSN \bibinfo{issn}{0092-640X},
  \urlprefix\url{https://www.sciencedirect.com/science/article/pii/0092640X88900095}.

\bibitem[{\citenamefont{Hammache et~al.}(2010)\citenamefont{Hammache, Heil,
  Typel, Galaviz, Sümmerer, Coc, Uhlig, Attallah, Caamano, Cortina
  et~al.}}]{Hammache10}
\bibinfo{author}{\bibfnamefont{F.}~\bibnamefont{Hammache}},
  \bibinfo{author}{\bibfnamefont{M.}~\bibnamefont{Heil}},
  \bibinfo{author}{\bibfnamefont{S.}~\bibnamefont{Typel}},
  \bibinfo{author}{\bibfnamefont{D.}~\bibnamefont{Galaviz}},
  \bibinfo{author}{\bibfnamefont{K.}~\bibnamefont{Sümmerer}},
  \bibinfo{author}{\bibfnamefont{A.}~\bibnamefont{Coc}},
  \bibinfo{author}{\bibfnamefont{F.}~\bibnamefont{Uhlig}},
  \bibinfo{author}{\bibfnamefont{F.}~\bibnamefont{Attallah}},
  \bibinfo{author}{\bibfnamefont{M.}~\bibnamefont{Caamano}},
  \bibinfo{author}{\bibfnamefont{D.}~\bibnamefont{Cortina}},
  \bibnamefont{et~al.}, \bibinfo{journal}{Physical Review C}
  \textbf{\bibinfo{volume}{82}} (\bibinfo{year}{2010}), ISSN
  \bibinfo{issn}{1089-490X},
  \urlprefix\url{http://dx.doi.org/10.1103/PhysRevC.82.065803}.

\bibitem[{\citenamefont{Goriely et~al.}(2008)\citenamefont{Goriely, Hilaire,
  and Koning}}]{Goriely08}
\bibinfo{author}{\bibfnamefont{S.}~\bibnamefont{Goriely}},
  \bibinfo{author}{\bibfnamefont{S.}~\bibnamefont{Hilaire}}, \bibnamefont{and}
  \bibinfo{author}{\bibfnamefont{A.~J.} \bibnamefont{Koning}},
  \bibinfo{journal}{Astron. Astrophys.} \textbf{\bibinfo{volume}{487}},
  \bibinfo{pages}{767} (\bibinfo{year}{2008}),
  \eprint{https://arxiv.org/abs/0806.2239}.

\bibitem[{\citenamefont{Coc et~al.}(2012{\natexlab{b}})\citenamefont{Coc,
  Goriely, Xu, Saimpert, and Vangioni}}]{Coc11}
\bibinfo{author}{\bibfnamefont{A.}~\bibnamefont{Coc}},
  \bibinfo{author}{\bibfnamefont{S.}~\bibnamefont{Goriely}},
  \bibinfo{author}{\bibfnamefont{Y.}~\bibnamefont{Xu}},
  \bibinfo{author}{\bibfnamefont{M.}~\bibnamefont{Saimpert}}, \bibnamefont{and}
  \bibinfo{author}{\bibfnamefont{E.}~\bibnamefont{Vangioni}},
  \bibinfo{journal}{Astrophys. J.} \textbf{\bibinfo{volume}{744}},
  \bibinfo{pages}{158} (\bibinfo{year}{2012}{\natexlab{b}}),
  \eprint{https://arxiv.org/abs/1107.1117}.

\bibitem[{\citenamefont{Nagai et~al.}(2006)}]{Nagai06}
\bibinfo{author}{\bibfnamefont{Y.}~\bibnamefont{Nagai}} \bibnamefont{et~al.},
  \bibinfo{journal}{Phys. Rev. C} \textbf{\bibinfo{volume}{74}},
  \bibinfo{pages}{025804} (\bibinfo{year}{2006}).

\bibitem[{\citenamefont{{Malaney} and {Fowler}}(1989)}]{Malaney89}
\bibinfo{author}{\bibfnamefont{R.~A.} \bibnamefont{{Malaney}}}
  \bibnamefont{and} \bibinfo{author}{\bibfnamefont{W.~A.}
  \bibnamefont{{Fowler}}}, \bibinfo{journal}{\apjl}
  \textbf{\bibinfo{volume}{345}}, \bibinfo{pages}{L5} (\bibinfo{year}{1989}).

\bibitem[{\citenamefont{Hashimoto et~al.}(2009)}]{Hashimoto09}
\bibinfo{author}{\bibfnamefont{T.}~\bibnamefont{Hashimoto}}
  \bibnamefont{et~al.}, \bibinfo{journal}{Phys. Lett. B}
  \textbf{\bibinfo{volume}{674}}, \bibinfo{pages}{276} (\bibinfo{year}{2009}).

\bibitem[{\citenamefont{Fukugita and Kajino}(1990)}]{Fukugita90}
\bibinfo{author}{\bibfnamefont{M.}~\bibnamefont{Fukugita}} \bibnamefont{and}
  \bibinfo{author}{\bibfnamefont{T.}~\bibnamefont{Kajino}},
  \bibinfo{journal}{Phys. Rev. D} \textbf{\bibinfo{volume}{42}},
  \bibinfo{pages}{4251} (\bibinfo{year}{1990}),
  \urlprefix\url{https://link.aps.org/doi/10.1103/PhysRevD.42.4251}.

\bibitem[{\citenamefont{Heil et~al.}(1989)\citenamefont{Heil, Ahrens, Andresen,
  Bornheimer, Conrath, Dietz, Gasteyer, Gessinger, Hartmann, Jethwa
  et~al.}}]{Heil89}
\bibinfo{author}{\bibfnamefont{W.}~\bibnamefont{Heil}},
  \bibinfo{author}{\bibfnamefont{J.}~\bibnamefont{Ahrens}},
  \bibinfo{author}{\bibfnamefont{H.}~\bibnamefont{Andresen}},
  \bibinfo{author}{\bibfnamefont{A.}~\bibnamefont{Bornheimer}},
  \bibinfo{author}{\bibfnamefont{D.}~\bibnamefont{Conrath}},
  \bibinfo{author}{\bibfnamefont{K.-J.} \bibnamefont{Dietz}},
  \bibinfo{author}{\bibfnamefont{W.}~\bibnamefont{Gasteyer}},
  \bibinfo{author}{\bibfnamefont{H.-J.} \bibnamefont{Gessinger}},
  \bibinfo{author}{\bibfnamefont{W.}~\bibnamefont{Hartmann}},
  \bibinfo{author}{\bibfnamefont{J.}~\bibnamefont{Jethwa}},
  \bibnamefont{et~al.}, \bibinfo{journal}{Nuclear Physics B}
  \textbf{\bibinfo{volume}{327}}, \bibinfo{pages}{1} (\bibinfo{year}{1989}),
  ISSN \bibinfo{issn}{0550-3213},
  \urlprefix\url{https://www.sciencedirect.com/science/article/pii/0550321389902848}.

\bibitem[{\citenamefont{Mendes et~al.}(2018)\citenamefont{Mendes,
  L\'epine-Szily, Descouvemont, Lichtenth\"aler, Guimar\~aes, de~Faria,
  Barioni, Pires, Morcelle, Pampa~Condori et~al.}}]{Mendes18}
\bibinfo{author}{\bibfnamefont{D.~R.} \bibnamefont{Mendes}},
  \bibinfo{author}{\bibfnamefont{A.}~\bibnamefont{L\'epine-Szily}},
  \bibinfo{author}{\bibfnamefont{P.}~\bibnamefont{Descouvemont}},
  \bibinfo{author}{\bibfnamefont{R.}~\bibnamefont{Lichtenth\"aler}},
  \bibinfo{author}{\bibfnamefont{V.}~\bibnamefont{Guimar\~aes}},
  \bibinfo{author}{\bibfnamefont{P.~N.} \bibnamefont{de~Faria}},
  \bibinfo{author}{\bibfnamefont{A.}~\bibnamefont{Barioni}},
  \bibinfo{author}{\bibfnamefont{K.~C.~C.} \bibnamefont{Pires}},
  \bibinfo{author}{\bibfnamefont{V.}~\bibnamefont{Morcelle}},
  \bibinfo{author}{\bibfnamefont{R.}~\bibnamefont{Pampa~Condori}},
  \bibnamefont{et~al.}, \bibinfo{journal}{Phys. Rev. C}
  \textbf{\bibinfo{volume}{98}}, \bibinfo{pages}{069901}
  (\bibinfo{year}{2018}),
  \urlprefix\url{https://link.aps.org/doi/10.1103/PhysRevC.98.069901}.

\bibitem[{\citenamefont{Efros et~al.}(1996)\citenamefont{Efros, Balogh, Herndl,
  Hofinger, and Oberhummer}}]{Efros96}
\bibinfo{author}{\bibfnamefont{V.}~\bibnamefont{Efros}},
  \bibinfo{author}{\bibfnamefont{W.}~\bibnamefont{Balogh}},
  \bibinfo{author}{\bibfnamefont{H.}~\bibnamefont{Herndl}},
  \bibinfo{author}{\bibfnamefont{R.}~\bibnamefont{Hofinger}}, \bibnamefont{and}
  \bibinfo{author}{\bibfnamefont{H.}~\bibnamefont{Oberhummer}},
  \bibinfo{journal}{Zeitschrift für Physik A Hadrons and Nuclei}
  \textbf{\bibinfo{volume}{355}}, \bibinfo{pages}{101} (\bibinfo{year}{1996}).

\end{thebibliography}
\bibliographystyle{apsrev}

%------------------------------------------------------------------------------

\end{document}